\documentclass{elsart}

\usepackage{natbib}

\usepackage{graphicx}

\usepackage{amssymb}

\begin{document}

\begin{frontmatter}



\title{Where, oh where has the $r$-process gone?}


\author[MN]{Y.-Z. Qian\corauthref{cor}},
\ead{qian@physics.umn.edu}
\corauth[cor]{Corresponding author.}
\author[Cal]{G.J. Wasserburg}
\ead{gjw@gps.caltech.edu}

\address[MN]{School of Physics and Astronomy, University of Minnesota,
Minneapolis, MN 55455, USA}

\address[Cal]{The Lunatic Asylum, Division of Geological and Planetary
Sciences, California Institute of Technology, Pasadena, CA 91125}

\begin{abstract}
We present a review of the possible sources for $r$-process nuclei
($r$-nuclei). It is known that there is as yet no self-consistent 
mechanism to provide 
abundant neutrons for a robust $r$-process in the neutrino-driven winds 
from nascent neutron stars. We consider that the heavy $r$-nuclei with
mass numbers $A>130$ (Ba and above) cannot be produced in the 
neutrino-driven winds. Nonetheless, the $r$-process and the 
neutrino-driven winds may be directly or indirectly related by some 
unknown additional mechanism, which, for example, could provide
ejecta with very short dynamic timescales of $\lesssim 0.004$~s. 
This undetermined mechanism must supply a
neutron source within the same general
stellar sites that undergo core collapse 
to produce the neutron star. Observational data on low-metallicity stars 
in the Galactic halo show that sites producing the heavy $r$-nuclei 
do not produce Fe or any other
elements between N and Ge. Insofar as a forming neutron star is key to
producing the heavy $r$-nuclei, then the only possible sources
are supernovae resulting from collapse of O-Ne-Mg cores or 
accretion-induced collapse of white dwarfs, neither of which produce the
elements of the Fe group or those of intermediate mass (above C and N).
Observational evidence on $s$ and $r$-nuclei in low-metallicity stars with
high C and N abundances shows that the $r$-process is also active 
in binary systems. 

The nuclei with $A\sim 90$--110 produced by charged-particle 
reactions (CPR) in the neutrino-driven winds are in general present 
in metal-poor stars with high or low abundances of heavy 
$r$-nuclei. The CPR nuclei and the heavy $r$-nuclei
are not strongly coupled. 
Some metal-poor stars show extremely high enrichments of heavy 
$r$-nuclei and have established that the abundance patterns 
of these nuclei are universally close to
the solar abundance pattern of heavy $r$-nuclei. 

Using a template star with high
enrichments of heavy $r$-nuclei and another with low
enrichments we develop a two-component model based on the 
abundances of Eu (from sources for heavy $r$-nuclei) and Fe 
(from Fe core-collapse supernovae). This model gives very good 
quantitative predictions for the abundances of all the other elements in 
those metal-poor stars with [Fe/H]~$\lesssim -1.5$
for which the Eu and Fe abundances are known. We attribute
the CPR elements such as Sr, Y, and Zr to reactions in
the neutrino-driven winds from a nascent neutron star and the heavy
$r$-nuclei to the hypothecated true ``$r$-process.''
The CPR nuclei should be produced whenever a neutron star is formed
regardless of whether heavy $r$-nuclei are produced or not. 
Using the two-component model we estimate the yield of the CPR element
Sr to be $\sim 10^{-6}\,M_\odot$ for a single neutron star formation 
event. Self-consistent astrophysical models are needed to establish that
the CPR nuclei are common to the neutron stars produced in
both sources for the heavy $r$-nuclei and those for Fe. We show that 
the observational data appear fully consistent with the two-component
model. The specific mechanism and site for the production of heavy
$r$-nuclei remains to be found.
\end{abstract}

\begin{keyword}
$r$-process nucleosynthesis \sep abundances in metal-poor stars \sep
Galactic chemical evolution \sep neutron star formation

\PACS 26.30.+k \sep 26.45.+h \sep 97.60.Jd
\end{keyword}

\end{frontmatter}

\section{Introduction}
\label{sec-intro}
This article is dedicated to Hans Bethe. The work reported here has its 
connection to Hans through a strange confluence of events. It has its 
origins in meteoritics, neutrino physics, and supernovae. Hans and Gerry 
Brown had been pursuing theories of hypernovae and their inferences 
on this problem led them to extensive talks with us and greatly stimulated 
our efforts. Their positive support of our exploration was a stimulating 
force in all of our work, even when possible direct connections disappeared. 

The clear demonstration that the early solar system (ESS) contained many 
radioactive nuclei with a short to intermediate lifetime required injection 
from various nucleosynthetic sources. Amongst these nuclei was 
$^{129}$I ($\bar\tau_{129}=23$ Myr) discovered by \citet{i129}. 
This nuclide cannot be produced by any slow (``$s$'') neutron-capture process 
and requires very high neutron densities for its production. It was clearly 
to be associated with the ``$r$''-process as enunciated by \citet{b2fh}
and \citet{al}. Measurements of many samples of ESS debris 
preserved in meteorites showed that the abundance of $^{129}$I 
relative to that of stable $^{127}$I was 
($^{129}$I/$^{127}$I)$_{\rm ESS}\approx 1.0\times 10^{-4}$ 
at the time of formation of the solar system. Upon recognizing that this was 
most plausibly the result of long-term Galactic nucleosynthesis, it followed 
that the time between the last addition of such $r$-process debris and
the formation of the solar system had to be $\sim 10^8$~yr
\citep{wfh,schwass}.

This matter rested almost two decades until clear evidence was found
for other short-lived nuclei in the ESS [see the recent review by
\citet{wess}]. The various 
possible stellar nucleosynthetic sources required to produce this inventory 
of short-lived nuclei led to theoretical models of a wide variety. One 
possibility, for some of these nuclei, was a low-mass star during its 
asymptotic giant branch (AGB) stage leading to the formation of a planetary 
nebula. This potential source had been extensively explored by \citet{gal}
and \citet{kap} in studying the $s$-process 
that is known to take place in such sites. In a collaboration between 
the Turino Group and the junior author, where he was educated by his Italian 
colleagues, an AGB model was explored and predictions made of the isotopic 
yields \citep{wagb}. These predictions 
included yields of $^{182}$Hf [$\bar\tau_{182}=12.8\pm 0.1$ Myr \citep{voc}]. 
It was shown that an AGB source could only produce very little $^{182}$Hf.
This nuclide was subsequently discovered to be also present in the ESS
\citep{lhhf,hjhf}. An AGB source for this nuclide was excluded 
by the level of its abundance relative to stable $^{180}$Hf of 
($^{182}$Hf/$^{180}$Hf)$_{\rm ESS}\sim 3\times 10^{-4}$. 
More recent measurements have now established a reliable value of 
($^{182}$Hf/$^{180}$Hf)$_{\rm ESS}=(1.00\pm 0.08)\times 10^{-4}$ 
\citep{yhf,khf}. However, the difficulty for an AGB source to account for 
the $^{182}$Hf in the ESS is only slightly alleviated. The major problem
persisted --- this nuclide must be produced in an $r$-process and 
cannot be made effectively in an $s$-process.

This led to the following dilemma: both $^{129}$I 
($\bar\tau_{129}=23$ Myr) and $^{182}$Hf ($\bar\tau_{182}=12.8$ Myr) 
could not have been produced by the same $r$-process source 
($r$-source) in order to
account for their measured relative abundances in the ESS. 
This results from the simple fact that there is a factor of 
1.8 difference in their lifetimes. For example, consider a model of continuous 
nucleosynthesis over a Galactic timescale $T_{\rm UP}\sim 10^{10}$ yr with 
uniform (number) production 
rates $P_i$ for nuclide $i$. In this case the interstellar medium (ISM) would 
have an inventory of $P_{\rm S}T_{\rm UP}$ for a stable nucleus (S) and 
$P_{\rm R}\bar\tau_{\rm R}$ for a radioactive one (R). If this uniform 
production is followed by a period of no production for a 
time $\Delta$ prior to the formation of the solar system, 
then the abundance of R in the ESS is diminished by a factor of 
$\exp(-\Delta/\bar\tau_{\rm R})$. For $^{129}$I we have 
\begin{equation}
\left(\frac{^{129}{\rm I}}{^{127}{\rm I}}\right)_{\rm ESS}=
\left(\frac{P_{129}}{P_{127}}\right)
\left(\frac{\bar\tau_{129}}{T_{\rm UP}}\right)\exp\left(-\frac{\Delta}
{\bar\tau_{129}}\right).
\end{equation} 
If the ratio of the production rates is of order unity, then the
above equation gives $\Delta\sim 72$~ Myr. For this $\Delta$ and
assuming $P_{182}/P_{180}\sim 1$, a similar equation for $^{182}$Hf would 
give ($^{182}$Hf/$^{180}$Hf)$_{\rm ESS}\sim 4.6\times 10^{-6}$,
much smaller than the measured value 
of $1.0\times 10^{-4}$. As there is no basis for considering a large production 
ratio $P_{182}/P_{180}\sim 22$ to make up for this gross difference, 
it follows that $^{129}$I and $^{182}$Hf, both $r$-process nuclei
($r$-nuclei), cannot be 
produced in the same type of event. As pointed out by \citet{wbg},
this fact requires at least two types
of $r$-process event and must be related to the differences between
the site of production for the light $r$-nuclei with mass numbers
$A\lesssim 130$ and that for the heavy ones with $A>130$. These authors
also noted that elemental abundances in low-metallicity stars formed in the
early Galaxy would be highly susceptible to the particular type of $r$-source 
contaminating their birth medium. 
The study of metal-poor stars in the Galactic halo, as was already being 
pursued by Chris Sneden and his colleagues, would clearly be of aid in
testing this inference. Such studies indeed prove to be of great importance 
in establishing the diversity of $r$-sources. 

For almost forty years there had been a general preference for considering
the $r$-process to be associated with a single type of astrophysical source
that has an appropriate dynamic timescale.\footnote{Neutron star mergers 
have often been cited as a site for the
$r$-process. Some workers have studied the possibility that
these events are the source for the heavy $r$-nuclei with $A>130$
[e.g., \citet{frt}]. However, the problem is
that these events are too infrequent (by a factor of $\gtrsim 1000$)
compared to core-collapse supernovae. If neutron star mergers were 
the source for heavy $r$-nuclei, then enrichment in these nuclei would 
not occur until the ISM had already been substantially enriched in
Fe by Fe core-collapse supernovae from progenitors of 
$\sim 12$--$25\,M_\odot$ \citep{qian,arg}. This is in contradiction 
to the existence of stars with very low Fe abundances but highly 
enriched in heavy $r$-nuclei (see Section~\ref{sec-fe-decoup}). 
In addition, the observations so far show that
high enrichments of heavy $r$-nuclei are always accompanied by
enrichments of the nuclei with $A\sim 90$--110 at a roughly
comparable level (see Sections~\ref{sec-cpr-h} and \ref{sec-ycpr}). 
However,
it has not been demonstrated that the latter nuclei can be produced 
in neutron star mergers. For the above reasons, we will not discuss
neutron star merger models for the $r$-process further.}
The site usually discussed was an explosive environment such as some 
region inside a core-collapse supernova. The evolution and 
nucleosynthesis of massive stars associated with such supernovae
is discussed in the book by
\citet{arnett} and is reviewed in a recent paper by \citet{whw}.
Following the seminal paper by
\citet{hans} on the neutrino-driven supernova mechanism, the material
ejected by neutrino heating from the vicinity of a nascent neutron star 
--- the neutrino-driven wind ---
was considered the most promising site for the $r$-process 
\citep{wb92,wh92,mey92,how93,tak94,woo94}. 
For many workers in the field, the possibility of multiple $r$-sources 
was not readily accepted. Although it was generally understood that a
superposition of yield patterns each covering a different mass range is 
required to reproduce the overall solar $r$-process abundance 
pattern ($r$-pattern) in the absence of fission cycling
[e.g., \citet{kra}], it was preferred to consider this superposition as the
result of mixing within a single astrophysical environment. For example, 
it was regarded as a major success when the model of 
\citet{woo94} appeared to produce an integrated yield pattern from
a single supernova that closely 
resembles the solar $r$-pattern with peaks at $A\sim 130$ and 195.
The compelling evidence obtained later from spectroscopic data 
of very high quality on several key halo stars of ultra-low metallicities has 
brought the issue of diverse $r$-sources to the forefront of theoretical 
consideration.

When the $^{129}$I and $^{182}$Hf dilemma appeared, the matter was 
a source of lunch-time discussions in the Athenaeum at Caltech with 
Hans and Gerry. It was also imposed on Rose who has gracefully endured 
this type of ``social'' conversation for many years. Hans, of course, on first 
hearing the dilemma would go on about the nuclear details of $^{129}$I 
and $^{182}$Hf without any books --- just from his incredible mind. 
Awesome!

The senior author was a research fellow in Petr Vogel's group at Caltech 
around the time when
the $^{129}$I and $^{182}$Hf dilemma was first being discussed. He had
just finished a detailed study on the neutrino-driven wind with Stan Woosley, 
which showed that the conditions in the wind fall short of what is required 
for an $r$-process \citep{qw96}. It did not appear that a straightforward
ab initio solution to the $r$-process problem was within reach. On
the other hand, a phenomenological approach based on the available
observational data might offer some helpful guidance. The $^{129}$I and 
$^{182}$Hf dilemma seemed to provide a good starting point for this
approach. In an initial effort, \citet{qvw} proposed two types of supernova
$r$-process event: the $H$ type producing mainly the heavy $r$-nuclei 
with $A>130$ and the $L$ type producing mainly the light
ones with $A\lesssim 130$. The overall solar $r$-pattern
was considered to have resulted from a mixture of these two types 
of event. The meteoritic
data on $^{129}$I and $^{182}$Hf could also be accounted for by requiring
that the last event contributing $r$-nuclei to the solar system be 
of the $H$ type. Based on later observations of abundances in metal-poor 
stars, the authors of this report developed a 
phenomenological model with more detailed description of the $H$ and 
$L$ events \citep{qw01,qw02}. This model makes quantitative predictions 
for elemental abundances that are in good agreement with the observations.
In this report, we will present inferences that we drew from the earlier 
data and then show some changes that must follow from the newer 
high-quality observational results obtained by many groups.
Our basic conclusions reached earlier appear to remain valid.

\section{The phenomenological $r$-process}
\label{sec-phenr}

\begin{figure}
\begin{center}
\vskip -1.5cm
\includegraphics*[scale=0.5]{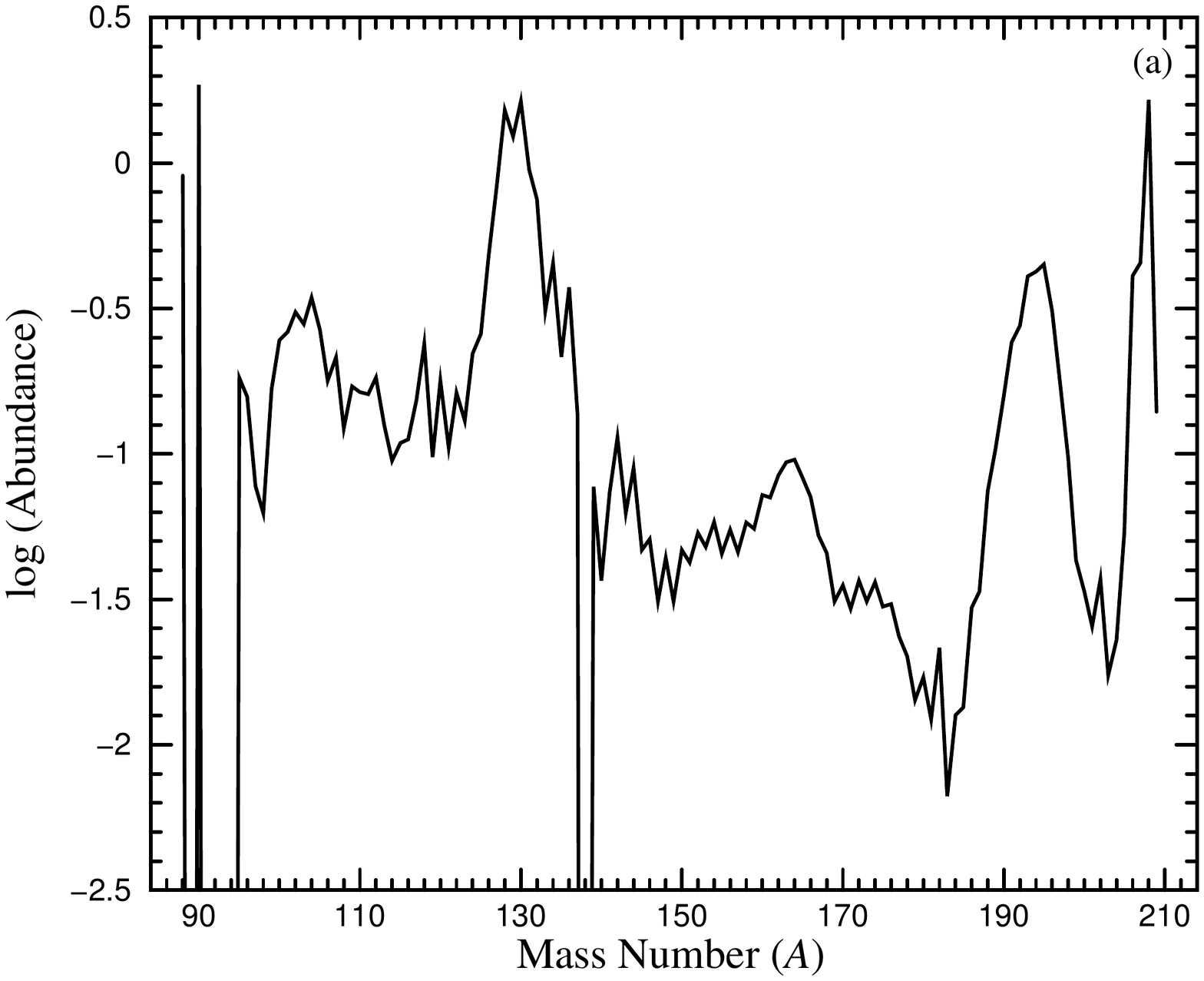}
\vskip -2.5cm
\includegraphics*[scale=0.5]{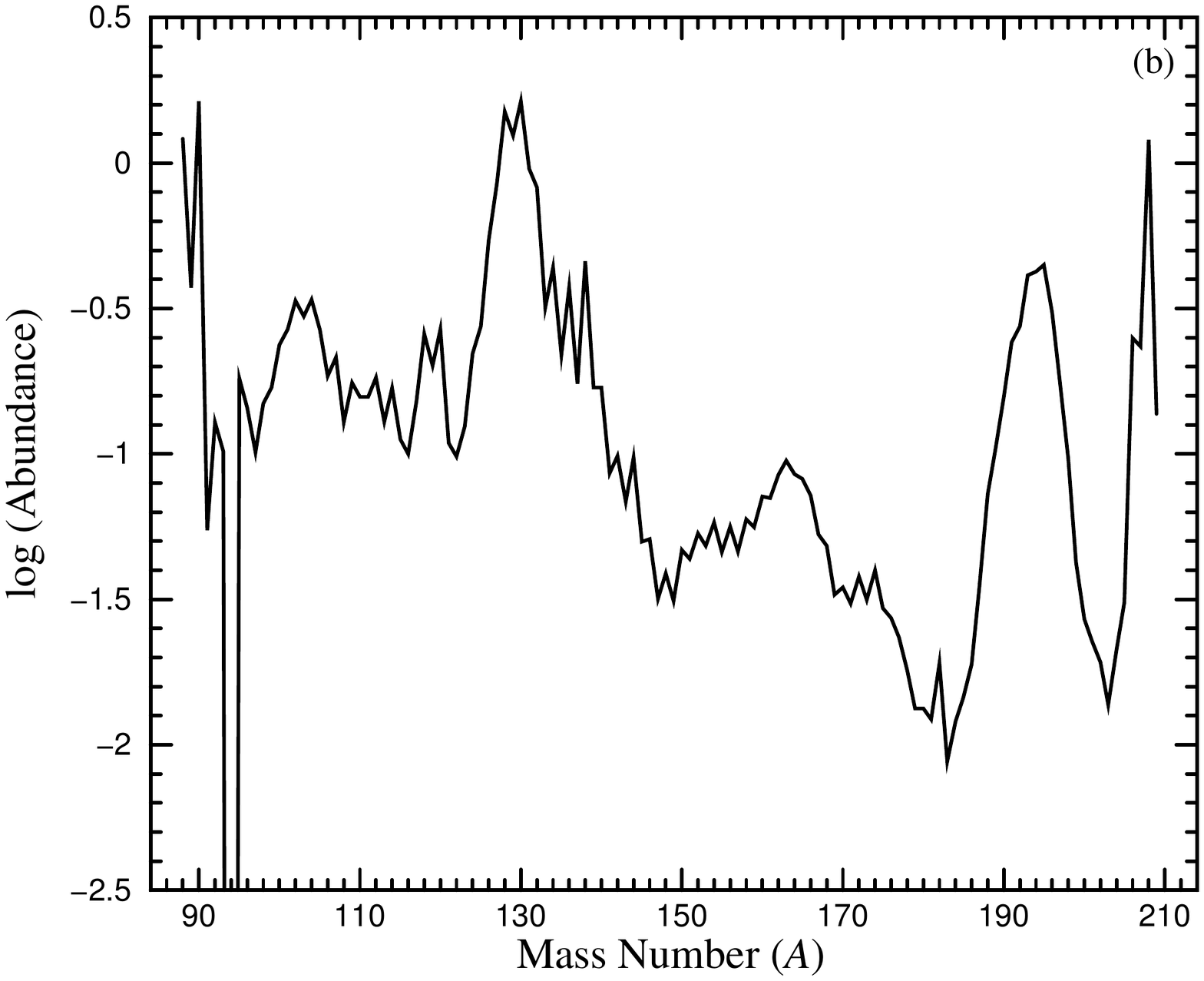}
\end{center}
\caption{The logarithm of the solar ``$r$''-abundance
$N_{\odot,r}=N_\odot-N_{\odot,s}$ (normalized so that the 
elemental abundance of Si is $10^6$) as a function of $A$. The $s$-process
contribution $N_{\odot,s}$ is taken from \citet{arl} and
calculated using (a) a phenomenological model or (b) a blend of $1.5$ and 
$3\,M_\odot$ AGB sources with an initial metallicity of half the solar value.}
\label{fig-rsun}
\end{figure}

The logarithm of the solar ``$r$''-process abundance (``$r$''-abundance) 
$N_{\odot,r}(A)$ of a nuclide with mass number $A$ is shown as a function
of $A$ in Figure~\ref{fig-rsun}. The value $N_{\odot,r}(A)$ is obtained from
\begin{equation}
N_{\odot,r}(A)=N_\odot(A)-N_{\odot,s}(A),
\end{equation}
where $N_\odot(A)$ is the total solar abundance of the nuclide derived from
meteoritic and spectroscopic measurements [normalized so that the 
elemental abundance of Si is $10^6$, see e.g., \citet{and}], and 
$N_{\odot,s}(A)$ is the $s$-process contribution to this nuclide 
(the solar main $s$-component) calculated 
using a phenomenological model (Fig.~\ref{fig-rsun}a) or a blend of $1.5$ 
and $3\,M_\odot$ AGB sources with an initial metallicity of half the solar 
value (Fig.~\ref{fig-rsun}b). The $s$-process contributions used are taken
from \citet{arl}. The total solar abundances are rather well determined,
and so are the majority of the solar ``$r$''-abundances. Neither sets of
abundances have changed drastically in forty years (except for Fe)
although both have
been regularly subject to revisions. There is a problem with 
attributing $N_{\odot,r}$ obtained by $s$-process subtraction for nuclei 
with $A\sim 90$--110 to the $r$-process. This will be discussed in 
Section~\ref{sec-wind}.

Clearly, the solar inventory is the result of contributions from many stellar 
sources over Galactic history. Due to mixing of nucleosynthetic products from 
different sources in the ISM over this long history, the patterns for the gross 
abundances of a wide range of elements in other stars with approximately
solar metallicity are observed to be close to the solar pattern. It is well
recognized that this apparent ``universality'' does not imply the existence
of a single process for making all the nuclei in nature. However, when
only production by rapid neutron capture is considered, it is tempting to
regard the solar $r$-pattern as a truly universal yield pattern of every
single $r$-process event. Theoretically speaking, it is possible to reproduce
the solar $r$-pattern with peaks at both $A\sim 130$ and 195 
(see Fig.~\ref{fig-rsun}) by fission cycling in a single environment with
extremely abundant supply of neutrons [e.g., \citet{sfc}]. However, this
would most likely couple the production of $^{129}$I and $^{182}$Hf, 
in contradiction 
to the meteoritic data discussed in Section~\ref{sec-intro}. In addition, so far
there is no self-consistent astrophysical model that can demonstrate the
occurrence of fission cycling [the model of \citet{frt} is of parametric nature]. 
There have also been studies aiming to reproduce the overall solar 
$r$-pattern from a single site without fission cycling [e.g., \citet{woo94}]. 
However, these studies are either parametric or unsuccessful [the 
nonparametric model of \citet{woo94} has several serious deficiencies, 
see e.g., \citet{mey95,qw96}]. 

An $r$-process requires an environment of high neutron (number) density
with a large abundance of neutrons relative to the seed nuclei capturing 
them. The theory of the $r$-process has four components: (1) what are 
the seed nuclei and how are they provided?
(2) how do temperature, density, and neutron abundance evolve with 
time during the $r$-process? (3) what are the properties of the nuclei 
participating in the $r$-process and the rates of the relevant reactions?
(4) how many types of astrophysical sources with distinct $r$-process
yield ($r$-yield) patterns are there to account for the chemical evolution 
of $r$-nuclei in the Galaxy and elsewhere? The third component is
in the domain of nuclear physics and is crucial to the determination of 
the exact $r$-yield pattern produced by a specific astrophysical source.
We will not discuss this component here. Instead, we address some
aspects of the first and second components and focus on the fourth
considering observations of metal-poor stars and data on the solar 
system.

Nearly all $r$-process models are of parametric nature and fall into
two categories: one is self-contained in that the first and second
components listed above are addressed within the model 
[e.g., \citet{hwq,mb97,fre99}],
while the other assumes a seed nucleus and subjects it to
neutron bombardment at constant neutron density and temperature
for some time [e.g., \citet{kra}]. In the absence of fission cycling,
both categories of models require
a superposition of $r$-yield patterns produced under different 
conditions to reproduce the overall solar $r$-pattern, but
neither can determine whether this superposition is achieved
by combining the $r$-nuclei produced in different regions
at different times within a single astrophysical source or by mixing
the $r$-nuclei produced by different types of source in the ISM.

\begin{figure}
\begin{center}
\includegraphics*[scale=0.5,angle=270]{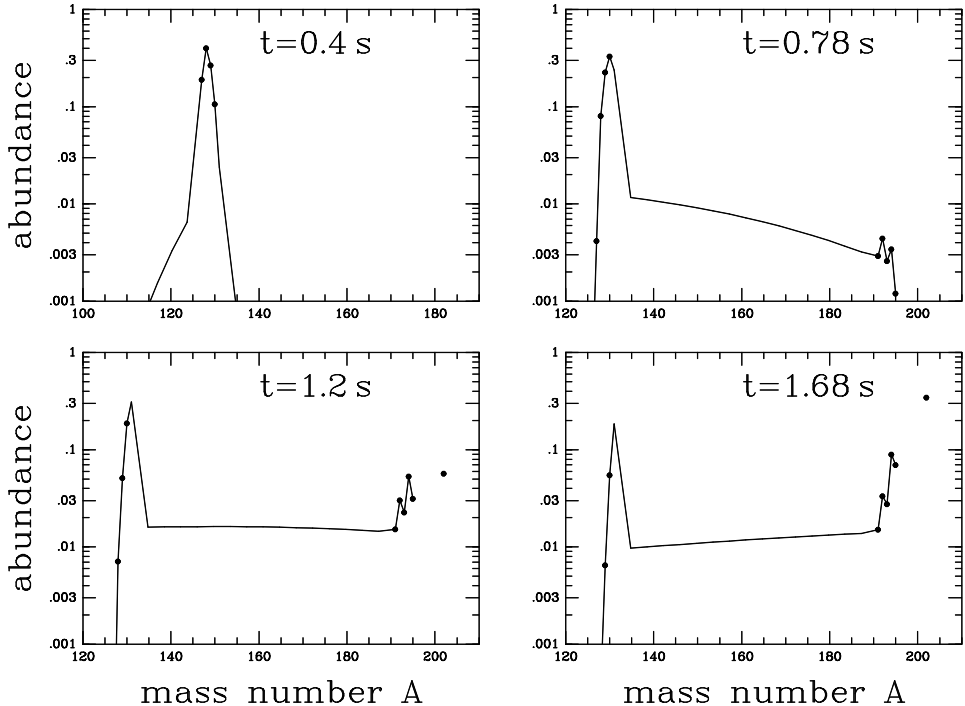}
\end{center}
\caption{Illustration of the progress of the $r$-process. It is assumed
that a single kind of seed nucleus with $A_s=90$ is present at $t=0$
and that the $r$-process passes through a fixed set of nuclei. Rapid
neutron capture stops at $t=0.78$~s for $n/s=44$ and at $t=1.68$~s
for $n/s=86$. The panels labeled by these times give the yield patterns 
for these two $n/s$ values. Other panels represent intermediate
stages before all neutrons are captured.}
\label{fig-rfo}
\end{figure}

The requirement of superposition to reproduce the solar $r$-pattern
without fission cycling can be understood from
the following simplified description of the $r$-process.
Consider an $r$-process starting with a ratio $n/s$ of the total 
number of free neutrons to that of the seed nuclei. As all the neutrons 
will be captured, the equation governing this $r$-process is
\begin{equation}
\langle A_s\rangle+ n/s= \langle A_r\rangle,
\label{eq-n/s}
\end{equation}
where $\langle A_s\rangle$ is the average mass number of the 
seed nuclei and $\langle A_r\rangle$ that of the final $r$-nuclei 
produced. It follows that for $\langle A_s\rangle\sim 90$, $n/s\sim 40$ 
is required to produce an abundance peak at $A\sim 130$. From the 
same seed distribution, to produce an abundance peak at $A\sim 195$
would require $n/s\sim 100$. It is immediately evident that the solar
$r$-pattern with peaks at both $A\sim 130$ and 195 
(see Fig.~\ref{fig-rsun}) can only be
reproduced by a superposition of $r$-yield patterns corresponding to
a range of $n/s$. 
For illustration, we assume that a single kind of seed
nucleus with mass number $A_s=90$ is present at time $t=0$ and that 
the $r$-process subsequently passes through a fixed set of nuclei. 
We show the progress of the $r$-process for different $n/s$ values
in Figure~\ref{fig-rfo}. For $n/s=44$ all the neutrons have been captured 
by $t=0.78$~s, resulting in an $r$-yield pattern dominated by the peak 
at $A\sim 130$ (panel labeled ``$t=0.78$~s''). By comparison, 
for a higher $n/s=86$ many neutrons remain to be captured at
$t=0.78$~s  and rapid neutron capture continues until
$t=1.68$~s, leading to major production
of the nuclei in the peak at $A\sim 195$ and beyond (panel labeled 
``$t=1.68$~s''). The solar $r$-pattern with peaks at both $A\sim 130$ 
and 195 can then be reproduced, for example, by an appropriate 
superposition of the $r$-yield patterns for $n/s=44$ and 86
\citep{qvw}.

To summarize, the solar $r$-pattern represents a superposition of
$r$-yield patterns produced with a range of $n/s$ (some variation
in the seed nuclei is also possible) in the absence of fission cycling. 
Meteoritic data on $^{129}$I
and $^{182}$Hf suggest that this superposition was accomplished
by mixing the $r$-nuclei produced by two distinct types of source 
in the ISM: one has high $n/s$ and produces mainly the heavy 
$r$-nuclei with $A>130$ while the other has lower $n/s$ and
produces mainly the light $r$-nuclei with $A\lesssim 130$. 
The issue of two distinct $r$-sources will be discussed
further in Section~\ref{sec-obs} in connection with observations of 
metal-poor stars. As mentioned above,  
it is possible to produce simultaneously the peaks at both
$A\sim 130$ and 195 in the solar $r$-pattern with fission cycling. 
In this case, the average mass number
$\langle A_r\rangle$ of the nuclei produced is 
determined by the $r$-process flow between the fission fragments
and the nuclei undergoing fission. The total abundance of the nuclei
produced is essentially governed by the available neutron supply.
Clearly, Equation~(\ref{eq-n/s}) does not apply to the case of
fission cycling as
the mass numbers of the seed nuclei are no longer pertinent.
The possible role of fission cycling in $r$-process
nucleosynthesis will be discussed in Section~\ref{sec-song}. 
We now turn to a class of astrophysical models 
for heavy element synthesis that may be closely related to the 
$r$-process.

\section{Heavy  element synthesis in neutrino-driven winds}
\label{sec-wind}

The formation of a neutron star in a core-collapse supernova results in the
emission of intense flux of $\nu_e$, $\bar\nu_e$, $\nu_\mu$, $\bar\nu_\mu$,
$\nu_\tau$, and $\bar\nu_\tau$ as the neutron star releases its enormous
gravitational binding energy of several $10^{53}$~erg during the first
$\sim 20$~s of its life. 
As $\nu_e$ and $\bar\nu_e$ pass through the ``atmosphere'' of free neutrons 
and protons surrounding the neutron star, some of them are captured via
the reactions 
\begin{eqnarray}
\nu_e+n&\to&p+e^-,
\label{eq-nuen}\\
\bar\nu_e+p&\to&n+e^+.
\label{eq-nuebarp}
\end{eqnarray}
The heating of the atmosphere mainly by the above reactions makes it
expand, thereby forming a neutrino-driven wind from the neutron star.
The heating occurs 
mostly within a few neutron star radii and becomes inefficient
at larger distance due to the decrease in the neutrino flux and
the increase in the wind velocity. Therefore, the wind material expands 
approximately adiabatically above several neutron star radii.
The profuse production of neutrons via the reaction in 
Eq.~(\ref{eq-nuebarp}) in the vicinity of the neutron 
star, although countered by destruction via the reaction in
Eq.~(\ref{eq-nuen}), suggests the possibility that the neutrino-driven
wind may be associated with the $r$-process.

The adiabatic expansion of a mass element in the wind is governed by 
three key parameters: the entropy $S$, the dynamic timescale 
$\tau_{\rm dyn}$, and the electron fraction $Y_e$.  The electron
fraction specifies the initial number fractions of free neutrons and
protons as $Y_n=1-Y_e$ and $Y_p=Y_e$, respectively, when the mass 
element is near the neutron star. The dynamic timescale $\tau_{\rm dyn}$ 
controls how fast
the temperature decreases as the element expands, and $S$ determines
the density at a given temperature. These parameters are determined 
in turn by the state of the neutron star that is defined by its mass 
$M_{\rm NS}$ and radius $R_{\rm NS}$ as well as its neutrino emission 
characteristics when the mass element is ejected [e.g., \citet{qw96,tbm}]. 
The amount of material 
ejected with a specific set of $S$, $\tau_{\rm dyn}$, and $Y_e$ depends 
on how long the neutron star stays in the corresponding state. During the 
period of interest, $M_{\rm NS}$ remains essentially constant but 
$R_{\rm NS}$ decreases substantially, the neutrino luminosity decreases 
drastically, and the neutrino energy spectra can change substantially. 
In any case, given the evolution of $R_{\rm NS}$ and the neutrino emission 
characteristics of a neutron star, one can calculate the amounts of material
ejected with various sets of $S$, $\tau_{\rm dyn}$, and $Y_e$ during this
evolution. For a mass element with a specific set of these parameters, 
nuclear reactions during its expansion produce a particular set of heavy 
nuclei with specific abundances. The total production of heavy nuclei in the 
neutrino-driven wind is then an integral over the evolution of the neutron star.

We now consider some examples of nucleosynthesis associated with the
expansion of a mass element in the wind from a neutron star with
$M_{\rm NS}=1.4\,M_\odot$. The evolution of the radius and neutrino
emission characteristics of this neutron star is taken from \citet{woo94}.
Consider the epoch for which $R_{\rm NS}=10$~km and the neutrino 
energy spectra change slowly. It is convenient to use the neutrino 
luminosity $L_\nu(t)$ as a function of time $t$ to represent the evolution 
of the neutron star during this epoch. At a given time, a mass element is 
ejected and processed as it moves outward until all nuclear reactions 
cease. At a later time, another element is ejected and moves outward 
being processed at a different neutrino luminosity
until all nuclear reactions cease. There are two timescales
involved: the dynamic timescale $\tau_{\rm dyn}$ for a mass element
to move out and the evolutionary timescale $\tau_\nu$ over which 
$L_\nu$ changes significantly. Typically we have 
$\tau_{\rm dyn}\ll\tau_\nu$. Thus, the conditions and the resulting
nucleosynthesis would be similar for mass elements ejected within 
an interval $\Delta t\lesssim\tau_\nu$.

We pick two times $t_A$ and $t_B$ during the evolution of the above 
neutron star. 
For $L_\nu(t_A)=3\times 10^{51}$~erg/s per species, the mass element
is ejected with $S\sim 74$ (in units of Boltzmann constant per baryon),
$\tau_{\rm dyn}=0.024$~s, and $Y_e=0.465$; for
$L_\nu(t_B)=10^{51}$~erg/s per species, the conditions change to
$S\sim 87$, $\tau_{\rm dyn}=0.066$~s, and $Y_e=0.372$ \citep{qw96}. 
The difference in $\tau_{\rm dyn}$ is mainly due to the factor of 3 difference 
in $L_\nu$ while that in $Y_e$ mostly results from the change in the 
neutrino energy spectra. For the mass element ejected at $t_A$, 
nucleosynthesis predominantly produces $^{88}$Sr, $^{89}$Y, and 
$^{90}$Zr [see Fig.~1 in \citet{hwq}]. For the mass element ejected at $t_B$, 
the dominant products
are $^{96}$Zr, $^{98,100}$Mo, $^{101,104}$Ru, $^{105,110}$Pd, and 
$^{107}$Ag [see Fig.~2 in \cite{hwq}. Note that $^{107}$Ag was made as 
$^{107}$Mo, which decays through $^{107}$Pd with 
$\bar\tau_{107}=9.4$~Myr, see \citet{wh92}]. 
However, for both cases the heavy nuclei are produced by
the so-called $\alpha$-process involving charged-particle reactions
\citep{wh92}, not by rapid neutron capture. In fact, by the 
time charged-particle reactions cease due to the prohibitive Coulomb 
barrier at a temperature $T\sim 2$--3$\times 10^9$~K, few
neutrons are left for each heavy nucleus so that no significant
processing by rapid neutron capture can occur subsequently
\citep{hwq}.

The above results are representative of nucleosynthesis in the 
neutrino-driven wind from a $1.4\,M_\odot$ neutron star. While nuclei with
$A\sim 90$--110 can be produced by the $\alpha$-process, there are
too few neutrons left at the end of the $\alpha$-process to drive a 
subsequent $r$-process. As the higher neutrino luminosity $L_\nu(t_A)$ is 
more efficient in ejecting the material than the lower $L_\nu(t_B)$, 
the total production integrated over the evolution of the neutron star
is dominated by $^{88}$Sr, $^{89}$Y, and $^{90}$Zr \citep{woo94}.
If we consider a more massive neutron star with $M_{\rm NS}=2\,M_\odot$,
the production at earlier times during its evolution is similar to the 
$1.4\,M_\odot$ case. However, the wind has significantly higher $S$
at later times. For example, the neutron star state with $R_{\rm NS}=10$~km 
and $L_\nu=6\times 10^{50}\,{\rm erg/s}$ per species corresponds 
to $S\sim 140$, $\tau_{\rm dyn}=0.11$~s, and $Y_e=0.354$ \citep{qw96}. For 
these parameters, the $\alpha$-process dominantly produces $^{97,100}$Mo,
$^{102,104}$Ru, $^{103}$Rh, $^{105,108,110}$Pd, $^{107,109}$Ag, and
$^{111}$Cd. In addition, a significant part of the nuclear flow has reached 
$A=123$--125, signaling an incipient $r$-process 
[see Fig.~6 in \citet{hwq}]. However, even in this case there are also only
$\lesssim 10$ neutrons available for each heavy nucleus produced by
the $\alpha$-process.

\begin{figure}
\begin{center}
\includegraphics*[scale=0.6]{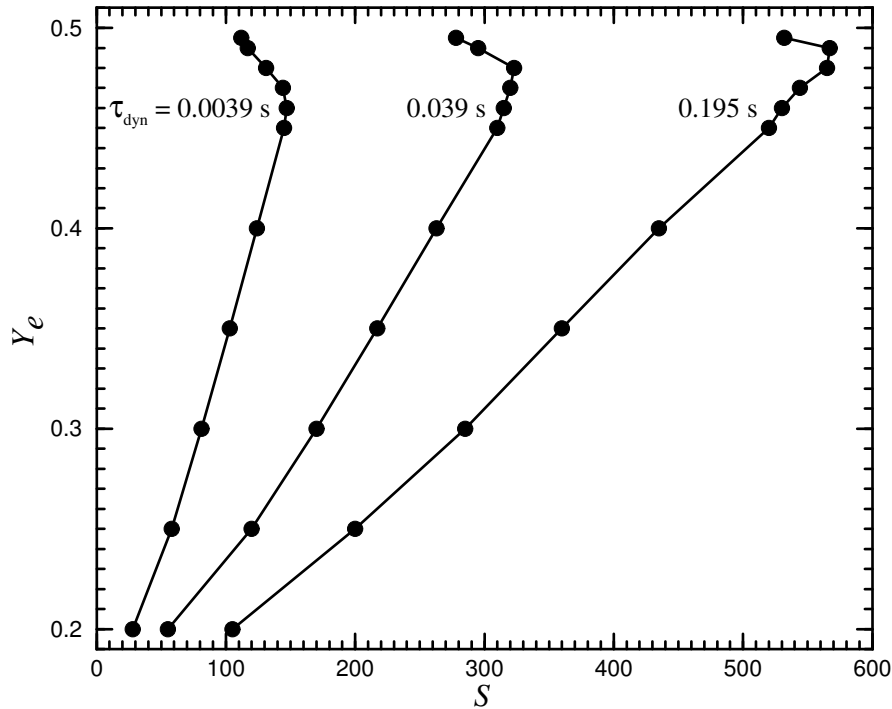}
\end{center}
\caption{Combinations of $S$, $Y_e$, and $\tau_{\rm dyn}$
required to produce $r$-nuclei with $A\sim 195$ in an adiabatically
expanding mass element such as that in the neutrino-driven wind.
Three choices of $\tau_{\rm dyn}$ are shown [see \citet{hwq} for
discussion and more results]. Note that for $S\lesssim 140$ and
$0.35\lesssim Y_e<0.5$ typical of the neutrino-driven wind, the
high $n/s$ required for production of $r$-nuclei with $A\sim 195$ 
can only be obtained with very short dynamic timescales 
$\tau_{\rm dyn}\lesssim 0.0039$~s. However, typical values for
the wind are $\tau_{\rm dyn}\sim 0.01$--0.1~s [e.g., \citet{qw96,tbm}].}
\label{fig-syetau}
\end{figure}

One caveat for the $2\,M_\odot$ neutron star case is that there is no 
self-consistent calculation of either its neutrino emission characteristics 
or the evolution of its radius. In the above example, we have simply 
assumed $R_{\rm NS}=10$~km and taken the neutrino energy spectra 
corresponding to $L_\nu=6\times 10^{50}\,{\rm erg/s}$ per species 
for the $1.4\,M_\odot$ case. Many studies of the $r$-process 
[see the recent review by \citet{wan}] are of this
parametric nature and thus lack self-consistency. It may be quite possible
that a self-consistent calculation of the evolution of a $2\,M_\odot$ neutron 
star may give rise to an $r$-process producing nuclei with $A\sim 130$
during some part of its evolution. For production of $r$-nuclei with
$A\sim 195$, there must be $\sim 100$ neutrons left for each heavy
nucleus produced when the $\alpha$-process ceases at 
$T\sim 2$--3$\times 10^9$~K. This
requires much more extreme conditions in either entropy or dynamic
timescale as shown in 
Figure~\ref{fig-syetau}. Given the typical conditions of $S\lesssim 140$,
$\tau_{\rm dyn}\sim 0.01$--0.1~s, and $0.35\lesssim Y_e<0.5$
in the neutrino-driven wind  [e.g., \citet{qw96,tbm}],
it appears that there is no clear case for sufficient neutrons 
to produce $r$-nuclei with $A\sim 195$ in the wind. We here consider that
the neutrino-driven wind is not directly responsible for the production of
the heavy $r$-nuclei with $A>130$.

Insofar as the neutrino-driven wind model is basically correct, then the 
need for a bridge between the $\alpha$-process in the wind and the real 
$r$-process remains the key issue. There is some confusion in the literature 
with regard to the ``light $r$-nuclei'' with $A\sim 90$--110.
Insofar as they mostly represent 
the products of the $\alpha$-process involving charged-particle reactions 
that would precede any true $r$-process, they should not be considered or 
lumped with the true $r$-nuclei produced by rapid neutron capture. 
We will henceforth refer to the nuclei with $A\sim 90$--110 produced by
charged-particle reactions (the $\alpha$-process) in
the neutrino-driven wind as ``CPR'' nuclei. The corresponding elements
are Sr, Y, Zr, Nb, Mo, Ru, Rh, Pd, Ag, and Cd.

\begin{figure}
\begin{center}
\includegraphics*[scale=0.25]{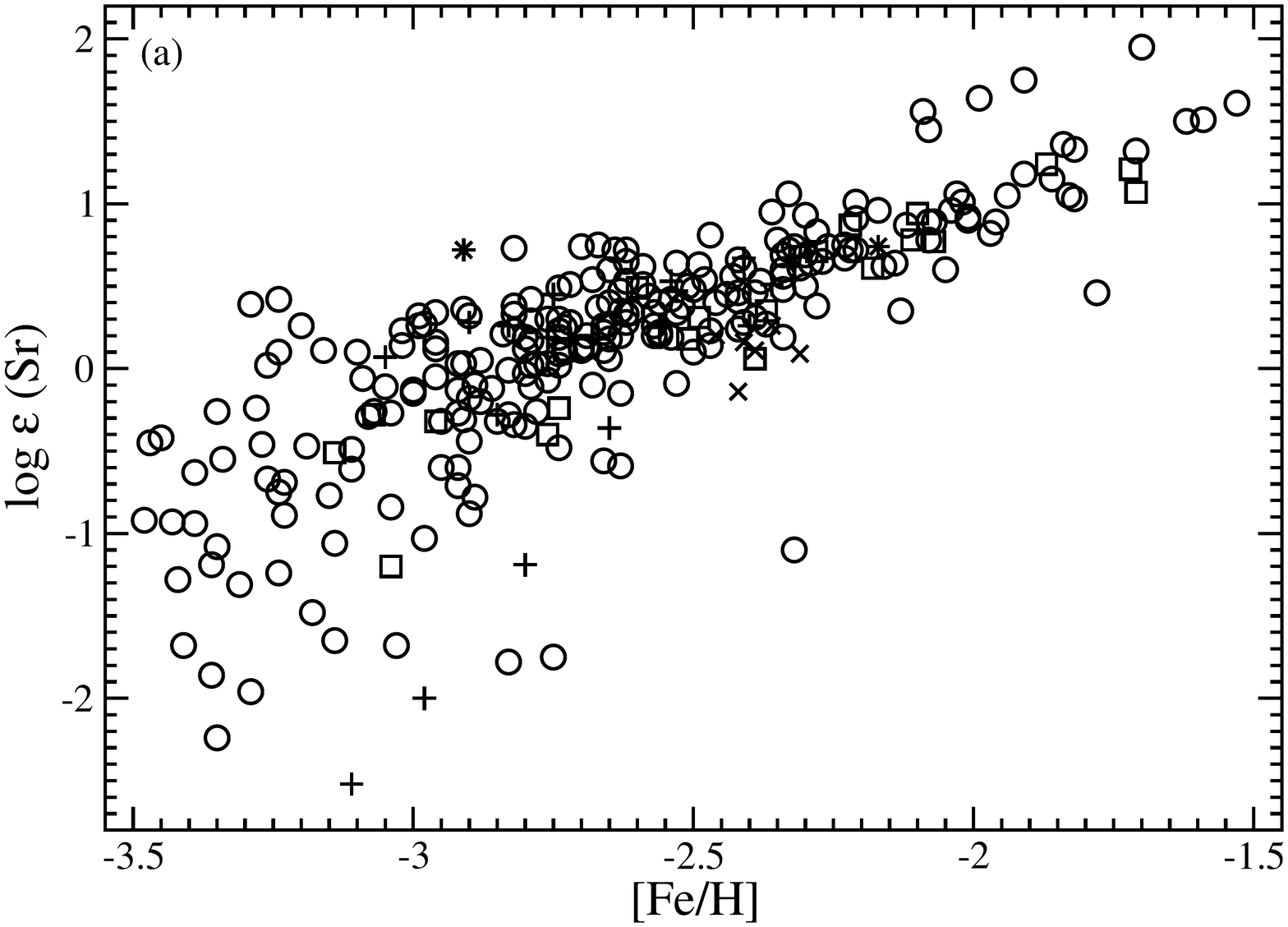}
\includegraphics*[scale=0.25]{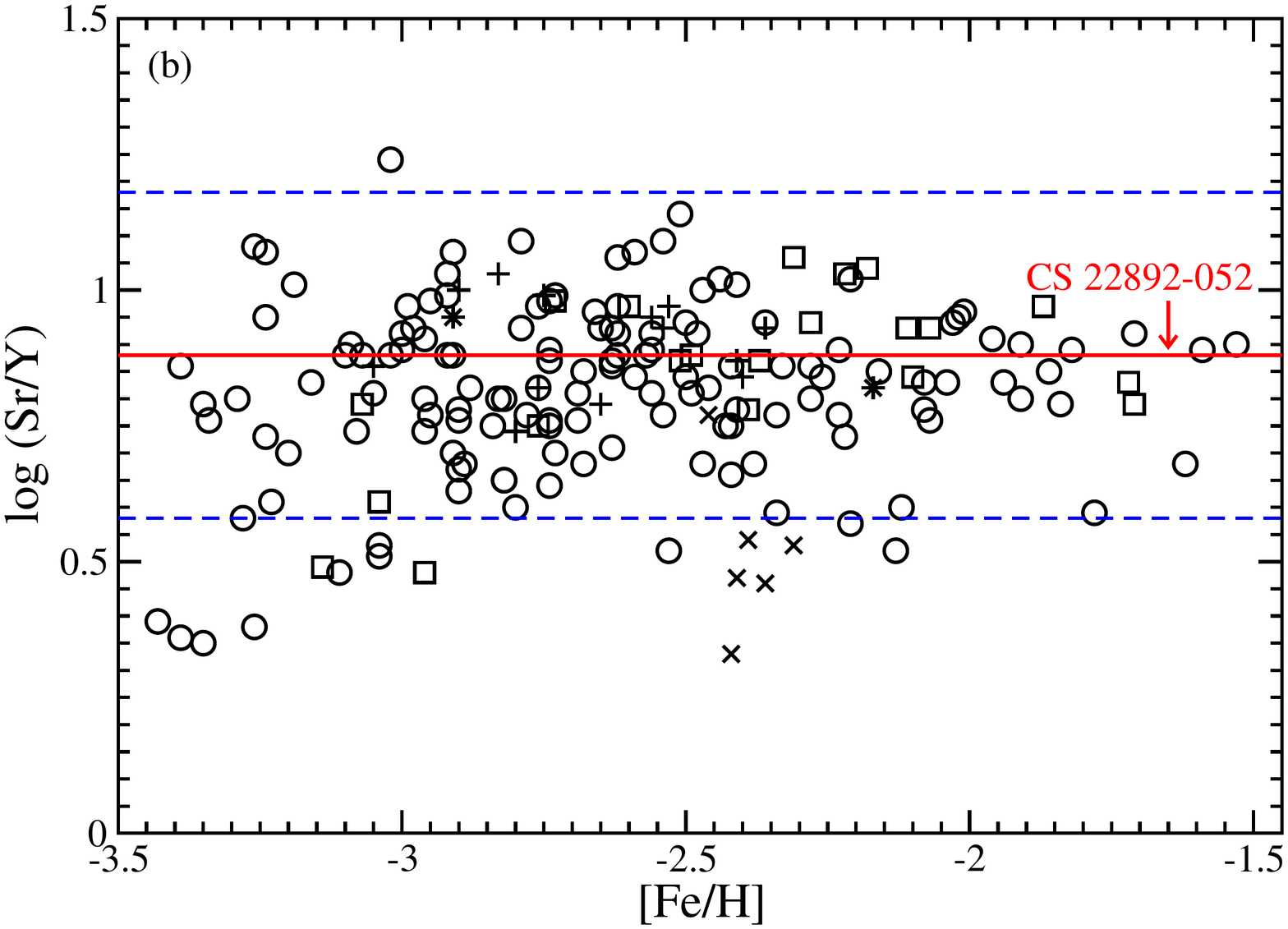}
\includegraphics*[scale=0.25]{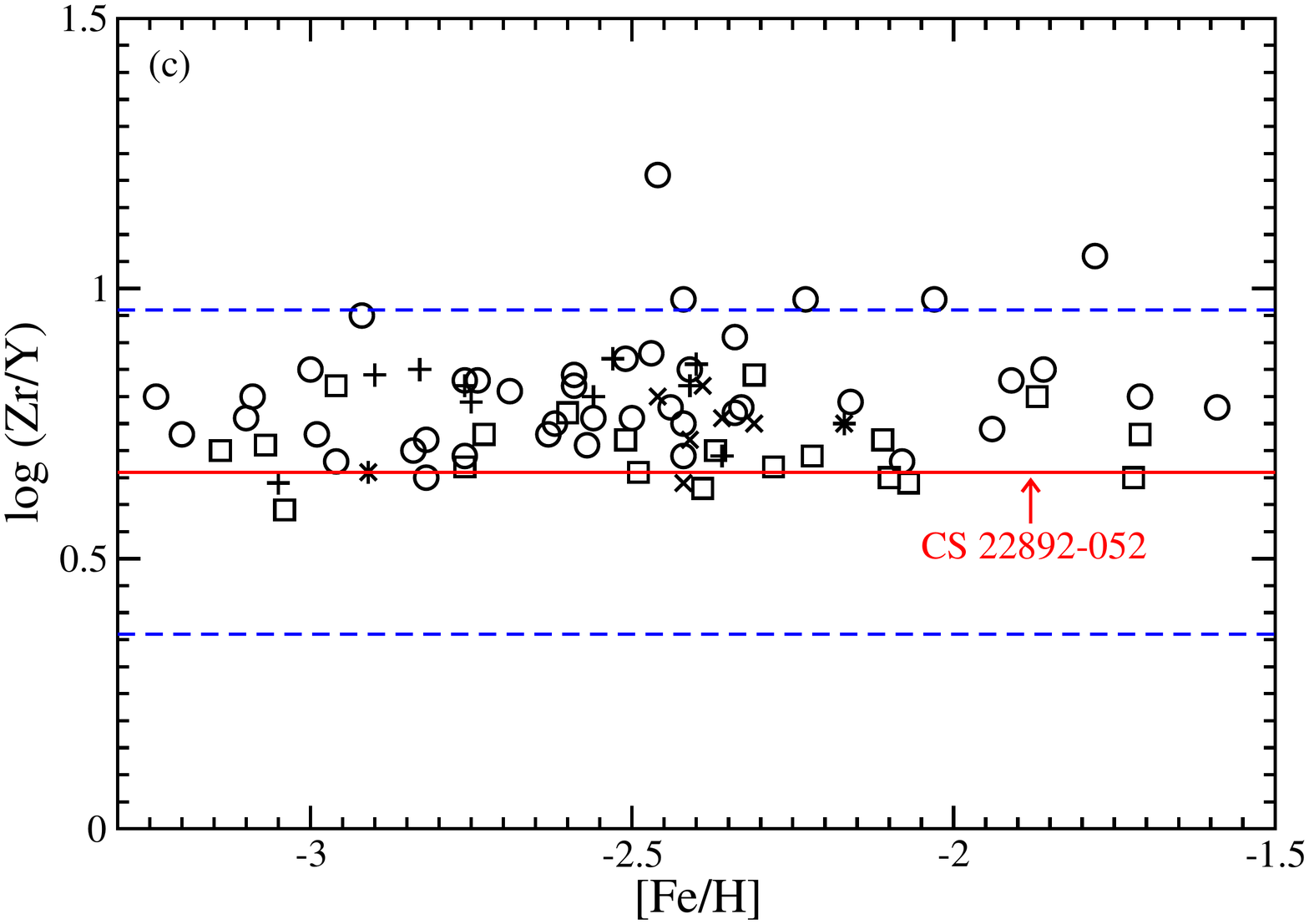}
\end{center}
\caption{(a) Data on $\log\epsilon({\rm Sr})$ versus [Fe/H] 
for metal-poor stars [squares: \citet{jb02}; pluses: \citet{aok05};
circles: \citet{he05}; crosses: \citet{ots06}; asterisks: \citet{hill,iv06}].
There is a large scatter in $\log\epsilon({\rm Sr})$ at any fixed [Fe/H]
over the wide range $-3.5\lesssim{\rm [Fe/H]}<-2.5$. This
can be accounted for if CPR nuclei are produced in the neutrino-driven
wind whether a neutron star is made in Fe core-collapse SNe,
or O-Ne-Mg core-collapse SNe, or AIC events. Note that Fe is produced
by the first source but not the latter two. (b) Data on $\log({\rm Sr/Y})$
versus [Fe/H] for the same stars shown in (a). Almost all the data lie within
0.3 dex of the value for CS~22892--052, a star with [Fe/H]~$=-3.1$
but highly enriched in heavy $r$-nuclei \citep{sn03}. (c) Same as (b)
but for $\log({\rm Zr/Y})$. Note the very narrow range for the abundance
ratio of this pair of
elements. In consideration of observational uncertainties (especially 
for Sr that is more difficult to measure), it appears from (b) and (c)
that CPR nuclei are rather robust products from the neutrino-driven
wind in all core-collapse SNe.}
\label{fig-sryzr-fe}
\end{figure}

The mechanism of the $\alpha$-process for producing CPR nuclei in the 
neutrino-driven wind is quite well understood \citep{wh92}. However, the 
quantitative yields of these nuclei depend on the detailed evolution of the 
radius and neutrino emission characteristics of 
nascent neutron stars as discussed
above. To our knowledge, so far there is only one calculation that follows 
the evolution of a $1.4\,M_\odot$ neutron star during the first $\sim 20$~s 
of its life \citep{woo94}. There are also significant uncertainties in calculating
the neutrino luminosity and energy spectra, which have a major impact
on $Y_e$ in the wind \citep{qian93}. Due to these issues, there are no
robust predictions for the yields of CPR nuclei from the neutrino-driven
wind. However, it seems clear from the available calculations that Sr, Y,
and Zr must be the dominant products [e.g., \citet{woo94}]. Using the
available data on metal-poor stars \citep{hill,jb02,aok05,he05,ots06,iv06},
we show $\log\epsilon({\rm Sr})\equiv\log({\rm Sr/H})+12$, where (Sr/H)
is the number ratio of Sr to H observed in a star, as a function of 
${\rm [Fe/H]}\equiv\log({\rm Fe/H})-\log({\rm Fe/H})_\odot$ in 
Figure~\ref{fig-sryzr-fe}a. It is clear that there is a large scatter in 
$\log\epsilon({\rm Sr})$ at any fixed [Fe/H] over the wide range 
$-3.5\lesssim{\rm [Fe/H]}<-2.5$. Therefore, there is no correlation
between Sr and Fe abundances at such low metallicities. In contrast,
$\log({\rm Sr/Y})$ and $\log({\rm Zr/Y})$ are rather constant over the
entire range $-3.5\lesssim{\rm [Fe/H]}\lesssim -1.5$ for the data shown
and cluster closely around the values for CS~22892--052
[\citet{sn03}, see Figs.~\ref{fig-sryzr-fe}b and \ref{fig-sryzr-fe}c]. 
As will be discussed in Section~\ref{sec-obs}, CS~22892--052 has
[Fe/H]~$=-3.1$ but is greatly enriched in heavy $r$-nuclei. 
The relationship between Sr, Y, and Zr relative to Fe shown in 
Figures~\ref{fig-sryzr-fe}b and \ref{fig-sryzr-fe}c can be understood 
if CPR nuclei are produced with an approximately fixed yield 
pattern in the neutrino-driven wind whenever a neutron star 
is formed. However, it follows from Figure~\ref{fig-sryzr-fe}a
that the CPR elements are not always produced with Fe 
(see Sections~\ref{sec-hl} and \ref{sec-ycpr}).

There are three possible venues for making a neutron
star: (1) collapse of an Fe core produced by progenitors of
$\sim 11$--$25\,M_\odot$, (2) collapse of an O-Ne-Mg core produced
by progenitors of $\sim 8$--$10\,M_\odot$, and (3) accretion-induced
collapse (AIC) of a white dwarf in binary systems. These venues will
be referred to as Fe core-collapse supernovae (SNe), 
O-Ne-Mg core-collapse SNe, and AIC events, respectively
(we also refer to all three generically as core-collapse SNe for
convenience). Only the first venue --- Fe core-collapse SNe --- 
can produce Fe (see Sec.~\ref{sec-rsource}). 
If a similar amount of CPR nuclei is produced per event
for all three venues, there will be no correlation
between these nuclei and Fe at low [Fe/H] values
for which the abundances
in stars would be highly susceptible to the particular type
of event contaminating their birth medium. In addition, if CPR nuclei
are always produced with nearly fixed yield ratios, then
the approximate constancy of $\log({\rm Sr/Y})$ and $\log({\rm Zr/Y})$ 
over the wide range of [Fe/H] shown in Figures~\ref{fig-sryzr-fe}b and
\ref{fig-sryzr-fe}c can also be understood. In summary, the data on
metal-poor stars provide some strong indirect evidence for the role
of the neutrino-driven wind in producing CPR nuclei. The relationship
between these nuclei and the true $r$-nuclei will be discussed in
Sections~\ref{sec-cpr-h}, \ref{sec-hl}, and \ref{sec-ycpr}.

\section{Critical astronomical observations on the $r$-rocess sites}
\label{sec-obs}
It is most plausible that stars of very low metallicities (as determined by 
[Fe/H]) would have formed when the ISM was only slightly enriched in 
the products of stellar nucleosynthesis. In particular, the $s$-process
contributions from low-mass AGB stars to the ISM
would not have occurred during the
first $\sim 1$~Gyr after the big bang due to the timescale required for 
the evolution of stars of $\sim 2\,M_\odot$. Likewise,
the contributions to the Fe group nuclei from Type Ia SNe (SNe Ia)
that are associated with white dwarfs left behind by low-mass stars
would also be mostly absent during this epoch. The dominant 
contributions to the ISM at low metallicities would be from massive 
stars of the early generations. Study of low-metallicity stars thus 
should provide a clear opportunity of observing contributions from 
these sources. Insofar as $r$-sources are associated with 
core-collapse SNe from massive stars, then contributions from 
individual (or few) SN events should be discernable. Taking a
typical Fe yield of $\sim 0.1\,M_\odot$ per SN [e.g., \citet{ww95}] 
and a dilution mass of $\sim 3\times 10^4\,M_\odot$ to mix
with the SN ejecta [e.g., \citet{thorn}], we estimate that an ISM
of pure big bang debris would be enriched with [Fe/H]~$\sim-2.6$
by a single Fe-producing SN using a solar Fe mass fraction of 
$X_\odot({\rm Fe})=1.27\times10^{-3}$ \citep{and}. Stars with such
low [Fe/H] values would provide important clues to whether and
how core-collapse SNe are associated with the $r$-process.

As almost all stellar observations yield elemental, but not isotopic 
abundances, it is necessary to choose diagnostic elements produced 
solely or predominantly by the $r$-process. Among these elements 
are Eu, which is commonly observed, and Re, Os, Ir, and Pt, which are 
rarely observed. In the absence of any contributions from the 
$s$-process, Ba (which is more easily measured) is also an 
important diagnostic. In the case where Th and U can be measured, 
this would be clear evidence for contributions from an $r$-process 
with large $n/s$ (possibly involving fission cycling)
as these nuclei can only be made in such a process.

\subsection{Decoupling of the sources for the heavy $r$-nuclei from
those for the elements above {\rm N} through the {\rm Fe} group}
\label{sec-fe-decoup}

\begin{figure}
\begin{center}
\includegraphics*[scale=0.5]{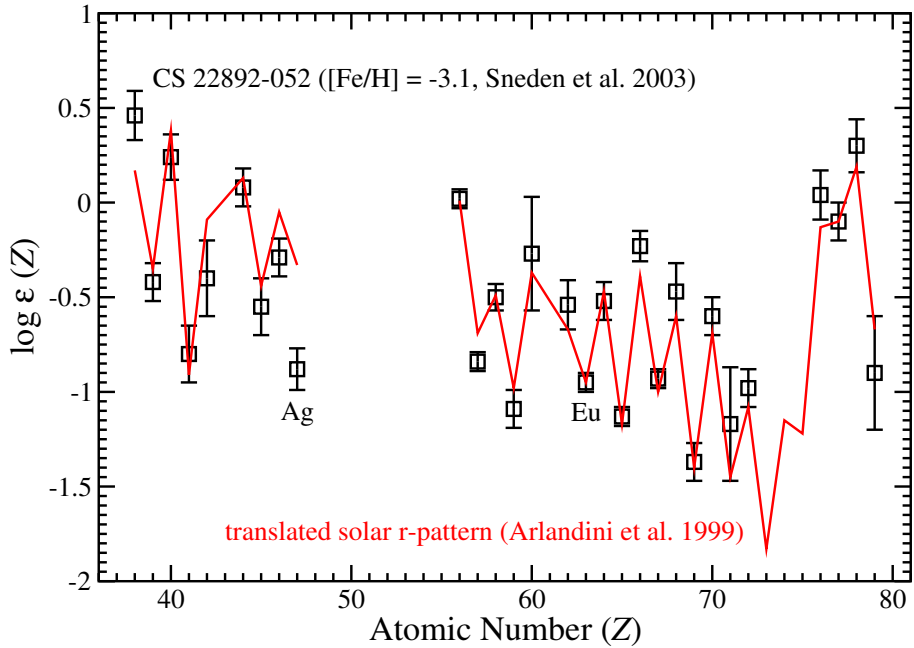}
\end{center}
\caption{Data (squares) on abundances of CPR 
elements and heavy $r$-elements
in CS~22892--052 \citep{sn03}. The curves give the overall
solar ``$r$''-pattern (shown in Fig.~\ref{fig-rsun}b) translated to 
pass through the data on Eu. Note that the heavy $r$-elements
($Z\geq 56$) 
in CS~22892--052 closely follow the corresponding part of the solar
$r$-pattern. There is also gross accord for the CPR elements 
at lower atomic numbers
but with clear discrepancies, especially for Ag.}
\label{fig-sned}
\end{figure}

As discussed above, the study of metal-poor stars is basic to the 
understanding of contributions from individual SNe. The H-K survey
\citep{bps} provided an extensive list of metal-poor star candidates.
A very important paper by \citet{mcw} presented high-resolution 
analyses of 33 extremely metal-deficient stars. This study clearly 
showed large scatter in Sr abundance at fixed [Fe/H] over a range 
of [Fe/H] (see Fig.~\ref{fig-sryzr-fe}a for the same result from
more recent data)
and large scatter in Ba abundance at [Fe/H]~$\sim -3$. 
In an earlier paper \citep{sn94}, the authors of this study reported 
the discovery of a star, CS~22892--052, with [Fe/H]~$=-3.1$ but 
greatly enhanced in heavy $r$-elements. They showed that the 
abundance pattern of heavy $r$-elements in this star is
indistinguishable from the corresponding part of the solar $r$-pattern. 
As this star possessed an abundance ratio of heavy $r$-element to 
Fe in excess of 40 times the solar value, \citet{mcw} concluded that 
``the heavy elements are not produced in the same quantities by all 
type II SN'' and that ``the SN largely responsible for CS~22892--052 
material must be an uncommon occurrence.'' The possibility discussed 
below that the production of heavy $r$-nuclei is completely separate 
from that of the Fe group was not considered. Later Th was also
detected in CS~22892--052 \citep{sn96}. Further work 
\citep{sn00,sn03} then showed 
that there was a clear deficiency in Ag relative to the overall 
solar ``$r$''-pattern (see Fig.~\ref{fig-sned}). As discussed
in Section~\ref{sec-wind}, Ag is one of the CPR elements produced
by the $\alpha$-process and is not of proper $r$-process origin. 

\begin{figure}
\begin{center}
\vskip-1cm
\includegraphics*[scale=0.35]{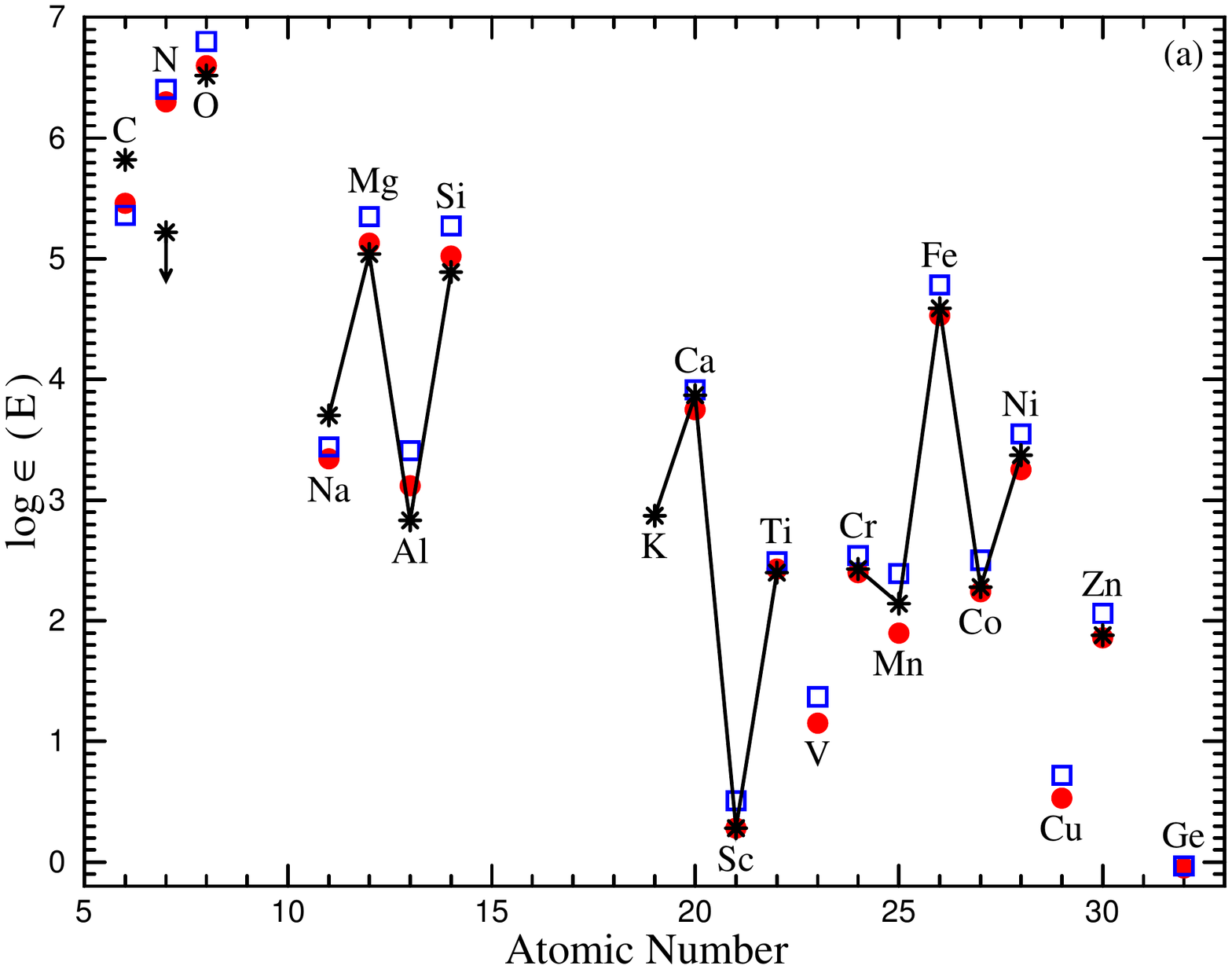}
\vskip-1cm
\includegraphics*[scale=0.35]{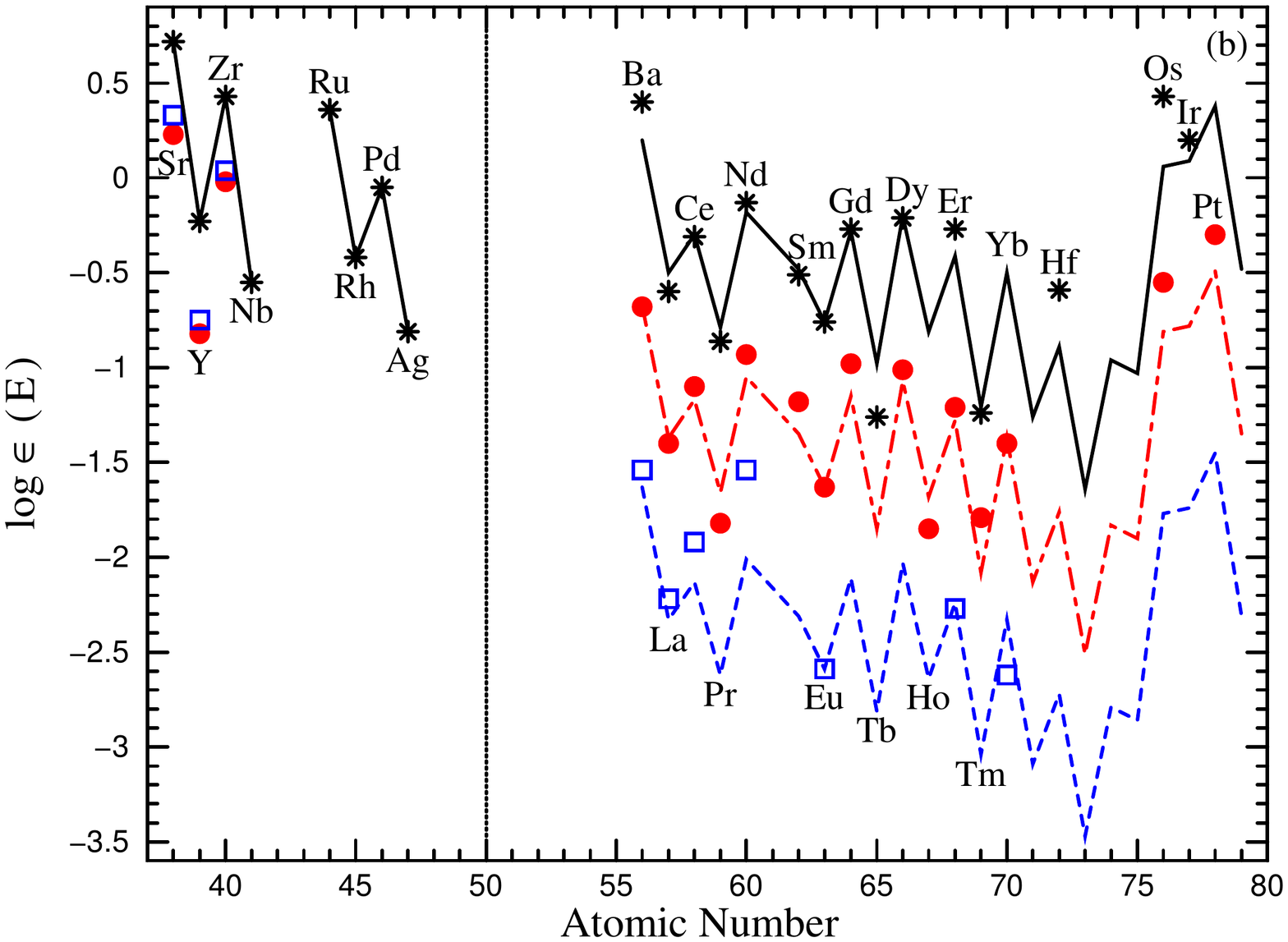}
\end{center}
\caption{Data on the elements from C to Pt in CS~31082--001 [asterisks
\citep{hill}], HD~115444 (filled circles), and HD~122563 [squares \citep{west}]
with [Fe/H]~$=-2.9$, $-2.99$, and $-2.74$, respectively.
(a) The $\log\epsilon$ values for the elements from C to Ge. The data
on CS~31082--001 are connected by solid line segments as a guide.
Missing segments mean incomplete data. The downward arrow at the
asterisk for N indicates an upper limit. Note that the available abundances 
for the elements from O to Ge are almost indistinguishable
for the three stars. (b) The $\log\epsilon$ values for 
the elements from Sr to Pt. The data on the CPR elements are shown
in the region to the left of the vertical dotted line, with those for
CS~31082--001 again connected by solid line segments as a guide. 
There is a rather small range
($\sim 0.5$~dex) in the abundances of the CPR elements for the three
stars. In the region to the right of the vertical dotted line, the data on
the heavy $r$-elements are compared with the solid, dot-dashed, and 
dashed curves, which are the solar ``$r$''-pattern translated to
pass through the Eu data for CS~31082--001, HD~115444, and
HD~122563, respectively. Note the general agreement between the
data and these curves. There is a range of $\sim 2$~dex in the 
abundances of the heavy $r$-elements. Combined with the almost
identical abundances of the elements from O to Ge for the three stars,
this strongly suggests that the sources for the heavy $r$-nuclei and
those for the elements from O to Ge are completely decoupled
\citep{qw02,qw03}.}
\label{fig-fe-decoup}
\end{figure}

\begin{figure}
\begin{center}
\includegraphics*[scale=0.4]{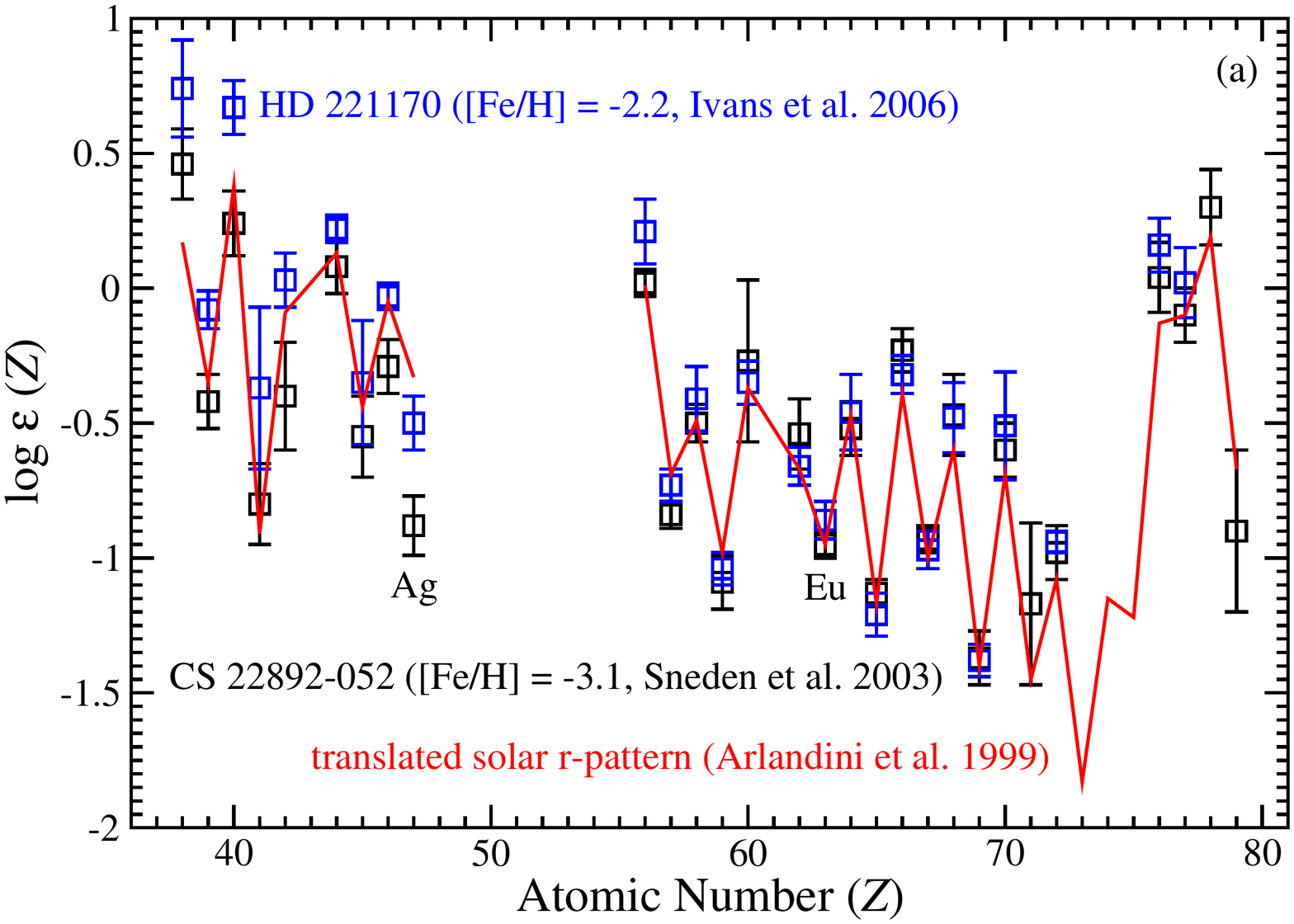}
\includegraphics*[scale=0.4]{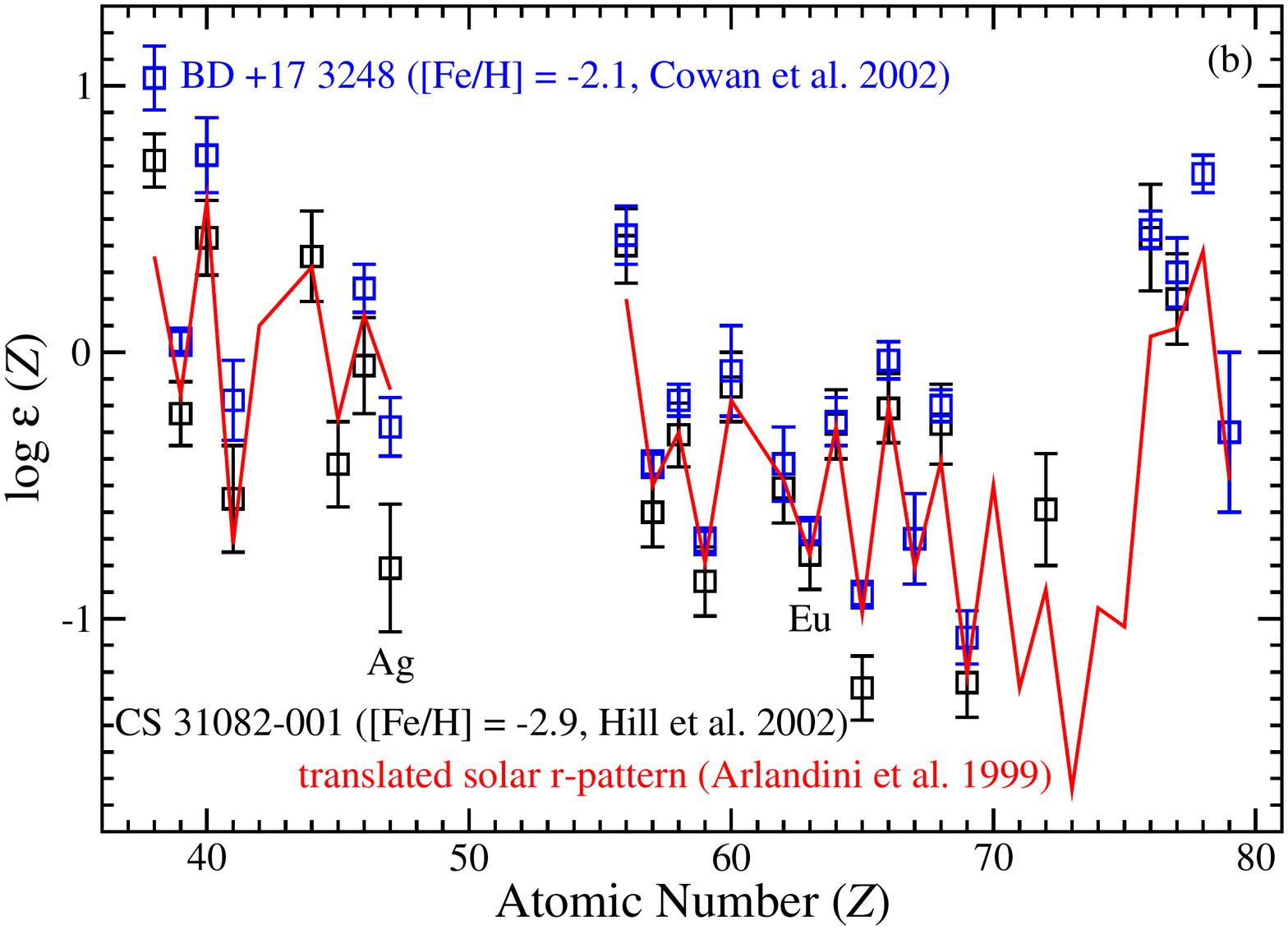}
\end{center}
\caption{Comparison of abundances in two pairs of metal-poor stars.
(a) CS~22892--052 with [Fe/H]~$=-3.1$ [black squares
\citep{sn03}] and HD~221170 with [Fe/H]~$=-2.2$ [blue squares 
\citep{iv06}]. (b) CS~31082--001 with [Fe/H]~$=-2.9$
[black squares \citep{hill}] and BD~$+17^\circ3248$
with [Fe/H]~$=-2.1$ [blue squares \citep{cow02}].
The solid curves in (a) and (b) give the solar ``$r$''-pattern translated
to pass through the Eu data for CS~22892--052 and
CS~31082--001, respectively. Note that each pair of stars
have nearly identical abundances of heavy $r$-elements
closely following the solar $r$-pattern but the (Fe/H) values differ
by a factor of 8 and 6 for the pair in (a) and (b), respectively. There
are also some differences in the abundances of the CPR elements,
especially Ag, for each pair of stars.}
\label{fig-fe-decoup2}
\end{figure}

The discovery of a star, CS~31082--001, with [Fe/H]~$=-2.9$ but
greatly enriched in heavy $r$-elements including both Th and U 
(both radioactive elements but with very different lifetimes)
was reported by \citet{cay}. The abundance ratio 
(Th/H) in this star is $\approx 1/12$ the solar value but the ratio (Fe/H) is
only $\approx 1/800$ the solar value. The data on this star  \citep{hill}
are shown as asterisks in Figure~\ref{fig-fe-decoup} with the heavy 
$r$-elements displayed in the region to the right of the vertical dotted 
line in Figure~\ref{fig-fe-decoup}b. The solid curve in this region
represents the solar $r$-pattern translated to pass through the Eu data.
It can be seen that there is general agreement between the data 
on the heavy $r$-elements and the solar $r$-pattern. Studies by several 
groups provide further examples of stars with a wide range in the
abundances of heavy $r$-elements that again approximately
follow the solar $r$-pattern. The data on two such stars, HD~115444 
with [Fe/H]~$=-2.99$ and HD~122563 with [Fe/H]~$=-2.74$ \citep{west},
are shown as filled circles and open squares, respectively, in 
Figure~\ref{fig-fe-decoup}. The three stars shown in this figure have 
almost identical abundances of the elements from O to Ge including Fe
(see Fig.~\ref{fig-fe-decoup}a), but their heavy $r$-elements differ in 
abundance by a factor of $\sim 100$ (see Fig.~\ref{fig-fe-decoup}b).
This led us to conclude that the production of heavy $r$-nuclei is 
completely separate from the production of the elements of the Fe group
and those of intermediate mass above C and N \citep{qw02,qw03}.

In addition, we may compare some stars with the same enrichment 
in heavy $r$-nuclei but with different [Fe/H]. Figure~\ref{fig-fe-decoup2}a 
shows the data on CS~22892--052 with [Fe/H]~$=-3.1$ [black squares
\citep{sn03}] and the recent results on HD~221170 
with [Fe/H]~$=-2.2$ [blue squares \citep{iv06}]. It can be 
seen that the abundances from Ba ($Z=56$) and above are 
indistinguishable although there is a factor of 8 difference in (Fe/H)
for these two stars. There are some differences in the abundances of
the CPR elements, in particular Ag. A similar comparison can be seen in 
Figure~\ref{fig-fe-decoup2}b for two other stars, CS~31082--001
[black squares \citep{hill}] and BD~$+17^\circ 3248$
[blue squares \citep{cow02}], which also have nearly the same 
abundances of heavy $r$-elements but very different (Fe/H)
(by a factor of 6) with some differences in the abundances of the 
CPR elements.

All of the results presented above demonstrate clearly that the heavy 
$r$-nuclei are not produced in conjunction with the Fe group elements.
Further, we note again that the abundances for all the elements from 
O to Ge appear to be almost identical for the three 
stars with very different enrichments of heavy $r$-elements shown in 
Figure~\ref{fig-fe-decoup}. This suggests that the heavy $r$-nuclei 
cannot be produced by massive stars of $>11\,M_\odot$, which 
result in Fe core-collapse SNe and are sources for the elements from 
O to Ge [see the review by \citet{whw} and Section~\ref{sec-rsource}].

\subsection{Regularity and variability of the yield pattern of heavy 
$r$-nuclei}
\label{sec-var-h}
It appears that the heavy $r$-elements Ba
and above exhibit an abundance pattern close to the corresponding
part of the solar $r$-pattern. This was recognized in all of the 
observational studies cited above. Two stars discussed above,
CS~22892--052 and CS~31082--001, have [Fe/H]~$\sim -3$ but 
extremely high enrichments of heavy $r$-elements 
[(Eu/H)~$\sim (1/30$--$1/20)({\rm Eu/H})_\odot$]. These stars must 
represent contributions from single $r$-process sources.
However, there is no observational basis 
for believing that there is a single universal yield pattern even for the 
heavy $r$-nuclei. In fact, observations show that CS~22892--052
and CS~31082--001 have $\log({\rm Th/Eu})=-0.62$ 
\citep{sn03} and $-0.22$ \citep{hill}, respectively. This difference of
0.4~dex is much larger than the observational error of $\sim 0.05$~dex 
\citep{hill}. Further, it cannot be attributed to the possible 
difference in age between the two stars as $^{232}$Th (the only 
long-lived isotope of Th) has an extremely long lifetime of
$\tau_{232}=20.3$~Gyr ---
even if the two stars were born 13.5~Gyr (age of the universe) apart, 
this would only give a difference of 0.3~dex in $\log({\rm Th/Eu})$.
Therefore, there is good reason to believe that the yields of Th and U 
relative to those of heavy $r$-elements (e.g., Eu) below
$A\sim 195$ should be variable. This variation renders calculations of
stellar ages from (Th/Eu) rather uncertain. Even a $30\%$ shift in 
the yield ratio of Th to Eu would give a shift of 6.1~Gyr in age due to 
the long lifetime of $^{232}$Th.

\begin{figure}
\vskip -1cm
\begin{center}
\includegraphics*[scale=0.38]{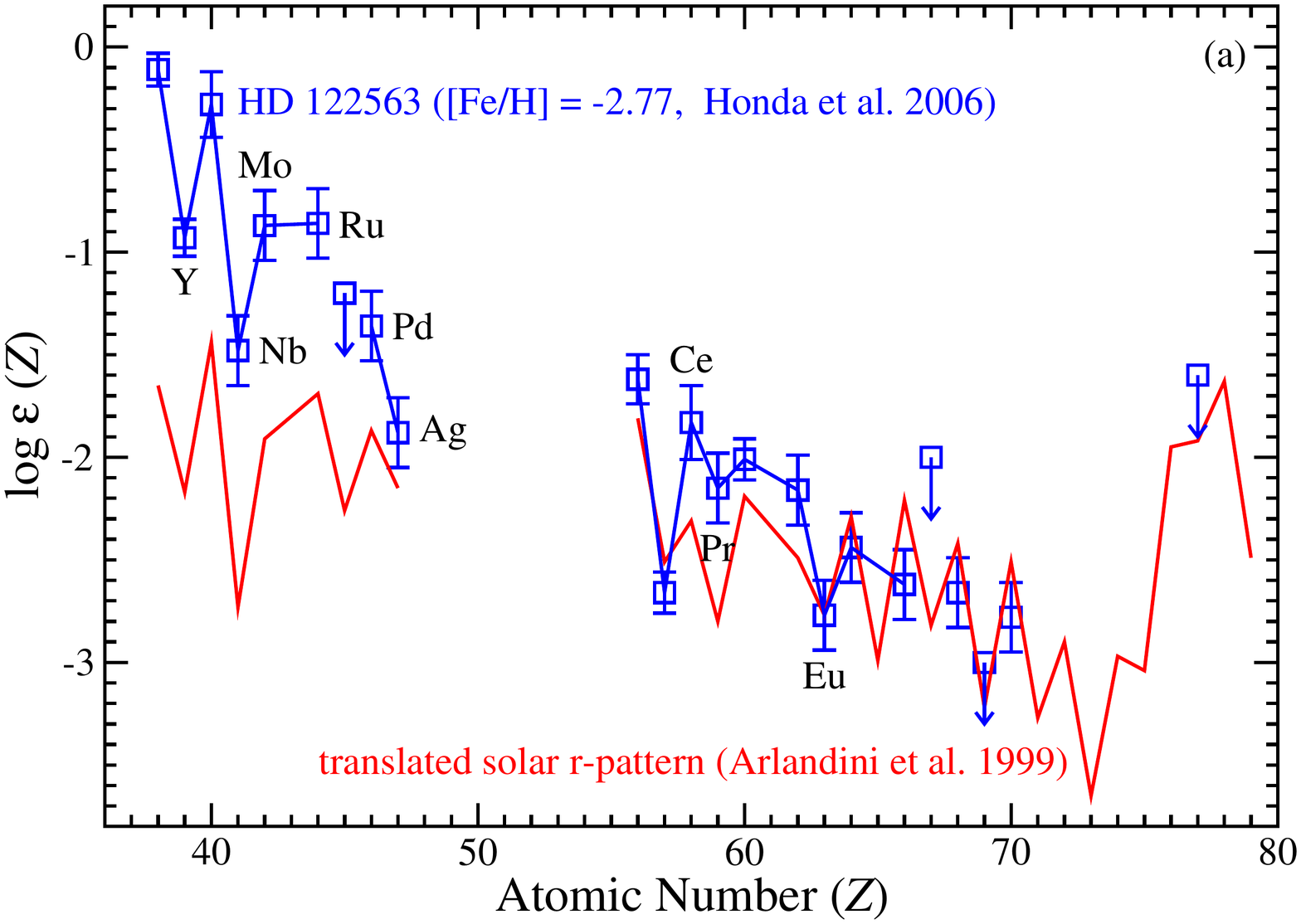}
\includegraphics*[scale=0.38]{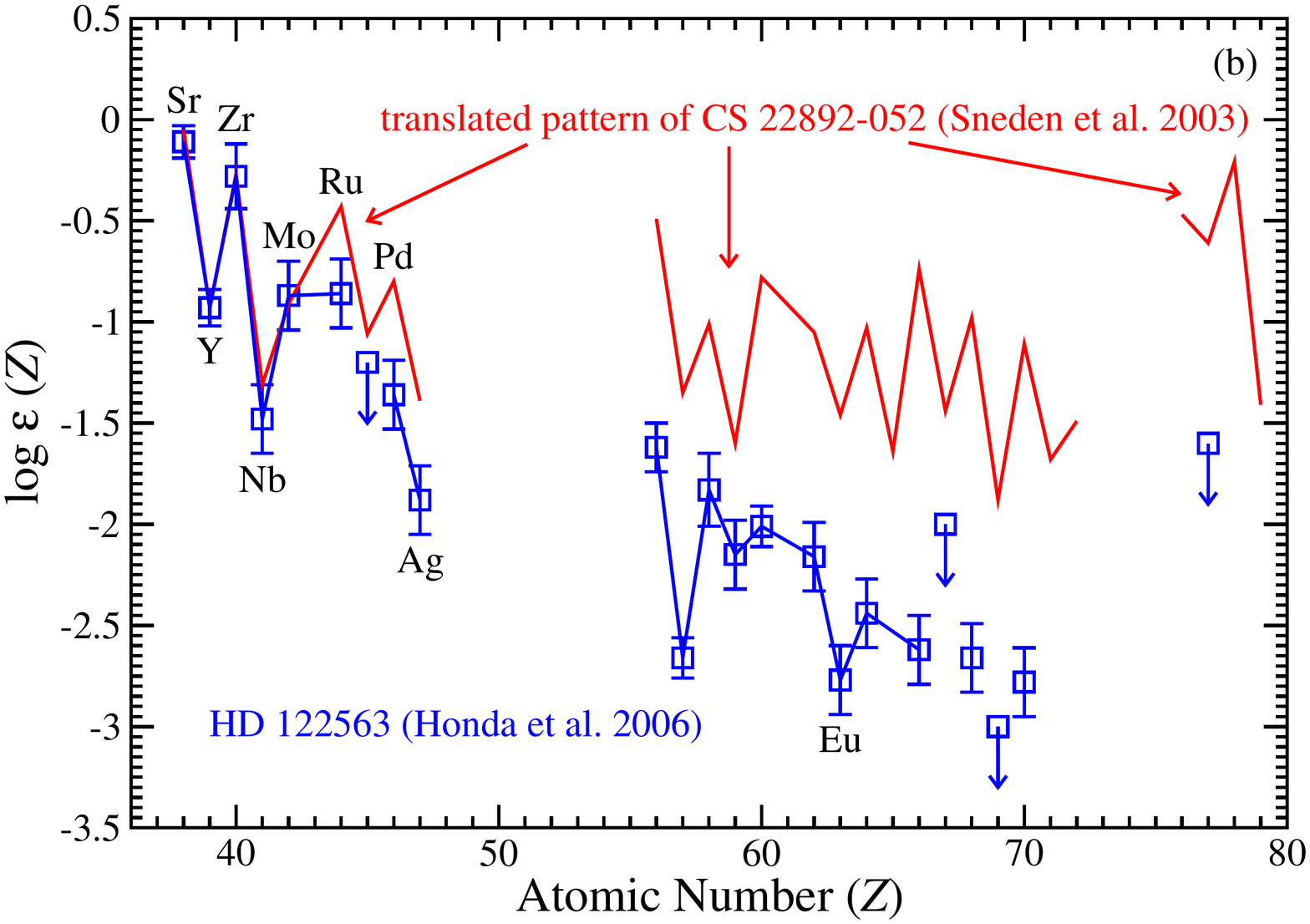}
\end{center}
\caption{(a) Data on HD~122563
[squares \citep{hon06}] compared with  the solar ``$r$''-pattern
translated to pass through the Eu data (red curves).
The squares are connected with blue line segments as a guide. 
Squares with downward arrows indicate upper limits. 
The abundance pattern of 
heavy $r$-elements [Ba ($Z=56$) and above] in HD~122563
shown in (a) is similar to the corresponding part of the solar 
``$r$''-pattern but with substantial differences, especially for Ce and Pr.
Note that (Ba/Eu)$_{\rm HD122563}\approx({\rm Ba/Eu})_{\odot,r}$.
There are also gross differences between the data and the 
translated solar ``$r$''-pattern for the CPR elements.  Specifically,
HD~122563 has much larger proportions of CPR elements relative to
heavy $r$-nuclei as compared to the solar ``$r$''-pattern.
(b) Comparison of the data on HD~122563
[squares \citep{hon06}] with those on
CS~22892--052 [red curves \citep{sn03}]
normalized to the same $\log\epsilon({\rm Y})$ as for HD~122563.
The large 
difference (by a factor of $\sim 20$) in the production of heavy 
$r$-nuclei relative to CPR
elements shown in (b) suggests that only some sources for CPR 
elements can also produce heavy $r$-nuclei.}
\label{fig-cpr-h}
\end{figure}

Further evidence for variations in the yield pattern of heavy $r$-nuclei
has been found in HD~122563 with [Fe/H]~$=-2.77$ \citep{hon06}.
The data on this star (squares) are 
compared with the solar ``$r$''-pattern (red curves) translated to pass 
through the Eu data in Figure~\ref{fig-cpr-h}a. It can be seen that there
is approximate accord between the data and the solar $r$-pattern for
the heavy $r$-elements ($Z\geq 56$), but there are also large 
discrepancies, especially for Ce and Pr (see also the comparison 
between the squares and the
dashed curve in Fig.~\ref{fig-fe-decoup}b). Clearly, it is important
for future measurements to determine whether such discrepancies
extend to other heavy $r$-elements in HD~122563, particularly 
Os, Ir, and Pt in the peak at $A\sim 195$ of the solar $r$-pattern. 
Discovery of other stars of this kind would help establish the range
of variations in the yield pattern of heavy $r$-nuclei.

\subsection{Relationship between heavy $r$-nuclei and CPR elements}
\label{sec-cpr-h}
Figure~\ref{fig-cpr-h}a also shows that the CPR elements from Sr 
($Z=38$) to Ag in HD~122563 clearly lie above the solar ``$r$''-pattern 
translated to pass through the Eu data and extended into this region. 
This is in sharp contrast to the cases for CS~22892--052, 
CS~31082--001, BD~$+17^\circ3248$, and HD~221170 shown in 
Figure~\ref{fig-fe-decoup2}. Therefore, we must conclude that there is
large variation in the production of heavy $r$-nuclei relative to CPR 
elements for different sources or that CPR elements sometimes may be 
produced independently of heavy $r$-nuclei. 

All the available observations show that 
high enrichments of heavy $r$-nuclei are always accompanied 
by enrichments of CPR elements. If some stars were found to be 
highly enriched in heavy $r$-nuclei but without any significant
abundances of CPR elements, this would bring serious doubt to 
any core-collapse SN model for the production of heavy $r$-nuclei.
In the absence of such observations, we may assume that 
a roughly fixed 
amount of CPR elements is produced in the neutrino-driven wind 
whenever a neutron star is formed in a core-collapse SN as 
discussed in Section~\ref{sec-wind} (see also Section~\ref{sec-ycpr}). 
Under this assumption,
we can normalize the data on different stars to the same abundance
of a typical CPR element (e.g., Y) to see the variation in the production 
of heavy $r$-nuclei relative to CPR elements. The data on 
CS~22892--052 [red curves \citep{sn03}] normalized to the same 
$\log\epsilon({\rm Y})$ as for HD~122563 are compared with those
on the latter star [squares \citep{hon06}] in Figure~\ref{fig-cpr-h}b. 
The large difference (by a factor of $\sim 20$)
in the production of heavy $r$-nuclei
[Ba ($Z=56$) and above] relative to
CPR elements shown in this figure (see also 
Fig.~\ref{fig-fe-decoup}b) can be accounted for if only some
core-collapse SNe can produce the heavy $r$-nuclei along with 
the CPR elements but others can only produce the CPR elements
without any heavy $r$-nuclei. This is then fully consistent with the
conclusion reached in Section~\ref{sec-fe-decoup} that heavy 
$r$-nuclei cannot be produced by Fe core-collapse SNe from 
progenitors of $>11\,M_\odot$, which are sources for the elements
above N through the Fe group.

\subsection{Implications for sources of heavy $r$-nuclei}
\label{sec-rsource}

\begin{figure}
\begin{center}
\vskip-2cm
\includegraphics*[scale=0.5]{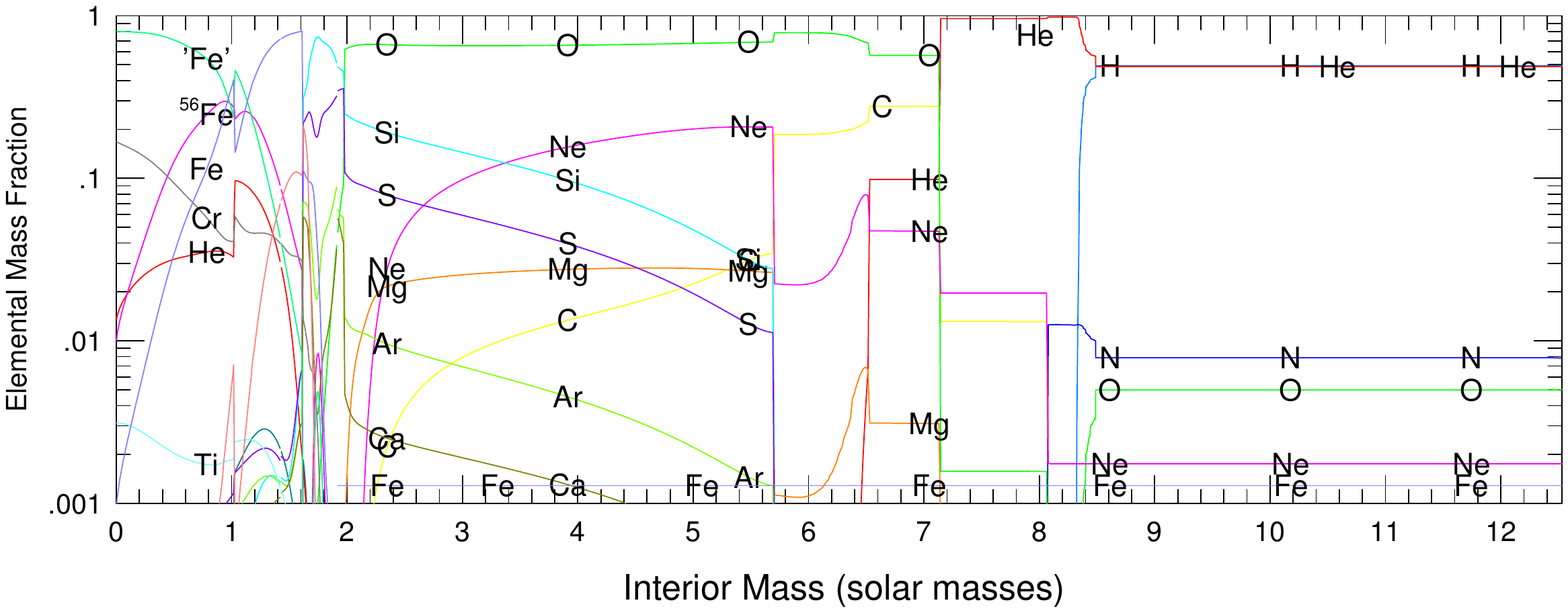}
\vskip-4.5cm
\includegraphics*[scale=0.5]{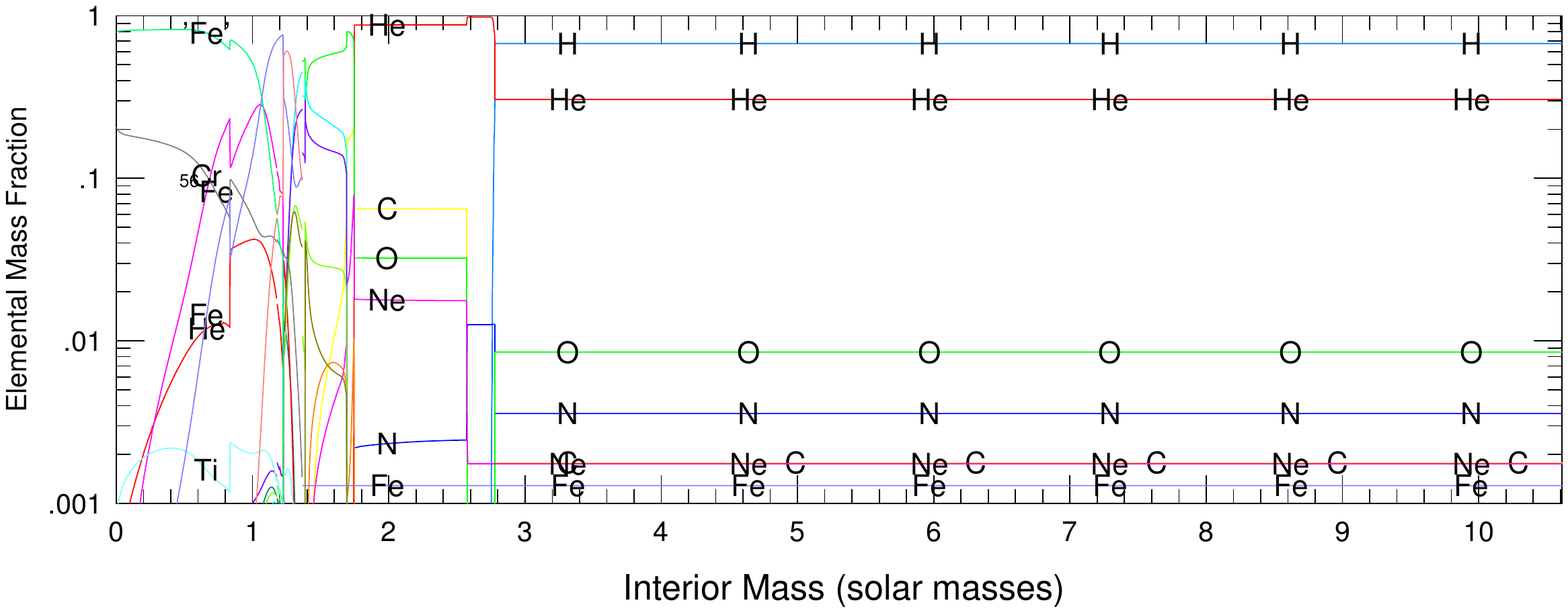}
\end{center}
\vskip-2.5cm
\caption{Composition structure at the onset of core collapse in
terms of the mass fractions of various elements as functions of
the mass coordinate for
two stars with initial masses of 25 (top panel) and $11\,M_\odot$ 
(bottom panel), respectively. Both stars had solar metallicity initially 
and were evolved from the main sequence with mass loss 
\citep{whw}. While both stars have developed an Fe core, the 
$25\,M_\odot$ star has a clear Si shell and an extensive O shell 
but the $11\,M_\odot$ star only has a He shell between the Fe
core and the H envelope.}
\label{fig-fe-core}
\end{figure}

Whether the elements of the Fe group and those of intermediate 
mass above N can be produced by a core-collapse SN is closely 
related to the pre-SN structure of the progenitor. 
Figure~\ref{fig-fe-core} compares the composition structure at the 
onset of core collapse for two stars with initial masses of 
25 (top panel) and $11\,M_\odot$ (bottom panel), respectively. 
Both stars had solar metallicity initially and were evolved from the 
main sequence with mass loss \citep{whw}. It can be seen 
that while both stars have developed an Fe core, the $25\,M_\odot$
star has a clear Si shell and an extensive O shell but the 
$11\,M_\odot$ star does not. The transition between these two kinds
of composition structure occurs at $\sim 12\, M_\odot$ 
[e.g., \citet{ww95}]. 
The elements from Si to the Fe group are produced by explosive 
burning as the SN shock propagates through the Si 
and O shells. (Note that the original Fe nuclei in the inner core become
free nucleons at nuclear density and those in the outer core are 
dissociated by the shock. Therefore, none of the Fe nuclei in the
initial core are ejected in the SN.)
It is considered that Fe core-collapse SNe from 
progenitors of $\lesssim 25\,M_\odot$ eject most of their 
nucleosynthetic products and leave behind neutron stars 
[e.g., \citet{whw}]. Thus, we expect that the products of explosive 
nucleosynthesis are ejected along with the elements
produced by pre-SN hydrostatic burning in the outer shells for 
Fe core-collapse SNe from progenitors of $\sim 12$--$25\,M_\odot$. 
In other words, a host of nuclei from C to the Fe group would be
ejected from such core-collapse SNe. 
In contrast, Fe core-collapse SNe from progenitors of 
$\sim 11\,M_\odot$ eject very little of the elements of the Fe group
or those of intermediate mass above N as these stars only have a He
shell between the Fe core and the H envelope (note that the Fe 
outside the core shown in Fig.~\ref{fig-fe-core} is due to the 
assumed initial metallicity of the star).

\begin{figure}
\begin{center}
\vskip-4cm
\includegraphics*[scale=0.6]{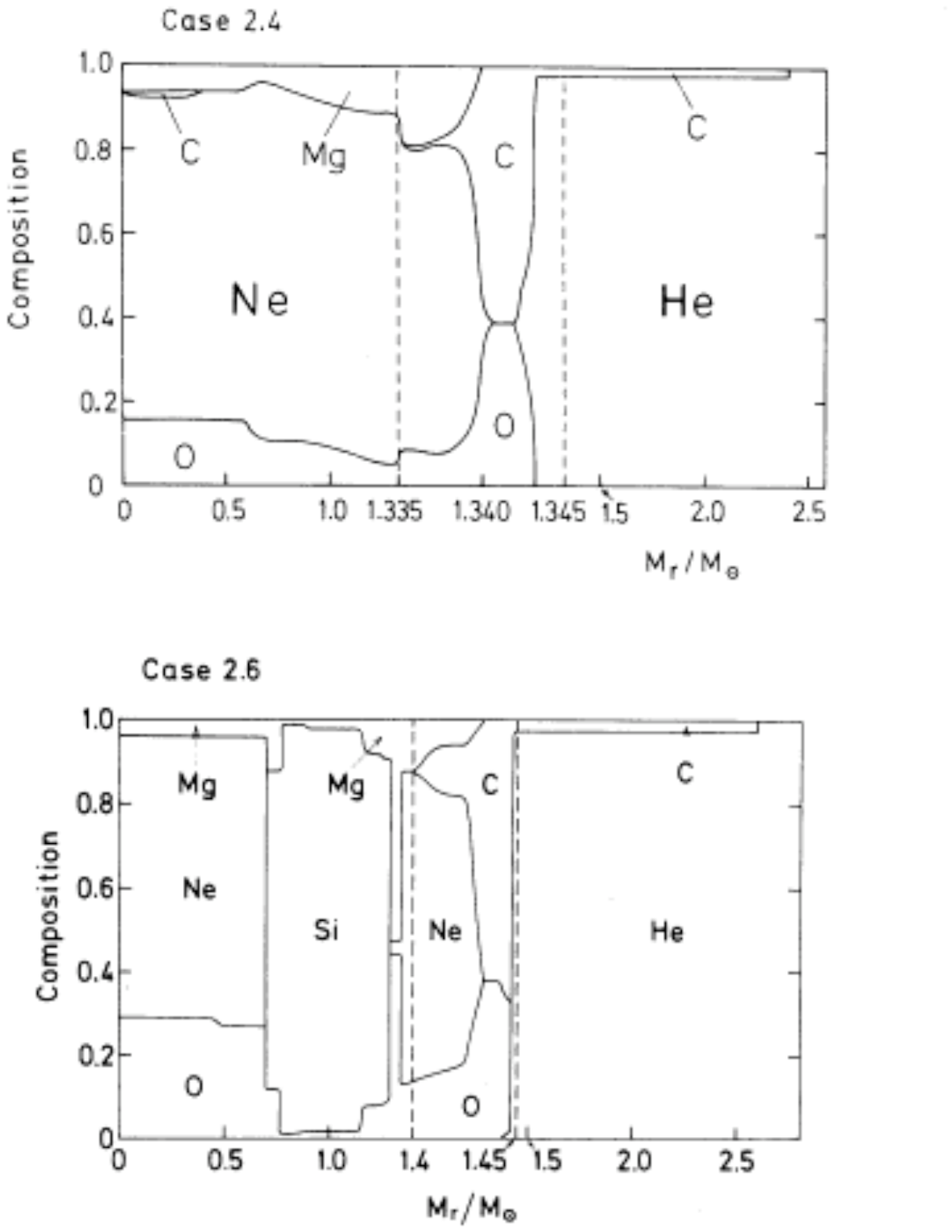}
\end{center}
\vskip-1.5cm
\caption{Composition structure prior to core collapse for He
cores of 2.4 (case 2.4) and $2.6\,M_\odot$ (case 2.6)
corresponding to stars of 9.6 and $10.4\,M_\odot$ \citep{no84}.
At a given mass coordinate, the vertical spacing between the 
boundaries of the region marked with an element gives
the mass fraction of this element. 
Note that case 2.4 has a simple O-Ne-Mg core but 
case 2.6 has a more evolved outer core outside the inner 
O-Ne-Mg core. In both cases, there is only a very thin C-O shell 
between the core and the He shell. The mass coordinate in this
region (between vertical dashed lines) is greatly magnified.}
\label{fig-onemg-core}
\end{figure}

The evolution of stars of $\sim 8$--$10\,M_\odot$ is more
complicated [see reviews by \citet{noha,whw,herw}]. 
\citet{no84,no87} studied the evolution of He cores of 2.2, 2.4, 
and $2.6\,M_\odot$ corresponding to stars of 8.8, 9.6, and
$10.4\,M_\odot$ and showed that these stars develop O-Ne-Mg
cores. The composition structure prior to core collapse for
He cores of 2.4 (case 2.4) and $2.6\,M_\odot$ (case 2.6)
is shown in the top and bottom panels of
Figure~\ref{fig-onemg-core}, respectively.
Note that case 2.4 has a simple O-Ne-Mg core but case 2.6
has a more evolved outer core outside the inner O-Ne-Mg core.
In both cases, there is only a very thin C-O shell between the
core and the He shell. While the further evolution of case 2.6 is 
rather uncertain, \citet{no84,no87} concluded that He cores of
2--$2.5\,M_\odot$ corresponding to progenitors of 
8--$10\,M_\odot$ result in O-Ne-Mg core-collapse SNe that
leave neutron stars but produce very little of the elements of 
the Fe group or those of intermediate mass above N.

An intermediate-mass star of $\lesssim 8\,M_\odot$ would become
a C-O or O-Ne-Mg white dwarf after passing through the AGB phase
and losing its envelope. If such a star is in a binary system with 
a low-mass companion, the white dwarf produced by this star
may accrete sufficient material from the companion and collapse
into a neutron star \citep{nk91}. Accretion is limited by the supply 
from the companion. As the white dwarf must exceed the Chandrasekhar
mass of $\approx 1.4\,M_\odot$ in order to collapse, this AIC
scenario may work only for the more massive white dwarfs produced 
by stars of $\sim 5$--$8\,M_\odot$. The neutron star produced by
an AIC event would have a neutrino-driven wind where CPR 
elements are made. However, with any overlying envelope being 
very small, such an event would not produce the elements above N 
through the Fe group.

From the above discussion and the considerations in the preceding 
subsections, we are left 
with a limited number of scenarios for the production of the heavy
$r$-nuclei. Insofar as this production is directly or indirectly related 
to the neutrino-driven wind from a nascent neutron star, there are
only three possibilities: Fe core-collapse SNe from progenitors of
$\sim 11\,M_\odot$, O-Ne-Mg core-collapse SNe from progenitors
of $\sim 8$--$10\,M_\odot$, and AIC of white dwarfs left behind by
stars of $\sim 5$--$8\,M_\odot$ in binary systems. None of these 
sources produce the elements above N through the Fe group, but all of
them produce CPR elements in the neutrino-driven wind. We will refer
to the first two possibilities collectively as low-mass core-collapse 
SNe. The nucleosynthetic signatures of the AIC scenario will be 
discussed in more detail in Section~\ref{sec-rs}. 
Other workers have proposed a low-mass core-collapse SN source 
for the $r$-process based on considerations of Galactic evolution of 
Eu with Fe [e.g., \citet{maco,iw99}] or parametric models of the 
$r$-process [e.g., \citet{wan03}]. These considerations and models are 
not related to the critical and definitive observations and arguments 
laid out in the preceding subsections. The arguments outlined here 
[see also \citet{qw02,qw03}] are based on the observed decoupling of 
the production of heavy $r$-nuclei from that of the elements above N
through the Fe group as well as the relationship between heavy 
$r$-nuclei and CPR elements.

One might argue that Fe core-collapse SNe with progenitors of 
$>25\,M_\odot$ may not eject the elements above N through the
Fe group due to ``fallback'' [e.g., \citet{ww95}],
and therefore could be a source for the heavy $r$-nuclei. There
are two problems with this argument. First, unlike the elements
from Si to the Fe group, which are produced by explosive burning in 
the inner
shells and therefore most susceptible to fallback for progenitors of 
$>25\,M_\odot$, the elements O, Na, Mg, and Al are produced by 
hydrostatic burning in the outer shells during pre-SN evolution and 
will not be as easily eliminated by fallback [e.g., \citet{ww95}]. Further, 
the heavy $r$-nuclei are usually considered to be made in the inner 
most part of an SN. Were they made in Fe core-collapse SNe with 
progenitors of $>25\,M_\odot$, it is very difficult to see how they 
can survive the fallback that prevents the ejection of those elements
produced in the outer regions.

It follows that low-mass core-collapse SNe and AIC events are
the only possible sources for heavy $r$-nuclei.
In reaching this conclusion, we have assumed
that stars of $\sim 8$--$11\,M_\odot$ would lead to low-mass
core-collapse SNe that do not produce the elements of the
Fe group or those of intermediate mass above N. Investigations
by several groups indicate that the evolution of stars in this narrow mass
interval depends on the initial metallicity and is especially sensitive to the 
treatment of convection and convective overshoot \citep{whw}. 
In particular, the exact mass range resulting in O-Ne-Mg core-collapse 
SNe is rather uncertain.
For example, \citet{rgi} found that for solar metallicity, only stars of 
$11\,M_\odot$ lead to O-Ne-Mg core-collapse SNe while those of
9, 10, and $10.5\,M_\odot$ become O-Ne-Mg white dwarfs after ejecting
their H envelope [see \citet{gbi,rgi96,gri,irg}]. 
This is in conflict with the results shown in
Figures~\ref{fig-fe-core} and \ref{fig-onemg-core}. Because stars of
$\sim 8$--$11\,M_\odot$ play such important roles in understanding the 
sources for heavy $r$-nuclei, it is highly desirable that further work
be pursued to follow the evolution of these stars starting
from the main sequence and considering the effects of the initial 
metallicity and mass loss.
 
\subsection{Occurrence of $s$ and $r$-processes in binary systems}
\label{sec-rs}
The AIC scenario discussed in the preceding subsection may have a
rather complicated history of mass transfer between the primary
intermediate-mass star of $\sim 5$--$8\,M_\odot$ and its low-mass
companion. The companion may accrete material from the primary star
during the red giant branch (RGB) and AGB stages of the latter, thus
acquiring typical products of RGB and AGB evolution, in particular the
$s$-process nuclei ($s$-nuclei). The primary star would
eventually eject its envelope and become a white dwarf.   
If this white dwarf were then
to accrete sufficient material from the expanded 
low-mass companion, it could collapse into a neutron star. While no
elements above N through the Fe group would be produced in this AIC
event, CPR elements would be made in the neutrino-driven wind from
the neutron star. If heavy $r$-nuclei could also be made in this event,
then the low-mass companion would acquire these nuclei as well as
the CPR elements through contamination of its surface by
the ejecta containing such
products. In this scenario, a low-mass star in a binary system may 
become highly enriched first with AGB products including C, N, and 
$s$-nuclei and then with CPR elements and heavy $r$-nuclei \citep{qw03}.
Extensive reviews of nucleosynthesis in AGB stars have been given by
\citet{bgw} and \citet{herw}.

\begin{figure}
\vskip -1cm
\begin{center}
\includegraphics*[scale=0.4]{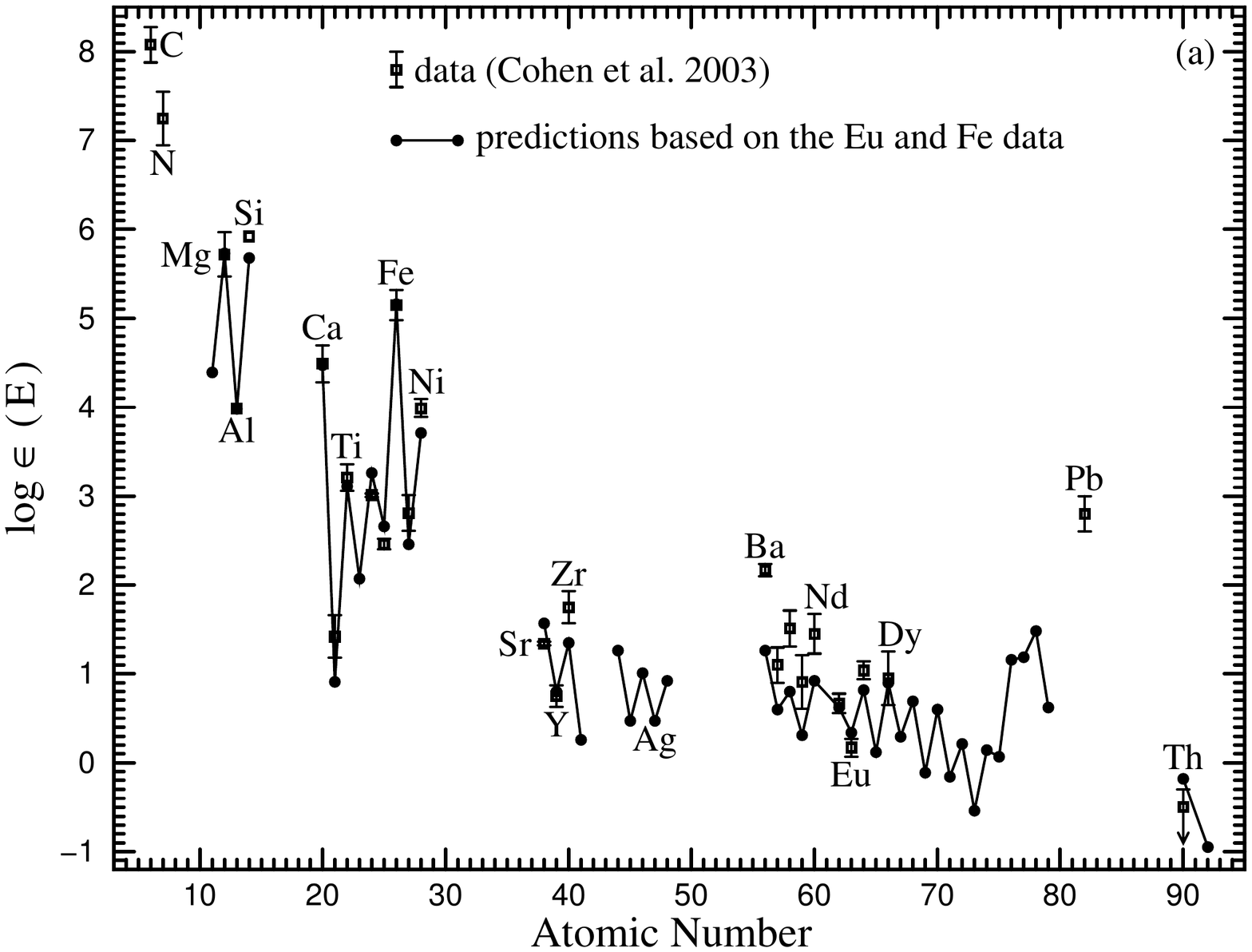}
\vskip -1cm
\includegraphics*[scale=0.4]{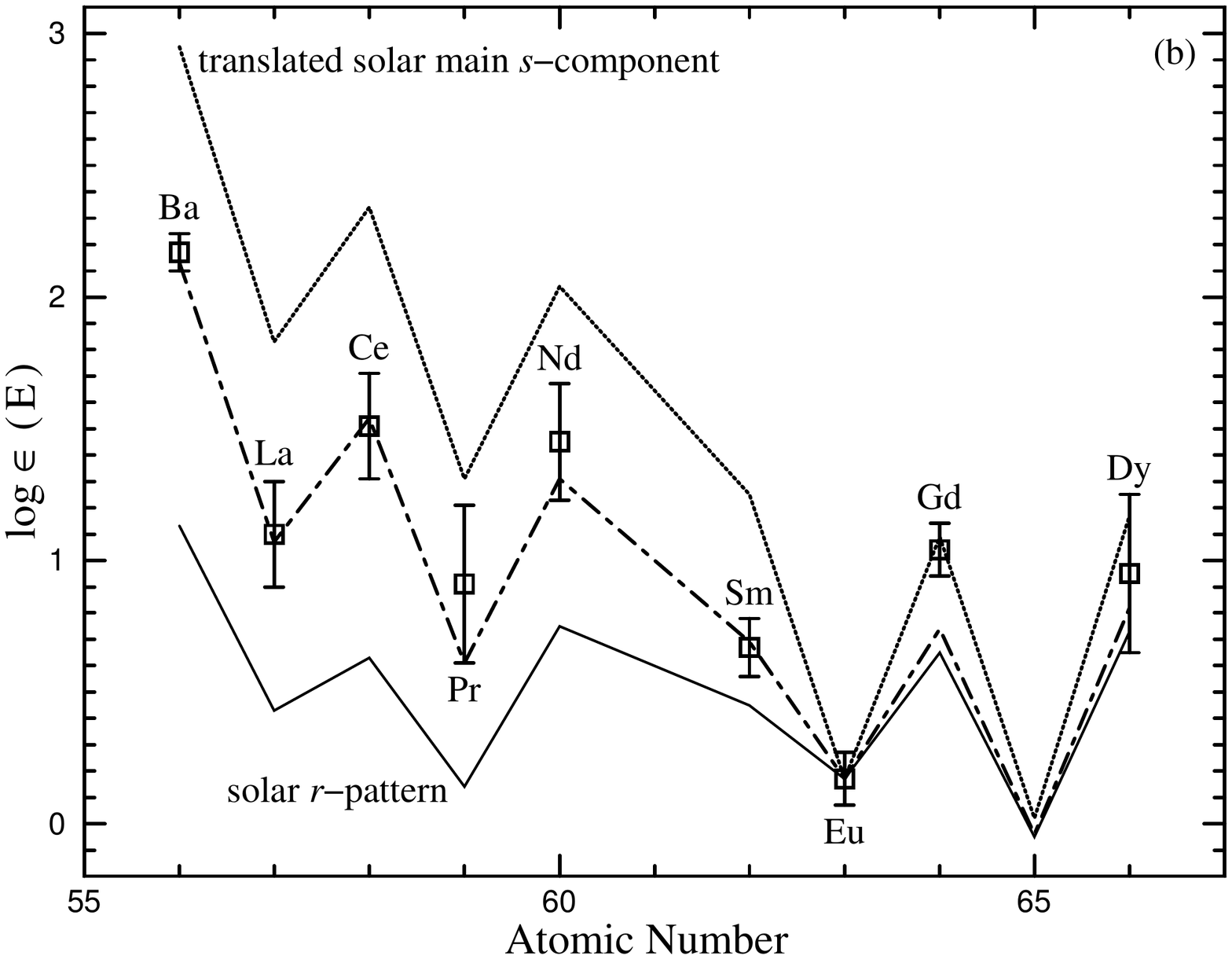}
\end{center}
\caption{Data (squares) on the ``$s+r$'' star HE~2148--1247 with 
[Fe/H]~$=-2.3$ discovered by \citet{cohen}. (a) All data from
C to Th are shown. The filled circles connected with solid
line segments are predictions based on the original ``LEGO-block''
model of \citet{qw01,qw02} using only the Eu and Fe abundances
for the star (see Section~\ref{sec-lego}). This model does not treat
the $s$-process contributions. Note that Ba is far above the
prediction and Pb stands very high. (b) The data on the elements
from Ba to Dy are compared with the solar $r$-pattern (thin solid
curve) and the solar main $s$-component (dotted curve), both of
which are translated to pass through the Eu data. It is clear
that the data must represent a mixture of the solar $r$-pattern
and the solar main $s$-component. The dot-dashed curve shows
such a mixture with 86\% of the Eu contributed by the $r$-process
and the corresponding fractional $r$-process contributions to
the other elements. This mixture describes the data very well
except for Gd.}
\label{fig-rs}
\end{figure}

The above scenario was motivated by the important discovery of a dwarf 
star, HE~2148--1247, which has [Fe/H]~$=-2.3$ but is highly enriched in
neutron-capture elements \citep{cohen}. The observed abundances in
this star are shown as squares in Figure~\ref{fig-rs}a. Specifically, this
star has nearly solar values of (Ba/H), (La/H), (Ce/H), (Pr/H), and (Nd/H),
half the solar value of (Eu/H), and 6.6 times the solar value of (Pb/H).
In addition, its (C/H) and (N/H) are 0.4 and 0.2 times the corresponding
solar values, respectively, with $^{12}$C/$^{13}$C~$\sim 10$ ($\sim 9$
times less than the solar ratio). 
Being a dwarf star, HE~2148--1247 could not have produced the 
observed high enrichments in C, N, and neutron-capture elements
by itself. 
These enrichments must represent the result of mass transfer from a 
more massive and rapidly evolving binary companion that had gone 
through the RGB and AGB stages. Indeed, this star is observed to be a 
radial velocity variable and is most likely a member of a long-period 
binary system \citep{cohen}.
The clear enhancement of Ba relative to Eu and gross enhancement of Pb 
demonstrate that the surface material of this star had undergone extensive 
$s$-processing. On the other hand, the Ba/Eu ratio is much smaller than
those typically obtained in $s$-process models [e.g., \citet{van}], thus
requiring significant contributions from the $r$-process. Except for Gd and
Pb, the data on other neutron-capture elements in HE~2148--1247
can be fitted very well by
a mixture of the solar main $s$-component and the solar $r$-pattern with 
86\% of the Eu contributed by the $r$-process 
[\citet{qw03}, see Fig.~\ref{fig-rs}b]. Metal-poor stars with both high
$s$ and $r$-enrichments appear to be quite common in binary systems. 
Shortly after the discovery of HE~2148--1247 by \citet{cohen}, similar
results on another star, CS~29497--030, with [Fe/H]~$=-2.16$ were 
reported by \citet{spc}. The latter star is also in a binary system
with evidence of mass transfer.

\begin{figure}
\vskip-4cm
\begin{center}
\includegraphics*[scale=0.5]{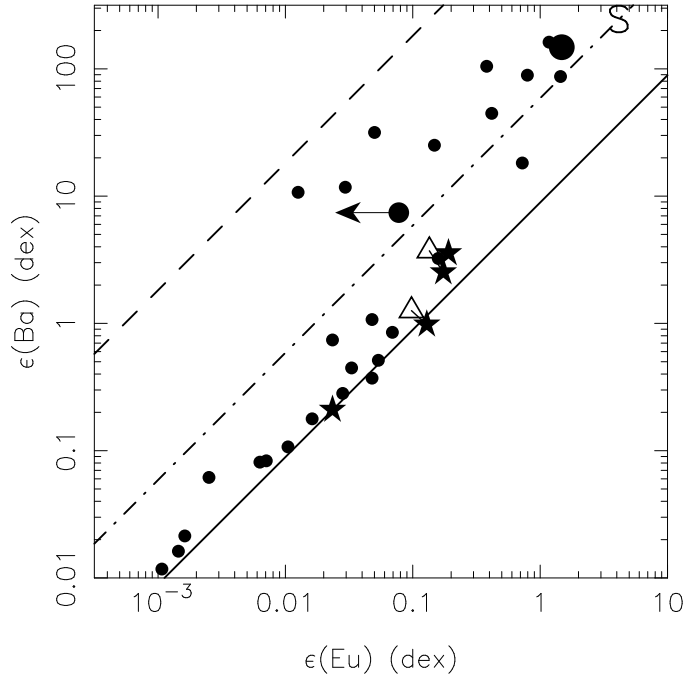}
\end{center}
\caption{Data on $\epsilon({\rm Ba})$ and $\epsilon({\rm Eu})$ 
for a sample of metal-poor stars with
[Fe/H]~$<-2$ compiled by \citet{cohen}. See that paper for
the sources of the data. The dot-dashed line 
represents the solar Ba/Eu ratio with the letter ``S'' on
this line indicating the position of the Sun. The solid line 
represents the Ba/Eu ratio for the solar $r$-pattern and 
the dashed line for the solar main $s$-component. 
HE~2148--1247 is shown as the large filled circle. 
The filled circle with the arrow represents a star with only 
an upper limit on Eu.
It can be seen that there are a number of other stars 
like HE~2148--1247 exhibiting both high $s$ and $r$-process 
enrichments.}
\label{fig-baeu-rs}
\end{figure}

The data on $\epsilon({\rm Ba})$ versus $\epsilon({\rm Eu})$ for a sample
of metal-poor stars with [Fe/H]~$<-2$ were shown in Figure~13 of
\citet{cohen}. This figure is reproduced here as Figure~\ref{fig-baeu-rs}.
The dot-dashed line represents the solar Ba/Eu ratio with the letter ``S'' on
this line indicating the position of the Sun. The solid line represents the
Ba/Eu ratio for the solar $r$-pattern and the dashed line for the solar main
$s$-component. HE~2148--1247 is shown as the large filled circle
in Figure~\ref{fig-baeu-rs}. It can be clearly seen that this low-metallicity 
star has near solar Ba and Eu
abundances with a Ba/Eu ratio slightly above that in the Sun. It can also
be seen that there are a number of other stars 
like HE~2148--1247 exhibiting both high $s$ and $r$-process enrichments. 
A more recent compilation of such stars, all of which appear to be greatly 
enhanced in C, can be found in \citet{jon06}. In addition to reporting the
new results on HE~0338--3945, these authors provided a comprehensive 
summary of both observations on similar stars in the literature 
[including those of \citet{cohen} and \citet{spc}] and theoretical 
interpretations of these results.

In the cases where both $s$ and $r$-process additions have been 
made to a low-mass star in a binary system, one may question whether 
the $r$-nuclei were already present in the material from which this star 
and its binary companion were formed. 
The viewpoint that the $r$-process 
material was already present in the ISM from which these stars were
formed has been pursued by several groups [e.g., \citet{iv05,aok06}].
We first note that in some of the stars with both $s$ and $r$-process 
enhancements, the level of $r$-enrichment is extremely high
(see Fig.~\ref{fig-baeu-rs}). For
example, HE~2148--1247 has half the solar value of (Eu/H). If we assume
that over the Galactic history of $\sim 10$~Gyr, $\sim 10^8$ core-collapse
SNe provided a solar (Eu/H) to a total of $\sim 10^{10}\,M_\odot$ of gas,
then to provide (Eu/H) at the solar level by a single SN would require that
the ejecta from this SN be mixed with only $\sim 100\,M_\odot$ of ISM. 
This mixing mass is much smaller than the typical value of 
$\sim 3\times 10^4\,M_\odot$ for SN ejecta [e.g., \citet{thorn}].
It is also known that the ``$s+r$'' combination is rather common in
metal-poor stars in binary systems but
extremely high $r$-enrichments are rare
among single metal-poor stars [e.g., \citet{jon06}]. Thus the above approach 
requires an arbitrary intimacy between a heavily $r$-enriched ISM and the 
formation of a binary system. We find it difficult to accept the notion that 
metal-poor binary systems were somehow preferentially selected to 
inherit very high $r$-enrichments from a precursor 
ISM or unrelated but intimate source. We consider the observational data to 
most strongly support the point of view that the ``$s+r$'' mixtures in
low-metallicity stars represent mass transfer involving both
$s$ and $r$-process products between these stars and their binary 
companions and that such binary systems were formed with initial 
compositions typical of the ISM. Note that in our preferred scenario,
the $s$-processing occurs during the AGB phase of the primary star
and precedes the $r$-processing proposed to occur during the later
AIC event. Thus, in contrast to the scenario assuming that the
binary system was formed from heavily $r$-enriched ISM, the heavy
$r$-nuclei are not exposed to $s$-processing in our scenario.

We again note that the ejecta from an
$r$-process event is diluted by a large amount of ISM so that stars
formed from an ISM enriched by a small number of events would be unlikely 
to exhibit high $r$-enrichments. Observations targeted toward metal-poor
stars with extreme $r$-enhancements would not be sampling the dominant
population of stars formed from such an ISM. Instead, such observations
would automatically be biased toward stars with extreme $s+r$ 
enhancements in binary systems. However, the results from these
observations do not imply that the $r$-process only occurs
in binary systems as they do not represent $r$-enrichment of the 
general ISM.

\begin{figure}
\begin{center}
\includegraphics*[scale=0.5]{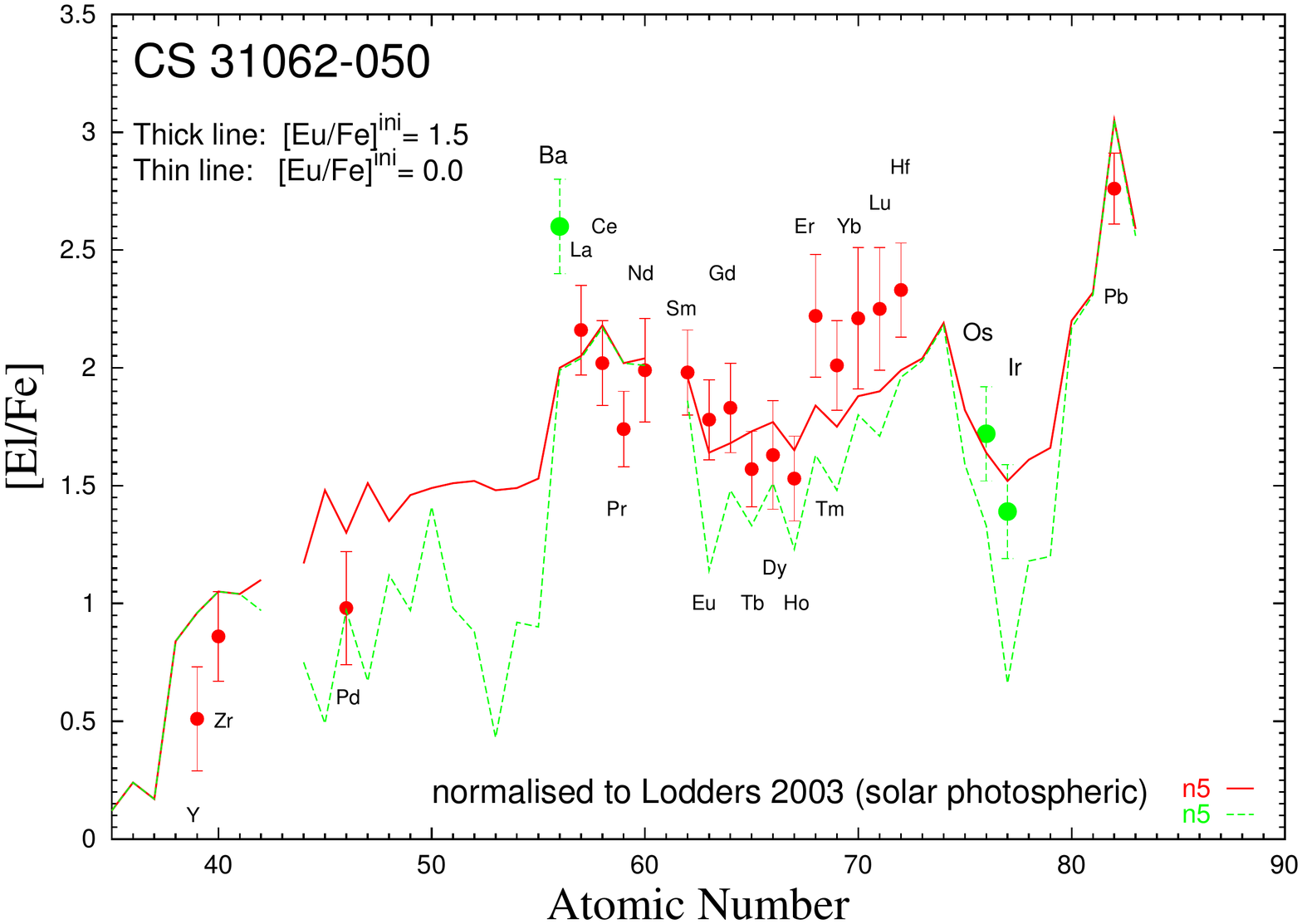}
\end{center}
\caption{Data [red filled circles: \citet{jb04};
green filled circles: \citet{aok06}]
on the elemental abundances relative to Fe 
in terms of [El/Fe]~$\equiv\log({\rm El/Fe})-\log({\rm El/Fe})_\odot$
for CS~31062--050 with [Fe/H]~$=-2.3$. As shown by \citet{jb04},
the very high abundances of Ba and Pb as well as Eu require
a mixture of $s$ and $r$-process contributions. 
The new measurements by \citet{aok06} on Os and Ir, both predominantly
made by the $r$-process, show that the abundances of these two elements
relative to Eu are compatible with the solar $r$-pattern. The solid
red curve gives the results from a theoretical model that calculated 
the $s$-process yields using the standard $^{13}$C pocket and assuming
an initial $r$-process inventory corresponding to 
[Eu/Fe]$^{\rm ini}=1.5$ inherited from the ISM. The green dashed
line is for a similar model with [Eu/Fe]$^{\rm ini}=0$. It can be
seen that a general overall agreement between the models and data 
was obtained for some elements using Pb as the key guide. However,
there is no particular agreement for many other elements including
Ba as noted by \citet{aok06}. The problem of full modelling of
``$s+r$'' mixtures is a matter for ongoing studies.}
\label{fig-aoki-rs}
\end{figure}

With regard to the calculations of $s$-processing in 
low-metallicity stars, there are several issues that require attention. 
First, calculations of the $s$-process in these stars have typically 
assumed that the neutron source is from a $^{13}$C pocket. It is not 
at all evident that this is applicable in the circumstances where 
AIC is involved. There the stars in which the $s$-process occurs 
must be of $\sim 5$--$8\,M_\odot$ and may have considerably different 
characteristics from the stars of $\sim 1$--$3\,M_\odot$ usually 
associated with the main $s$-process. In particular, at the more elevated 
temperatures in the intermediate-mass stars, the reaction
$^{22}{\rm Ne}(\alpha,n)^{25}{\rm Mg}$ will play a significant role. 
Calculations that just use the neutrons from this source may give 
significantly different $s$-yield patterns from those produced with
the standard $^{13}$C pocket.
\citet{aok06} recently reported for the first time the detection of
Os and Ir in CS~31062--050, a star with [Fe/H]~$=-2.3$ but with
high $s + r$ enrichments [see also \citet{jb04}]. 
The elements Os and Ir are predominantly
made by the $r$-process and their abundances
in CS~31062--050 relative to the Eu enhancement found
by \citet{jb04} are compatible with the solar $r$-pattern.
In accounting for the observed abundances of Eu, Os, and Ir,
\citet{aok06} assumed that these $r$-enrichments were inherited
by the star from the ISM (see discussion in the two preceding
paragraphs).
The highly enhanced Pb in CS~31062--050
again clearly shows the presence of an $s$-component produced
by a low-metallicity AGB star as predicted by \citet{gals}. However, 
as noted by \citet{aok06} and seen in Figure~\ref{fig-aoki-rs}, the
$s$-process model proposed to 
explain the Pb clearly fails to explain the Ba
and some of the rare-earth elements (e.g., Er to Hf). It does 
not appear that the observations 
can be described using a standard $s$-process model. Studies
of the $s$-process in low-metallicity intermediate-mass stars of  
$\sim 5$--$8\,M_\odot$ must be undertaken in 
order to better understand the problem at hand. 

The question of whether a higher-mass star of e.g., $\sim 10\,M_\odot$
could produce the $s$-nuclei and transfer them to a low-mass binary
companion also remains to be addressed. Such a binary system 
could lead to the same general results as for the AIC scenario outlined
above. With a primary star of $\sim 10\,M_\odot$, the $r$-nuclei
may be produced by the O-Ne-Mg core-collapse SN from this star
(assuming a neutron source).
This was preferred by some workers [e.g., \citet{wan06}] because it 
does not rely on the transfer of material from the low-mass companion 
to induce the collapse as in the AIC scenario. As noted in
Section~\ref{sec-rsource}, the evolution of stars of 
$\sim 8$--$11\,M_\odot$ is rather complex. This mass range is bounded
from below by the onset of O-Ne-Mg core formation and from above by
the onset of Fe core formation. Neither boundaries have been clearly 
determined [see e.g., the reviews by \citet{whw} and \citet{herw}].
Whether stars in the above mass range
leave an O-Ne-Mg white dwarf or undergo
O-Ne-Mg core collapse to produce a neutron star also requires further
investigation. A complete knowledge of the full stellar evolution for
stars of $\sim 8$--$11\,M_\odot$ is urgently needed. This has been 
mentioned by \citet{herw}, who referred to these as ``super-AGB'' 
stars. A new report on the evolution of such stars by \citet{sagb}
is in preparation.

We also note that some studies of the detailed evolution of single
intermediate-mass stars have been made [e.g., \citet{kl04}], but
none of them carried out $s$-process calculations. The same is
true of the higher masses of $\sim 10\,M_\odot$. For
further progress to be made, it is now evident that detailed calculations
for the evolution of binary systems with a primary member of 
$\sim 5$--$11\,M_\odot$ and a low-mass companion, including
$s$-processing and
various episodes of mass transfer, are necessary in order to tell whether 
mass transfer of all resulting
elements (C, N, and $s$-nuclei) can plausibly explain the observational
data at hand and the increasing quantity of such data. 

\subsection{Where has the $r$-process gone?}
\label{sec-song} 
We summarize the observations on metal-poor stars in terms of the
following rather strict rules on the $r$-process:

(1) The production of heavy $r$-nuclei with $A>130$ is fully decoupled
from that of Fe and all the other elements between N and Ge.

(2) The production of CPR nuclei with $A\sim 90$--110 is not tightly
coupled to that of heavy $r$-nuclei. (All the available observations show
that while metal-poor
stars highly enriched with heavy $r$-nuclei always have 
enrichments of CPR nuclei at a comparable level, there are also stars 
enriched with CPR nuclei but having very low abundances of
heavy $r$-nuclei.)

The meteoritic data on $^{129}$I and $^{182}$Hf discussed in
Section~\ref{sec-intro} give the third rule:

(3) The production of heavy $r$-nuclei including $^{182}$Hf is decoupled 
from that of light $r$-nuclei with $A\lesssim 130$ including $^{129}$I.

To infer the possible sites for the $r$-process, we combine the above
rules with the following aspects of stellar evolution and nucleosynthesis
discussed in Sections~\ref{sec-wind}, \ref{sec-rsource}, and
\ref{sec-rs}:

(a) Stars of $\sim 12$--$25\,M_\odot$ result in Fe core-collapse
SNe, which produce a neutron star of $\approx 1.4$--$2\,M_\odot$
and are sources for the elements between N and Ge.

(b) A neutron star of $\approx 1.4\,M_\odot$
is produced in: Fe core-collapse SNe from progenitors of 
$\sim 11\,M_\odot$; O-Ne-Mg core-collapse SNe from progenitors
of $\sim 8$--$10\,M_\odot$; and AIC of white dwarfs from progenitors 
of $\sim 5$--$8\,M_\odot$ in binary systems. None of these
core-collapse SNe produce the elements between N and Ge.

(c) The nascent neutron star produced by any of the core-collapse
SNe in (a) and (b) has a neutrino-driven wind, which is a natural source 
for CPR nuclei.

Table~\ref{tab-sum} summarizes aspects (a)--(c) and shows the logical 
alternatives from considering these and rules (1)--(3). If we assume that
there is an adequate neutron source directly or indirectly related to the 
neutrino-driven wind from all neutron stars by some unknown 
mechanism, then all the core-collapse SNe listed above could provide
both light and heavy $r$-nuclei. However, in consideration of the strict 
decoupling of heavy $r$-nuclei from Fe [rule (1)] , Fe core-collapse SNe 
from progenitors of $\sim 12$--$25\,M_\odot$ are excluded as a source
for heavy $r$-nuclei. If we also consider the meteoritic data [rule (3)]
and assume that $^{182}$Hf and other heavy $r$-nuclei
are produced by low-mass SNe from 
progenitors of $\sim 8$--$11\,M_\odot$ and AIC events, then these
sources cannot be the source for the light $r$-nuclei. This leaves
Fe core-collapse SNe from progenitors of $\sim 12$--$25\,M_\odot$
as a possible source for the light $r$-nuclei. As all the
core-collapse SNe listed above are sources for CPR nuclei but
only low-mass SNe and AIC events could provide heavy $r$-nuclei,
rule (2) on the relationship between these two groups of nuclei 
is also satisfied.

\begin{table}
\caption{Characteristics of possible sources for CPR and $r$-process nuclei}
\smallskip
\begin{tabular*}{\hsize}{@{\extracolsep{\fill}}*{4}l@{}}
\hline
\hline
Progenitor&AIC&8--$11\,M_\odot$&12--$25\,M_\odot$\\
Neutron star mass&$\approx 1.4\,M_\odot$&$\approx 1.4\,M_\odot$&
$\approx 1.4$--$2\,M_\odot$\\
Fe source?&no&no&yes\\
\hline
\multicolumn{4}{c}{considering neutrino-driven winds}\\
\hline
CPR source?&yes&yes&yes\\
\hline
\multicolumn{4}{c}{assuming neutron sources}\\
\hline
$A\lesssim 130$ source?&yes&yes&yes\\
$A>130$ source?&yes&yes&yes\\
\hline
\multicolumn{4}{c}{considering rule (1) on decoupling of $A>130$ from Fe}\\
\hline
$A>130$ source?&yes&yes&no\\
\hline
\multicolumn{4}{c}{considering rule (1) and rule (3) on 
decoupling of $^{182}$Hf from $^{129}$I}\\
\hline
$A\lesssim 130$ source?&no&no&yes\\
\hline
\end{tabular*}
\label{tab-sum}
\end{table}

Regarding the heavy $r$-nuclei, we again note the remarkable fact 
that their abundance pattern is rather constant. It follows that there
must be a rather fixed nuclear processing scheme for producing 
them. A plausible mechanism is
fission cycling. The meteoritic data require that $^{129}$I not be
produced together with $^{182}$Hf and other heavy $r$-nuclei.
This sets a lower bound of $A\sim 130$ on the nuclei produced 
by fission cycling if it is responsible for the yield pattern of
heavy $r$-nuclei. However, we have no direct observational evidence
on the nuclei with $A\sim 130$ from metal-poor stars that might 
elucidate the matter. If fission cycling is operative, it must permit
some variability in the yield pattern over $130<A\lesssim 195$ 
(as shown by some data sets, see Figs.~\ref{fig-fe-decoup} and 
\ref{fig-cpr-h}a) and in the yield of Th relative to Eu (as shown
by the data on CS~22892--052 and CS~31082--001,
see discussion in Section~\ref{sec-var-h}). Fission cycling
would be rather unlikely if some stars were observed to have high
enrichments in nuclei with $130<A\lesssim 195$ but very low
Th abundances.

In any case, the neutron source for producing the heavy $r$-nuclei 
is not known and does not appear to be in the neutrino-driven wind. 
If the wind is not the source of the neutrons, 
then some other mechanism must exist in the same stellar
environment but independent of the wind.
This undetermined mechanism appears to be restricted to 
stars of $\lesssim 11\,M_\odot$. We do not know the answer to this 
puzzle and would be delighted if anyone knew!!

There is a little children's song:
\begin{verse}
Oh where, oh where has my little dog gone,\\
Oh where, oh where can he be?\\
With his ears cut short and his tail cut long,\\
Oh where, oh where can he be?
\end{verse}

For the $r$-process that we know exists, we propose the following ditty:
\begin{verse}
Oh where, oh where has my $r$-process gone,\\
Oh where, oh where can it be?\\
With neutrons cut short and ``neutrino'' winds long,\\
Oh where, oh where can it be?
\end{verse}

We end this long section with a cautionary note. We here have only 
explored the consequences of the decoupling between Fe and 
heavy $r$-nuclei. The sharp decoupling 
of the production of Fe and other elements between N and Ge
from that of heavy $r$-nuclei is based upon a considerable body of 
observations on low-metallicity stars. However, such observations
would have missed low-mass stars enriched by the following hypothetical 
scenario. If somehow a low-metallicity star of $\sim 11\,M_\odot$ 
managed to produce heavy $r$-nuclei and ejected a substantial amount 
of Fe (e.g., from its original Fe core by some unknown mechanism) 
at the same time, and if it
was in a binary system with a low-mass companion, then the surface of
the companion would have been contaminated with large amounts of 
heavy $r$-nuclei and Fe. If the primary star had also gone through an 
AGB phase, it would have produced a large amount of Pb by the 
$s$-process at low-metallicity. Whether the binary system remains intact 
or was disrupted, the original 
low-mass companion would now appear to be highly enriched
in heavy $r$-nuclei and Pb but also be Fe rich. Such stars would not be 
sampled by the observations directed toward halo stars with 
low to very low [Fe/H] values. An effort to investigate halo stars with
high [Fe/H] values and very large Pb enhancements might clarify 
whether the above cautionary scenario is plausible. 

\section{A LEGO-block model of elemental abundances in stars and the ISM
at low metallicities}
\label{sec-lego}
We have outlined our consideration of the characteristics exhibited by
stellar sources of elements in the regime of [Fe/H]~$\lesssim -1.5$ where
only massive stars might contribute to the ISM. This led to a series of
rules (or regularities) for the presumed elemental yield patterns of
possible prototypical sources. Based on these rules we attempted
to use a simple three-component model to explain the abundances in stars 
and the ISM at low metallicities in general. This ``LEGO-block'' model
has three assumed building blocks: 

$H$ --- the heavy $r$-nuclei along with
some CPR nuclei from a source (low-mass core-collapse SNe and AIC
events) producing no elements between N and Ge,

$L$ --- the light $r$-nuclei and CPR nuclei from another source 
(Fe core-collapse SNe from progenitors of $\sim 12$--$25\,M_\odot$)
producing the elements between N and Ge including Fe,

$P$ --- a universal prompt inventory from the first very massive stars 
(VMSs).

The prompt inventory ($P$ inventory) in our original papers on the
LEGO-block model \citep{qw01,qw02} was motivated by the apparent jump 
in Ba and Eu abundances at [Fe/H]~$\sim -3$ shown by the data available
then, which suggested that in early epochs VMSs prevailed producing 
Fe and elements of lower atomic numbers but no heavy $r$-nuclei.  
The element Eu that is almost exclusively of $r$-process origin was 
used to serve as an index for contributions from the $H$ source. 
The element Fe was considered to be exclusively produced by the $L$ 
source at [Fe/H] exceeding the value [Fe/H]$_P$ for the $P$ inventory 
and used as an index for $L$ 
contributions. 

\begin{figure}
\begin{center}
A colorful photo showing many Lego blocks.
\end{center}
\caption{Early major stellar sources for the elements in the ISM
(and the intergalactic medium) after the initial contributions from 
the big bang. These are very massive first generation stars (VMSs)
and core-collapse SNe [mostly Type II SNe (SNe II)]. There would
not have been sufficient time for SNe Ia to occur at the early 
epochs. Two types of SNe II are considered with SNe II($H$)
producing heavy $r$-nuclei but no Fe and SNe II($L$) producing
Fe and related elements. Some intermediate-mass stars may also
have been involved (e.g., in producing heavy $r$-nuclei from
AIC events in binary systems). This simple approach is then used
to model stellar abundances for [Fe/H]~$\lesssim -1.5$.}
\label{fig-bucket}
\end{figure}

This approach led to the following rule for the abundance of element E 
relative to H for arbitrary stars formed from the ISM (not in binary systems
experiencing local $s$-process contamination) at [Fe/H]~$\lesssim -1.5$:
\begin{equation}
\left(\frac{\rm E}{\rm H}\right)=\left(\frac{\rm E}{\rm H}\right)_P+
\left(\frac{\rm E}{\rm Eu}\right)_H\left(\frac{\rm Eu}{\rm H}\right)+
\left(\frac{\rm E}{\rm Fe}\right)_L\left[\left(\frac{\rm Fe}{\rm H}\right)
-\left(\frac{\rm Fe}{\rm H}\right)_P\right].
\label{eq-lego}
\end{equation}
This rule was, of course, also applicable to the solar abundances
corrected for the $s$-process contributions from AGB stars and
the contributions to the Fe group elements from SNe Ia. In particular,
the solar $r$-abundance of E was calculated as
\begin{eqnarray}
\left(\frac{\rm E}{\rm H}\right)_{\odot,r}&=&
\left(\frac{\rm E}{\rm H}\right)_\odot(1-\beta_{\odot,s})\\
&=&\left(\frac{\rm E}{\rm H}\right)_P+
\left(\frac{\rm E}{\rm Eu}\right)_H\left(\frac{\rm Eu}{\rm H}\right)_{\odot,r}+
\left(\frac{\rm E}{\rm Fe}\right)_L\left[\left(\frac{\rm Fe}{\rm H}\right)_{\odot,L}
-\left(\frac{\rm Fe}{\rm H}\right)_P\right],
\label{eq-lego2}
\end{eqnarray}
where $\beta_{\odot,s}$ is the fraction of E in the Sun produced by the
net $s$-process and (Fe/H)$_{\odot,L}$ is the 
$L$ contribution to the solar Fe inventory. 

It follows that if (E/H)$_P$, (E/Eu)$_H$, and (E/Fe)$_L$ are known for 
all the elements, then the value of (E/H) for a metal-poor
star is determined by its
(Eu/H) and (Fe/H). For elements of the Fe group and lower atomic 
numbers that receive no $H$ contributions, 
the (E/H)$_P$ values were taken from the data on stars with 
[Fe/H]~$\sim -4$ to $-3$ and scaled to [Fe/H]$_P=-3$, and the 
(E/Fe)$_L$ values were calculated from Equation~(\ref{eq-lego}) using 
the data on a star with [Fe/H]~$=-2$. It turned out that for these 
elements, the $P$ inventory was almost identical to the $L$-yield pattern,
except for small but
significant shifts in some of the Fe group elements
\citep{qw02}. For the heavy $r$-nuclei produced only by the $H$ source, 
the (E/Eu)$_H$ values were
taken to be the same as those for the solar $r$-pattern. For the CPR
nuclei, the (E/H)$_P$ and (E/Eu)$_H$ were calculated from 
Equation~(\ref{eq-lego}) using the data on two stars with 
[Fe/H]~$\approx -3$ but high enrichments of heavy 
$r$-nuclei, and the (E/Fe)$_L$ values were calculated from 
Equation~(\ref{eq-lego2}) using
(Fe/H)$_P=10^{-3}{\rm (Fe/H)}_\odot$,
(Fe/H)$_{\odot,L}={\rm (Fe/H)}_\odot/3$, and the
(E/H)$_{\odot,r}$ values from \citet{arl}.

\begin{figure}
\vskip -1cm
\begin{center}
\includegraphics*[scale=0.25]{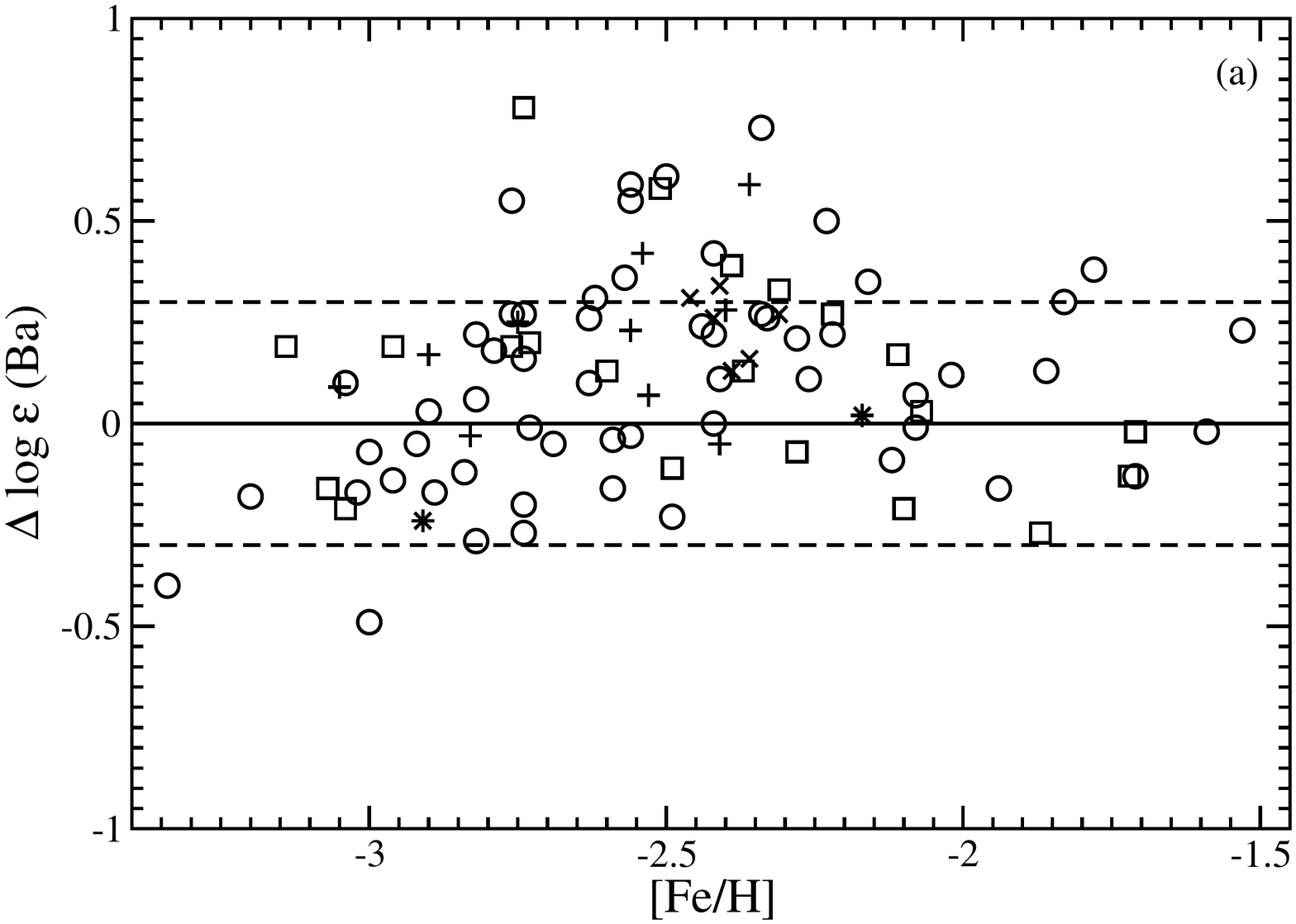}
\vskip -0.3cm
\includegraphics*[scale=0.25]{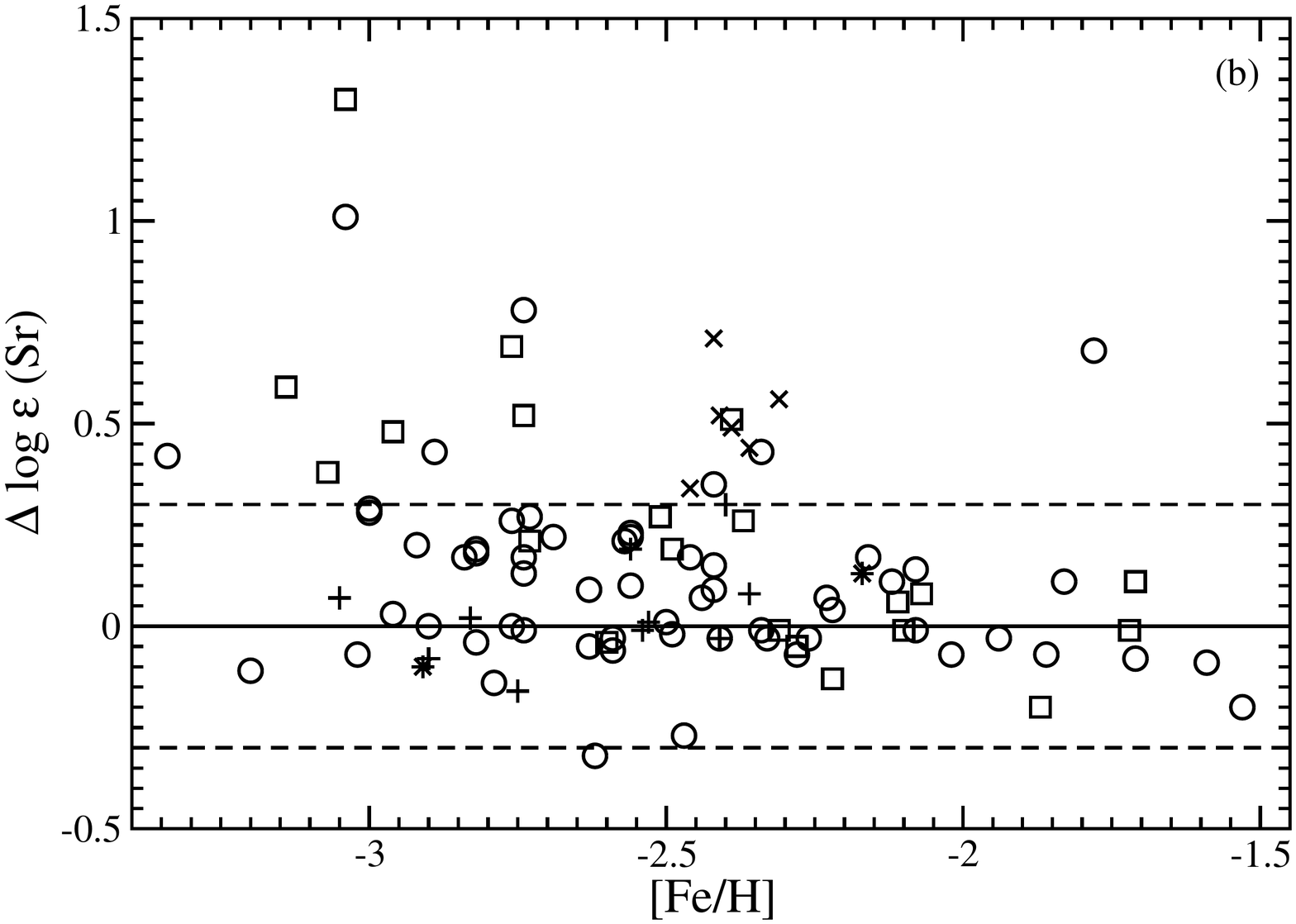}
\vskip -0.3cm
\includegraphics*[scale=0.25]{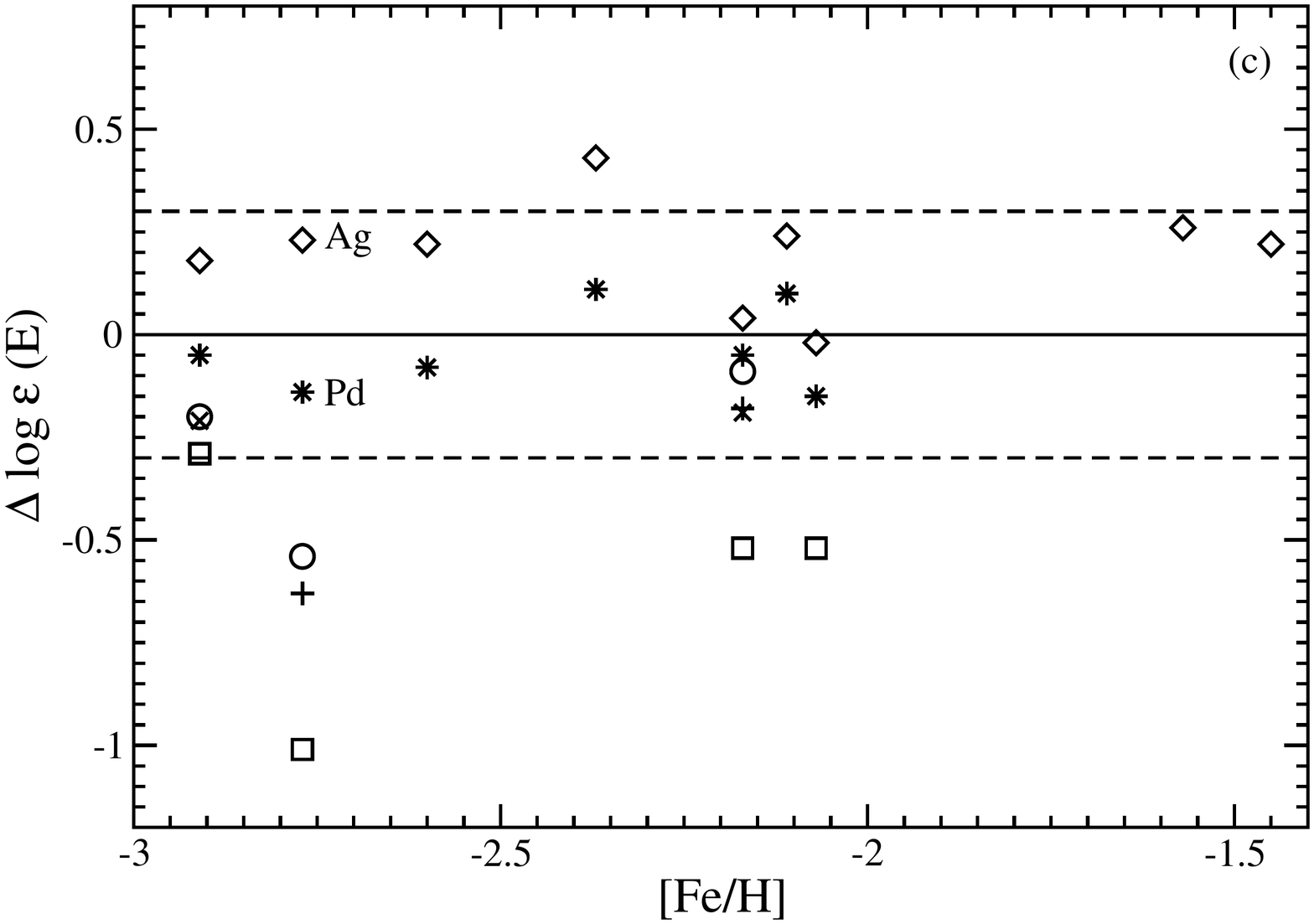}
\end{center}
\caption{Comparison of the original ``LEGO-block'' model with
the data for a large sample of stars.
(a) The difference between the calculated abundance of Ba
and the observed value is shown in terms of
$\Delta\log\epsilon({\rm Ba})\equiv
\log\epsilon_{\rm cal}({\rm Ba})-\log\epsilon_{\rm obs}({\rm Ba})$
as a function of [Fe/H]. The calculation uses only the observed
Eu and Fe
abundances. The symbols represent the following
data sets: \citet{jb02} (squares), \citet{aok05} (pluses),
\citet{he05} (circles), \citet{ots06} (crosses), 
\citet{hill} and \citet{iv06} (asterisks).
(b) Same as (a) but for the CPR element
Sr. (c) Comparison between the model
and data on the other 
CPR elements are shown. The limited data available
are taken from \citet{craw,hill,cow02,jb02,hon06,iv06}.
Only in (c) the symbols represent the CPR elements:
Nb (squares), Mo (pluses), Ru (circles), Rh (crosses),
Pd (asterisks), and Ag (diamonds). It can be seen from these
examples that the elemental abundances in a metal-poor star are 
often well estimated from the model using only the Eu and Fe 
abundances to identify the contributions from different
sources to the star. However, there are some serious 
discrepancies. In particular, the attribution of some Ba 
production to the $L$ source causes the large differences
for some stars shown in (a). This attribution is
eliminated in the upgraded model (see Section~\ref{sec-hl}
and Fig.~\ref{fig-baeu-fe}).}
\label{fig-phl}
\end{figure}

Using the $P$ inventory and the $H$ and $L$-yield patterns obtained by
the above procedure, we made a direct comparison between calculated 
and observed abundances for a large number of metal-poor stars
\citep{qw01,qw02}. Here we show in Figure~\ref{fig-phl} the comparison
for the stars with [Fe/H]~$\lesssim-1.5$ from a new set of
observations \citep{hill,jb02,aok05,he05,ots06,iv06}.
The results for Ba as a representative of the heavy $r$-elements are 
shown in terms of $\Delta\log\epsilon({\rm Ba})\equiv
\log\epsilon_{\rm cal}({\rm Ba})-\log\epsilon_{\rm obs}({\rm Ba})$ 
as a function of [Fe/H] in Figure~\ref{fig-phl}a. It can be seen that the
difference between the calculated and observed abundances lies 
within $\pm 0.3$ dex of zero for most of the stars. There are some clear 
outliers. The results for Sr as a typically observed CPR element are shown 
in Figure~\ref{fig-phl}b. For the CPR elements Nb, Mo, Ru, Rh, Pd, and Ag, 
there are only limited data available for a relatively small number of 
stars \citep{craw,hill,cow02,jb02,hon06,iv06}. 
The results for all these elements are
shown in Figure~\ref{fig-phl}c. As in the case of Ba, the agreement 
between the model and data for the CPR elements is reasonable for
most stars but with some extreme outliers. In this version of the LEGO-block
model, we have attributed some Ba production to the $L$ source 
\citep{qw01,qw02}.

As shown above, the predictive power for our simple LEGO-block model is 
quite surprising. This allowed us to win some bottles of wine of varying 
quality from some observer friends with whom we made wagers
\citep{hill,norb}. Further comparison between this model and observations 
have been made by other workers [e.g., \citet{ful,ven}]. We also note that
our original paper on this model \citep{qw01} showed that a significant 
revision to the solar $s$ and $r$-abundances of Sr, Y, Zr, and Ba as
given by \citet{kap} and \citet{arl} was 
required in order to obtain good agreement between the model and 
the data available then. This revision is reasonable as 
$\beta_{\odot,s}\sim 1$ for these elements and their solar
$r$-abundances derived from subtracting the $s$-process contributions 
are subject to large uncertainties. Some years later a study
of Galactic chemical evolution (GCE) by \citet{tra} confirmed the need 
for such a revision and obtained essentially the same quantitative 
results as ours (but without citation). These authors attributed the required 
increase in the $r$-process contributions to Sr, Y, and Zr to a new type of
stellar source. This attribution is, from our point of view, quite
unnecessary.

In all of our discussion we have not treated the stars with exceptionally
low [Fe/H] values of $<-5$ discovered by \citet{norb02} and
\citet{fre05}. It is clear that our model
does not explain these observations and we have deferred considering 
them within the framework of our study. These 
exceptionally metal-poor stars certainly are objects of
intense theoretical and observational interest and require understanding.
They are exceptions that do not fit into the three-component model
outlined here or its successor to be discussed in the following section. 
Some plausible scenarios for the origins of the abundance patterns in 
these exceptionally metal-poor stars have been discussed by \citet{iwa05}.

\section{An upgraded LEGO-block model}
\subsection{New yield patterns of $H$ and $L$ sources}
\label{sec-hl}
\begin{figure}
\begin{center}
\includegraphics*[scale=0.4]{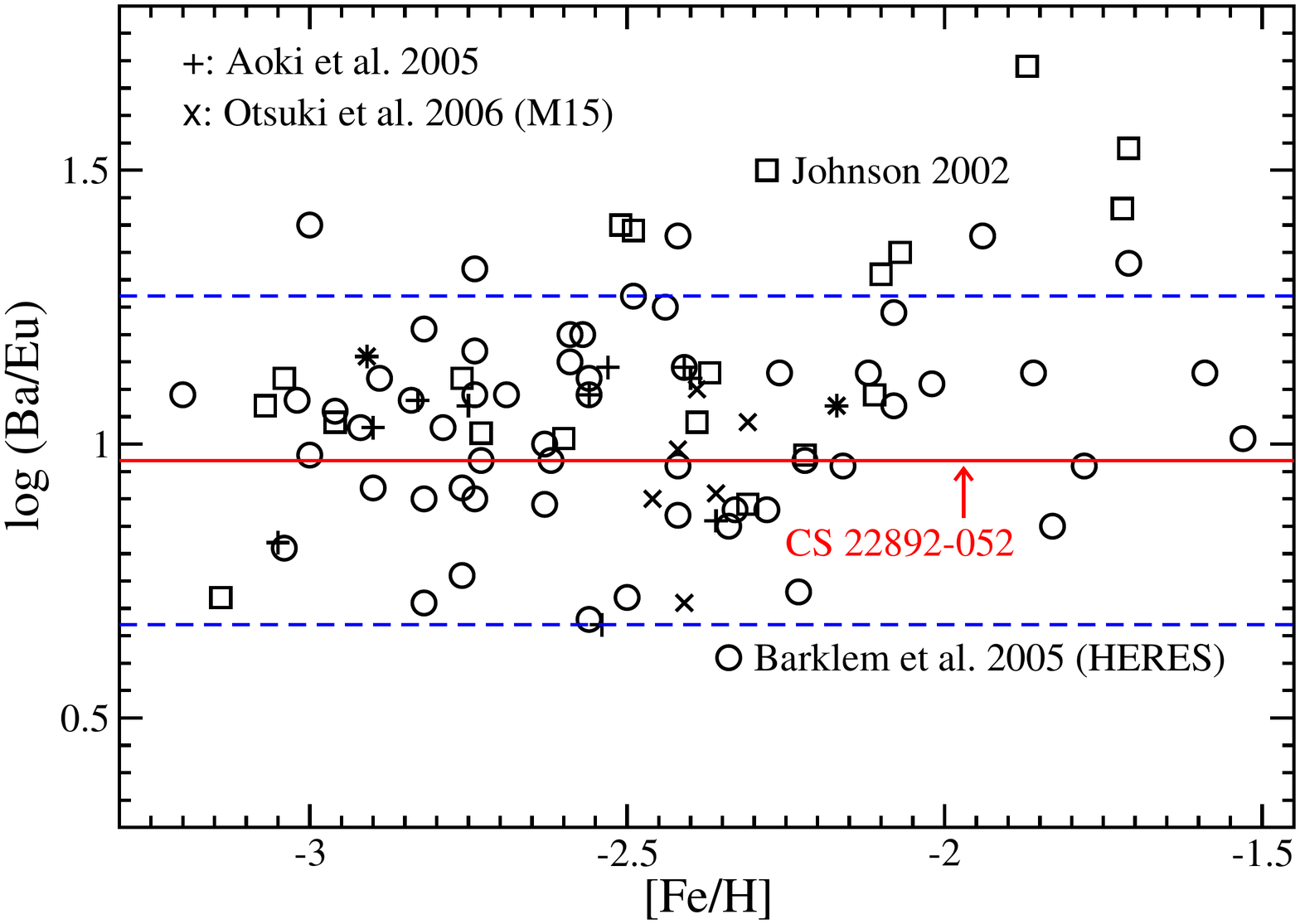}
\end{center}
\caption{Data [squares: \citet{jb02}; pluses: \citet{aok05};
circles: \citet{he05}; crosses: \citet{ots06}; asterisks: 
\citet{hill,iv06}] on $\log({\rm Ba/Eu})$ as a function of [Fe/H]
for a large sample of stars. Note that the $\log({\rm Ba/Eu})$
values are clustered around that for the solar $r$-pattern as
represented by CS~22892--052. These values are typically far below
$\log({\rm Ba/Eu})_\odot=1.7$. There are clearly some outliers.
These results show that Ba and Eu in the regime of
[Fe/H]~$\lesssim -1.5$ are governed by $r$-process contributions.}
\label{fig-baeu-fe}
\end{figure}

With the intense interest in elemental abundances in low-metallicity stars
there is a great increase in high-quality data available as compared with
the data sets used by us earlier to develop the original LEGO-block model. 
From the new results it is found that the Ba/Eu ratio
is almost constant between [Fe/H]~$\sim -3.5$ and $-2$ 
[\citet{hill,jb02,aok05,he05,ots06,iv06}; see Fig.~\ref{fig-baeu-fe}]. For the
yield patterns used in the original LEGO-block model, 
it was assumed that significant Ba production
occurred in the $L$ source. This is no longer valid and both
Ba and Eu must be assigned solely to the $H$ source. 

Further, in
proposing the $P$ inventory VMSs were assumed. As noted in the
preceding section, the $P$ inventory for the elements of the Fe group 
and lower atomic numbers was almost identical to the $L$-yield pattern
except for small but
significant shifts in some of the Fe group elements
[see Fig.~4 of \citet{qw02}]. The $P$
inventory then mostly affects estimates for the abundances of
Sr, Y, and Zr. The $P$ inventory for these elements was calculated
from the data on two stars with [Fe/H]~$\approx -3$ but high
enrichments of heavy $r$-nuclei. However, in contrast to the elements
of the Fe group and lower atomic numbers, the $P$ inventory of
which is in proportion to (Fe/H) for [Fe/H]~$\sim -4$ to $-3$,
the contributions to Sr attributed to the $P$ inventory do not follow
this scaling. The general nature of VMS contributions to metal-poor 
stars also changed as observations of the intergalactic medium
(IGM) showed that the cosmic mean abundance of Si is quite large with
[Si/H]$_{\rm IGM}=-2$ accompanied by a rather high value of
[Si/C]$_{\rm IGM}=0.77$ \citep{agu}. If we assume that VMSs and
SN-driven galactic outflows each contributed 50\% of the Si 
to account for the high [Si/C]$_{\rm IGM}$, 
then this would have resulted in [Fe/H]$_{\rm IGM}\sim -2.3$ 
within a few Gyr after the big bang \citep{qw05}.
This result on the average Fe abundance in the early IGM would not be 
changed if only galactic outflows contributed the metals. It follows that
the low-metallicity stars of concern here, 
especially those with [Fe/H]~$\lesssim -2.3$,
cannot be representative of the average IGM and the
formation of such stars must have been fed by the infall of gas from
a ``metal-poor'' IGM into the Galaxy. 
As a result, we cannot consider a $P$ inventory from
VMSs as a component in pursuing the LEGO-block model. 

\begin{table}
\caption{Yield patterns of heavy $r$-nuclei for the 
$H$ and $L$ sources in the upgraded LEGO-block model}
\smallskip
\begin{tabular*}{\hsize}{@{\extracolsep{\fill}}*{6}l@{}}
\hline
\hline
Element&$\log({\rm E/Eu})_H$&$\log({\rm E/Fe})_L$&
Element&$\log({\rm E/Eu})_H$&$\log({\rm E/Fe})_L$\\
\hline
Ba&0.96&$-\infty$&Tm&$-0.45$&$-\infty$\\
La&0.26&$-\infty$&Yb&0.26&$-\infty$\\
Ce&0.46&$-\infty$&Lu&$-0.50$&$-\infty$\\
Pr&$-0.03$&$-\infty$&Hf&$-0.13$&$-\infty$\\
Nd&0.58&$-\infty$&Ta&$-0.88$&$-\infty$\\
Sm&0.28&$-\infty$&W&$-0.20$&$-\infty$\\
Gd&0.48&$-\infty$&Re&$-0.27$&$-\infty$\\
Tb&$-0.22$&$-\infty$&Os&0.82&$-\infty$\\
Dy&0.56&$-\infty$&Ir&0.85&$-\infty$\\
Ho&$-0.05$&$-\infty$&Pt&1.14&$-\infty$\\
Er&0.35&$-\infty$&Au&0.28&$-\infty$\\
\hline
\end{tabular*}
\label{tab-h}
\end{table}

Based on the above discussion, we have changed the model to 
eliminate the $P$ inventory and only consider the $H$ and $L$ 
components. In this ``upgraded'' LEGO-block model, the $H$
source is solely responsible for all the heavy $r$-nuclei (see above
discussion on the Ba/Eu ratio and Fig.~\ref{fig-baeu-fe}).
The (E/Eu)$_H$ values for these nuclei are taken from the 
corresponding part of the solar $r$-pattern calculated by \citet{arl}.
The corresponding $\log({\rm E/Eu})_H$ values are given in
Table~\ref{tab-h}.
For the CPR nuclei, we obtain their $H$ and $L$-yield patterns
from the data on CS~22892--052 \citep{sn03} and HD~122563
\citep{hon06}. These two stars have similar Fe abundances but
the former star is highly enriched in Eu while the latter has
an Eu abundance lower by a factor of 66 (see Table~\ref{tab-hl}). 
Clearly, the abundances of CPR nuclei in HD~122563 are dominated
by $L$ contributions. So we use the data on this star to calculate the 
(E/Fe)$_L$ values for these nuclei as
\begin{equation}
\log\left(\frac{\rm E}{\rm Fe}\right)_L=\log\epsilon_{\rm HD}({\rm E})-
\log\epsilon_{\rm HD}({\rm Fe}),
\end{equation}
where $\log\epsilon_{\rm HD}({\rm E})$ refers to the data on
HD~122563. The calculated $\log({\rm E/Fe})_L$ values are given
in the fifth column of Table~\ref{tab-hl}. 
Based on these values and the Fe abundance of CS~22892--052, 
we find that the $L$ contributions to the abundances of CPR nuclei 
in this star are negligible. We then use the data on CS~22892--052 
to calculate the (E/Eu)$_H$ values for these nuclei as
\begin{equation}
\log\left(\frac{\rm E}{\rm Eu}\right)_H=\log\epsilon_{\rm CS}({\rm E})-
\log\epsilon_{\rm CS}({\rm Eu}),
\end{equation}
where $\log\epsilon_{\rm CS}({\rm E})$ refers to the data on
CS~22892--052. The calculated $\log({\rm E/Eu})_H$ values are given
in the fourth column of Table~\ref{tab-hl}.

\begin{table}
\caption{Input data and yield patterns of CPR nuclei for the 
$H$ and $L$ sources in the upgraded LEGO-block model}
\smallskip
\begin{tabular*}{\hsize}{@{\extracolsep{\fill}}*{5}l@{}}
\hline
\hline
&$\log\epsilon_{\rm CS}({\rm E})$&
$\log\epsilon_{\rm HD}({\rm E})$&&\\
Element&CS~22892--052&HD~122563&$\log({\rm E/Eu})_H$&
$\log({\rm E/Fe})_L$\\
\hline
Fe&4.40&4.74&$-\infty$&0\\
Eu&$-0.95$&$-2.77$&0&$-\infty$\\
Sr&0.46&$-0.11$&1.41&$-4.85$\\
Y&$-0.42$&$-0.93$&0.53&$-5.67$\\
Zr&0.24&$-0.28$&1.19&$-5.02$\\
Nb&$-0.80$&$-1.48$&0.15&$-6.22$\\
Mo&$-0.40$&$-0.87$&0.55&$-5.61$\\
Ru&0.08&$-0.86$&1.03&$-5.60$\\
Rh&$-0.55$&$<-1.20$&0.40&$<-5.94$\\
Pd&$-0.29$&$-1.36$&0.66&$-6.10$\\
Ag&$-0.88$&$-1.88$&0.07&$-6.62$\\
\hline
\end{tabular*}
\label{tab-hl}
\end{table}

\begin{figure}
\vskip -1cm
\begin{center}
\includegraphics*[scale=0.4]{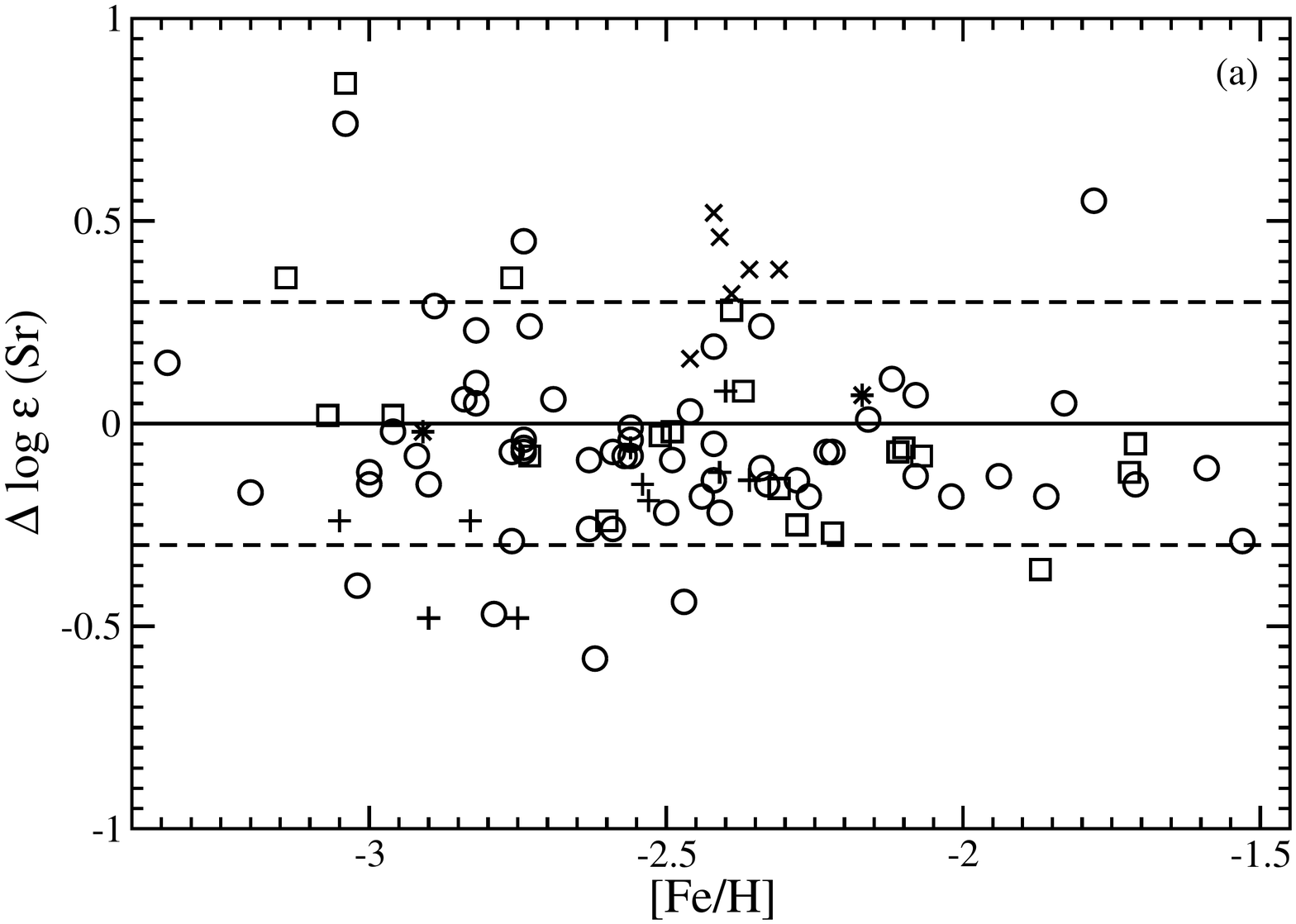}
\vskip -0.5cm
\includegraphics*[scale=0.4]{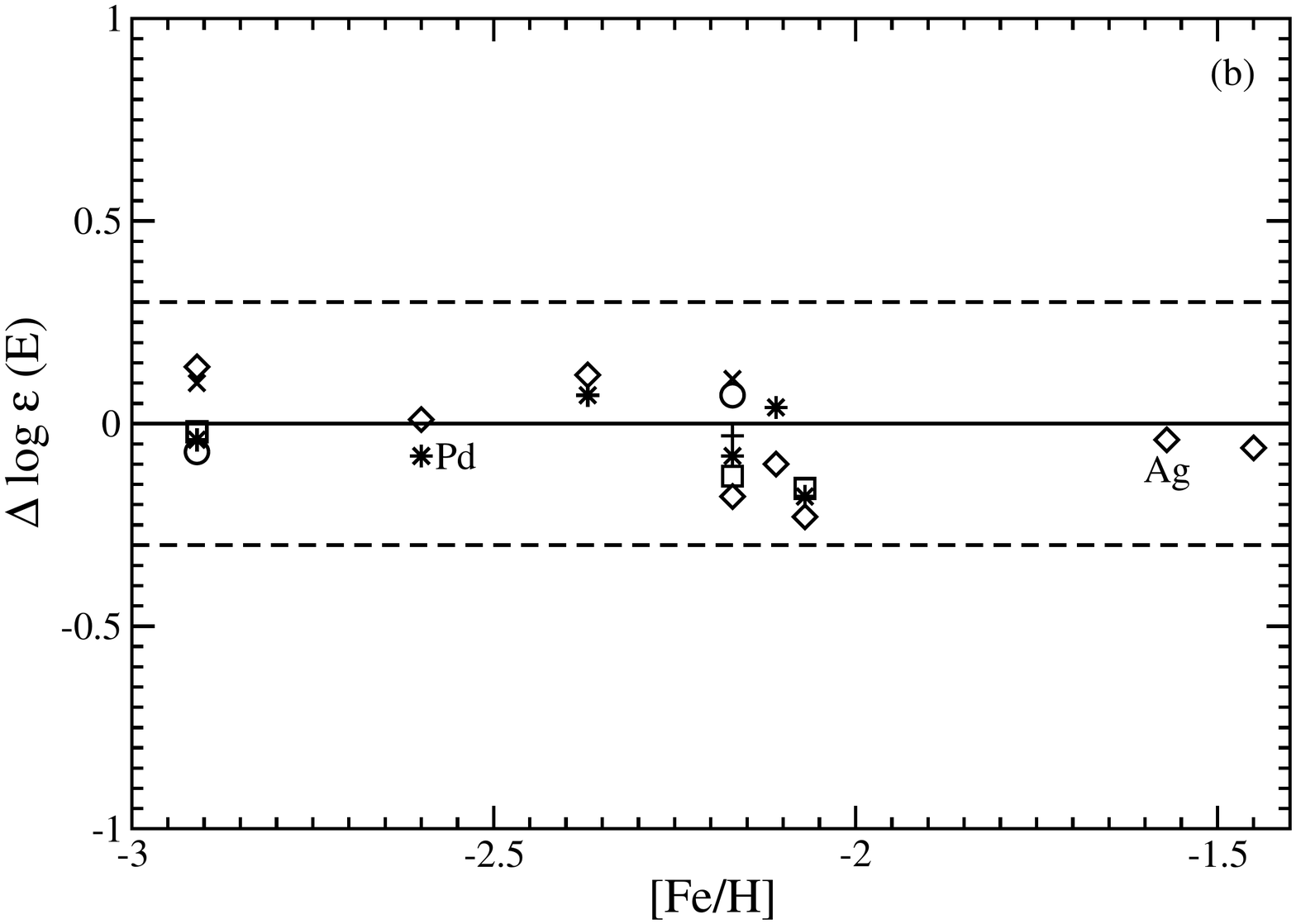}
\end{center}
\caption{Comparison of the ``upgraded'' LEGO-block model with
the data for a large sample of stars. (a) The difference 
between the calculated abundance of the CPR element Sr
and the observed value is shown in terms of
$\Delta\log\epsilon({\rm Sr})\equiv
\log\epsilon_{\rm cal}({\rm Sr})-\log\epsilon_{\rm obs}({\rm Sr})$
as a function of [Fe/H]. The calculation uses only the observed
Eu and Fe abundances. The symbols represent the following
data sets: \citet{jb02} (squares), \citet{aok05} (pluses),
\citet{he05} (circles), \citet{ots06} (crosses), 
\citet{hill} and \citet{iv06} (asterisks).
(b) Comparison between the model
and data on the other CPR elements are shown. 
The limited data available
are taken from \citet{craw,hill,cow02,jb02,iv06}.
The symbols represent the CPR elements:
Nb (squares), Mo (pluses), Ru (circles), Rh (crosses),
Pd (asterisks), and Ag (diamonds).
It can be seen from these examples that the elemental 
abundances in a metal-poor star are 
rather well estimated from the model using only the Eu 
and Fe abundances to identify the contributions from different
sources to the star.}
\label{fig-hl}
\end{figure}

Using the $H$ and $L$-yield patterns in Table~\ref{tab-hl}, we
calculate the abundance of an element E relative to H in a
star with [Fe/H]~$\lesssim-1.5$ as
\begin{equation}
\left(\frac{\rm E}{\rm H}\right)=
\left(\frac{\rm E}{\rm Eu}\right)_H\left(\frac{\rm Eu}{\rm H}\right)+
\left(\frac{\rm E}{\rm Fe}\right)_L\left(\frac{\rm Fe}{\rm H}\right).
\label{eq-lego-hl}
\end{equation}
The comparison between the calculated abundances of the
CPR nuclei and the data is shown in Figure~\ref{fig-hl}
for the same set of stars shown in Figure~\ref{fig-phl}.
It can be seen from Figure~\ref{fig-hl}
that the agreement is excellent for Nb, Mo,
Ru, Rh, Pd, and Ag for the limited data available. The agreement
for Sr is also much improved compared with Figure~\ref{fig-phl}b
although the few outliers still exist.

\subsection{Absolute yields of CPR nuclei}
\label{sec-ycpr}
Based on the comparison shown in Figure~\ref{fig-hl}, we consider 
that the upgraded LEGO-block model with the $H$ and $L$-yield 
patterns given in Tables~\ref{tab-h} and \ref{tab-hl} provides 
a good description of the 
observations on CPR nuclei. 
Further, we can use these patterns to estimate the
absolute yields of CPR nuclei from each $H$ and $L$ event.
As both $H$ and $L$ events involve the formation of neutron stars
that have neutrino-driven winds producing CPR nuclei
(see Section~\ref{sec-wind}), the yields of these nuclei should
be comparable in both cases.
We have argued that the $L$ source is Fe core-collapse SNe from
progenitors of $\sim 12$--$25\,M_\odot$. These SNe have a typical
Fe yield of $Y_L({\rm Fe})\sim 0.1\,M_\odot$ per event [e.g.,
\citet{ww95}]. With (Sr/Fe)$_L=10^{-4.85}$ (see Table~\ref{tab-hl}),
this gives an $L$-yield $Y_L({\rm Sr})$ of Sr as
\begin{equation}
Y_L({\rm Sr})=\left(\frac{\rm Sr}{\rm Fe}\right)_L
\left(\frac{A_{\rm Sr}}{A_{\rm Fe}}\right)Y_L({\rm Fe})
\sim 2.2\times 10^{-6}\,M_\odot,
\end{equation}
where $A_{\rm Sr}\approx 88$ and $A_{\rm Fe}\approx 56$ are
the mass numbers of Sr and Fe, respectively.
We have also argued that the $H$ source is low-mass core-collapse
SNe from progenitors of $\sim 8$--$11\,M_\odot$ and AIC events.
We estimate the Eu yield from an $H$ event by assuming that
over the Galactic history of $\sim 10$~Gyr, the occurrence of
$\sim 10^8$ such events at a rate of $\sim (100\,{\rm yr})^{-1}$
enriched $\sim 10^{10}\,M_\odot$ of gas with a solar mass fraction
$X_{\odot}({\rm Eu})=3.75\times 10^{-10}$ of Eu \citep{and}.
This gives $Y_H({\rm Eu})\sim 3.75\times 10^{-8}\,M_\odot$.
Using (Sr/Eu)$_H=10^{1.41}$ (see Table~\ref{tab-hl}), we obtain
\begin{equation}
Y_H({\rm Sr})=\left(\frac{\rm Sr}{\rm Eu}\right)_H
\left(\frac{A_{\rm Sr}}{A_{\rm Eu}}\right)Y_H({\rm Eu})
\sim 5.6\times 10^{-7}\,M_\odot,
\end{equation}
where $A_{\rm Eu}\approx 152$ is the mass number of Eu.
It can be seen that the yields of CPR nuclei are indeed
comparable for
an $H$ and $L$ event. This is entirely consistent with these nuclei
being produced in the neutrino-driven wind whenever a neutron star 
is made in any core-collapse SNe. Thus, from
the reasonable accord between
the predictions of the model and the available data over
a wide range of [Fe/H] as shown in Figure~\ref{fig-hl},
it appears that the production of CPR elements per neutron star
forming event is rather constant with yields as estimated above.

Using the $H$-yield of Eu estimated above, we can provide some
quantitative description of how the extreme $r$-enhancements in
HE~2148--1247 discussed in Section~\ref{sec-rs} could have occurred.
Taking the radius of this star to be $R$ and the distance from
its binary companion to be $d$ when the companion underwent
AIC or O-Ne-Mg core collapse
to produce the heavy $r$-nuclei, we estimate that 
a fraction $\sim (R/2d)^2$ of the $r$-process ejecta would be
intercepted by HE~2148--1247. 
The resulting Eu enrichment would be
\begin{equation}
\left(\frac{\rm Eu}{\rm H}\right)\sim\frac{Y_H({\rm Eu})/A_{\rm Eu}}
{X_{\rm H}M_{\rm dil}}\left(\frac{R}{2d}\right)^2,
\end{equation}
where $X_{\rm H}\approx 0.76$ is the mass fraction of H and
$M_{\rm dil}$ is the dilution mass for the intercepted material.
Using $Y_H({\rm Eu})\sim 3.75\times10^{-8}\,M_\odot$,
$M_{\rm dil}\sim 0.1\,M_\odot$, $R\sim R_\odot$, and
$\log\epsilon({\rm Eu})=0.17$ \citep{cohen},
we obtain $d\sim 23R_\odot$ from the above equation. 
This appears reasonable
and would be consistent with the observational indication that
HE~2148--1247 is now in a long-period binary system
\citep{cohen} if its orbit was widened greatly
following the AIC or O-Ne-Mg core-collapse event due to, e.g.,
the kick imparted to the neutron star remnant. In either case,
there should be a common-envelope phase during the RGB and AGB
evolution of the original primary star.

Finally, we note that the production of CPR nuclei and
heavy $r$-nuclei by $H$ events may make important contributions
to the inventory of some radioactive nuclei in the ESS.
As $^{182}$Hf is exclusively made in $H$ events 
and its inventory in the ESS is measured, we first discuss
$^{182}$Hf. Consider the simple model 
where uniform production of $^{182}$Hf had lasted for
a period $T_{\rm UP}$ and was then followed by
an interval $\Delta_{182}$ of no production prior to the
formation of the solar system (see Section~\ref{sec-intro}).
This would result in
\begin{equation}
\left(\frac{^{182}{\rm Hf}}{^{182}{\rm W}_r}\right)_{\rm ESS}=
\left(\frac{\bar\tau_{182}}{T_{\rm UP}}\right)\exp\left(-\frac{\Delta_{182}}
{\bar\tau_{182}}\right),
\end{equation} 
where $^{182}{\rm W}_r$ is the part of $^{182}{\rm W}$ in
the solar system contributed by the $r$-process.
Using $(^{182}{\rm Hf}/{^{180}{\rm Hf}})_{\rm ESS}=10^{-4}$
[\citet{yhf,khf}], 
$(^{180}{\rm Hf}/{^{182}{\rm W}_r})_\odot=0.0541/0.019$
\citep{and,arl}, and $\tau_{182}=12.8$~Myr, we obtain
$\Delta_{182}\sim 19.2$~Myr for $T_{\rm UP}\sim 10$~Gyr.

As mentioned in Section~\ref{sec-wind}, the neutrino-driven wind
makes $^{107}$Mo, which decays through $^{107}$Pd producing
$^{107}$Ag. The model for production of $^{182}{\rm Hf}$ by
$H$ events discussed above gives
\begin{equation}
\left(\frac{^{107}{\rm Pd}}{^{182}{\rm Hf}}\right)_{{\rm ESS},H}=
\left(\frac{P_{107}}{P_{182}}\right)_H
\left(\frac{\bar\tau_{107}}{\bar\tau_{182}}\right)
\exp\left(-\frac{\Delta_{182}}{\bar\tau_{107}}+
\frac{\Delta_{182}}{\bar\tau_{182}}\right),
\end{equation} 
where $(P_{107}/P_{182})_H$ is the (number) production ratio
of $^{107}$Pd to $^{182}$Hf for $H$ events and can be
estimated as 
$\sim(1/2)({\rm Ag/Eu})_H({\rm Eu}/^{182}{\rm W})_{\odot,r}$
(the factor of 1/2 is used as Ag has an additional isotope
$^{109}$Ag). Taking $({\rm Ag/Eu})_H=10^{0.07}$ 
(see Table~\ref{tab-hl}), 
$({\rm Eu}/^{182}{\rm W})_{\odot,r}=0.0917/0.019$
\citep{arl}, and $\bar\tau_{107}=9.4$~Myr,
we obtain $(^{107}{\rm Pd}/{^{182}{\rm Hf}})_{{\rm ESS},H}
\sim 1.21$. This gives 
\begin{equation}
\left(\frac{^{107}{\rm Pd}}{^{108}{\rm Pd}}\right)_{{\rm ESS},H}=
\left(\frac{^{107}{\rm Pd}}{^{182}{\rm Hf}}\right)_{{\rm ESS},H}
\left(\frac{^{182}{\rm Hf}}{^{180}{\rm Hf}}\right)_{\rm ESS}
\left(\frac{^{180}{\rm Hf}}{^{108}{\rm Pd}}\right)_\odot
\sim 1.78\times 10^{-5},
\end{equation}
where $(^{180}{\rm Hf}/{^{108}{\rm Pd}})_\odot=0.0541/0.368$
\citep{and} is used. The above result is close to
the measured value of 
$(^{107}{\rm Pd}/{^{108}{\rm Pd}})_{\rm ESS}=2\times 10^{-5}$
[\citet{kw}, see also the recent review by \citet{wess}].
Note that although $L$ events also produce $^{107}$Pd,
their contributions to the $^{107}$Pd in the ESS are
negligible as
the interval between the last $L$ contributions and
the formation of the solar system is $\sim 72$~Myr
based on the $^{129}$I data (see Section~\ref{sec-intro}).
Therefore, it appears that the $H$ events were responsible
for both $^{107}$Pd and $^{182}$Hf in the ESS. This eliminates
the conflict with the data on $^{129}$I, which is not 
produced by the $H$ events. However, the broader problems and 
conflicts regarding $^{53}$Mn, $^{60}$Fe, and $^{247}$Cm 
would not be resolved [see the recent review by \citet{wess}].

\subsection{Other models of Galactic chemical evolution}
In the LEGO-block model presented here we have restricted our
analysis to a regime where neither low-mass AGB stars nor SNe Ia
had contributed significant material to the ISM. The model claims
to predict the abundances of all the other elements for any star 
(without experiencing $s$-process contamination) in this regime 
given the Eu and Fe abundances of the star. 
It does not seek to provide a detailed or global study
of GCE. Models of GCE require 
assumptions on the rates of formation for a wide range of stellar
sources, their yields, and the amount of dilution by mixing with 
the ISM over time. Many of these models successfully demonstrate 
general trends over most of Galactic history [e.g., 
\citet{maco,tww,tra99}]. Some models describe the
stochastic process of enrichments and can provide a statistical
description of abundances in stars [e.g.,
\citet{iw99,rai,arg}]. To our knowledge, no GCE
models are capable of predicting the abundances of all the other
elements in an individual star from its Eu and Fe abundances. 
We are still open to wagers on the predictions of our LEGO-block
model. 

\section*{Acknowledgments}
The authors thank Hans for many stimulating and illuminating 
conversations and for being supportive of a strange new approach
to the science at hand. We thank Alex Heger and Ken Nomoto for
educating us on presupernova evolution and for providing us with
Figures~\ref{fig-fe-core} and \ref{fig-onemg-core}, respectively, 
John Lattanzio for pointing us to literature on AGB stars, and
Judy Cohen and Wako Aoki for permitting us to use 
Figures~\ref{fig-baeu-rs} and \ref{fig-aoki-rs}, respectively, from their 
published works. We also thank Marc Kamionkowski for his
patience with our unsuccessful efforts in writing a more extensive
article with cosmic views on the subject and ask for his forgiveness.
Hopefully, this contribution covers much of what we were to write.
This work was supported in part by the US Department of Energy
under grants DE-FG02-87ER40328 (Y.-Z.Q.) and DE-FG03-88ER13851 
(G.J.W.), Caltech Division Contribution 9176(1122).


\appendix
\section{The parable of the key: a story for Hans}
\label{sec-app}
Y.-Z.Q. was heavily involved with teaching in the fall of 2006. In trying to meet the 
deadline for this volume, G.J.W. traveled to Minneapolis to work on the manuscript. 
Bad weather conditions left him in Denver overnight. To get some rest, he went out 
to a more or less nearby Radisson Hotel and caught a few hours' sleep before
returning to the Denver Airport early next morning for the remainder of the trip to 
Minneapolis. Y.-Z.Q. had arranged for him to stay at the Radisson Hotel near the 
University of Minnesota and met him at the hotel. Upon checking in and getting a 
card key, they went up to the seventh floor to find Room 729. They searched the 
seventh floor intensely but could not find a room with that number. G.J.W. was getting 
quite exhausted so they asked a chambermaid where the room was. She searched 
and could not find it, either. She asked another maid, but to no avail. Much exercised, 
G.J.W. insisted that she call the front desk, but she did not know how to make the call. 
G.J.W. grabbed the phone, got the front desk, and told the clerk of the difficulty. 
It turned out that Room 729 was not the assigned room. (In fact, it did not exist!)
Instead, G.J.W. should go to Room 762. Both Y.-Z.Q. and G.J.W. read a clearly written 
``729'' on the envelope enclosing the key. Nonetheless, off they went to Room 762.
This room actually existed. G.J.W. then took the key from the envelope and promptly 
opened the door to what in fact was his room. 

Upon dropping the luggage in the room, G.J.W. (quite exhausted) and Y.-Z.Q. went 
down to get lunch. G.J.W. was rather ticked off that the clerk had clearly written the 
wrong room number on the key envelope and reported this to the manager. He 
apologized in the presence of Ben the clerk and asked to see the envelope. Upon 
seeing it, he immediately said, ``this envelope and this key do not belong to this hotel!'' 
He then destroyed the key and re-issued a new one. 

Having got no decent sleep, 
G.J.W. forgot to return the key for Room 729 where he stayed at the Radisson Hotel 
in Denver. Being rather disoriented, he randomly pulled the envelope enclosing this 
key out of a pocket and used it as a guide in search of his room in Minneapolis!
Now the question remains, how did the key for a room at the Radisson Hotel in 
Denver open another room at the Radisson Hotel in Minneapolis? Did 
the master key to all Radisson Hotel rooms happen to fall into the personal
possession of G.J.W.? If so, why was he so stupid as to let the manager take it 
away and destroy it? Was this just a random event? 

Maybe this is how we do science. We use what keys we have and try to get into 
different rooms and, maybe, sometimes we get into the correct room by accident. 
G.J.W. still wishes that he did not give up that old key --- however, if the new key is 
the same as the old one, maybe he still has the UNIVERSAL KEY. What luck!!!

Hans would have enjoyed this story.

P.S. Ben got a six-pack of Budweiser for enduring (unjustly) the critical wrath 
of the junior author.
\end{document}